\documentclass[twocolumn]{aastex631} 

\usepackage[T1]{fontenc} 
\usepackage[utf8]{inputenc} 
\usepackage{amssymb}
\usepackage{float}
\usepackage{placeins}

\newcommand{\feh}{\ensuremath{{\rm [Fe/H]}}}
\newcommand{\teff}{\ensuremath{T_{\rm eff}}}

\newcommand{\logg}{\ensuremath{\log{g}}}

\newcommand{\vsini}{\ensuremath{v \sin{i_\star}}}

\newcommand{\rstar}{\ensuremath{{\rm R}_{\star}}}
\newcommand{\mstar}{\ensuremath{{\rm M}_{\star}}}

\newcommand{\rsun}{\ensuremath{{\rm R}_{\odot}}}
\newcommand{\msun}{\ensuremath{{\rm M}_{\odot}}}

\setcitestyle{notesep={ }}

\begin{document}


\title{The Spin-Orbit Alignment of 8 Warm Gas Giant Systems\footnote{Based on observations made with ESO Telescopes at the La Silla Paranal Observatory under programs ID 108.22C0, 109.238M, 110.23Y8, and 113.26KH.}}

\correspondingauthor{Juan I. Espinoza-Retamal}
\email{jiespinozar@uc.cl}

\author[0000-0001-9480-8526]{Juan I. Espinoza-Retamal} 
\affiliation{Instituto de Astrof\'isica, Pontificia Universidad Cat\'olica de Chile, Av. Vicu\~na Mackenna 4860, 782-0436 Macul, Santiago, Chile}
\affiliation{Millennium Institute for Astrophysics, Santiago, Chile}
\affiliation{Anton Pannekoek Institute for Astronomy, University of Amsterdam, Science Park 904, 1098 XH Amsterdam, The Netherlands}

\author[0000-0002-5389-3944]{Andr\'es Jord\'an} 
\affil{Facultad de Ingenier\'ia y Ciencias, Universidad Adolfo Ib\'{a}\~{n}ez, Av. Diagonal las Torres 2640, Pe\~{n}alol\'{e}n, Santiago, Chile}
\affil{Millennium Institute for Astrophysics, Santiago, Chile}
\affil{El Sauce Observatory --- Obstech, Coquimbo, Chile}

\author[0000-0002-9158-7315]{Rafael Brahm} 
\affil{Millennium Institute for Astrophysics, Santiago, Chile}
\affil{Facultad de Ingenier\'ia y Ciencias, Universidad Adolfo Ib\'{a}\~{n}ez, Av. Diagonal las Torres 2640, Pe\~{n}alol\'{e}n, Santiago, Chile}

\author[0000-0003-0412-9314]{Cristobal Petrovich}
\affiliation{Instituto de Astrof\'isica, Pontificia Universidad Cat\'olica de Chile, Av. Vicu\~na Mackenna 4860, 782-0436 Macul, Santiago, Chile}
\affiliation{Millennium Institute for Astrophysics, Santiago, Chile}
\affiliation{Department of Astronomy, Indiana University, Bloomington, IN 47405, USA}

\author[0000-0002-7444-5315]{Elyar Sedaghati} 
\affiliation{European Southern Observatory (ESO), Av. Alonso de Córdova 3107, 763-0355 Vitacura, Santiago, Chile}

\author[0000-0001-7409-5688]{Guðmundur Stefánsson} 
\affil{Anton Pannekoek Institute for Astronomy, University of Amsterdam, Science Park 904, 1098 XH Amsterdam, The Netherlands}

\author[0000-0002-5945-7975]{Melissa J. Hobson} 
\affiliation{Millennium Institute for Astrophysics, Santiago, Chile}
\affiliation{Observatoire de Genève, Département d’Astronomie, Université de Genève, Chemin Pegasi 51b, 1290 Versoix, Switzerland}

\author[0009-0004-8891-4057]{Marcelo Tala Pinto} 
\affiliation{Millennium Institute for Astrophysics, Santiago, Chile}
\affil{Facultad de Ingenier\'ia y Ciencias, Universidad Adolfo Ib\'{a}\~{n}ez, Av. Diagonal las Torres 2640, Pe\~{n}alol\'{e}n, Santiago, Chile}

\author[0000-0003-2186-234X]{Diego J. Muñoz} 
\affiliation{Department of Astronomy and Planetary Science, Northern Arizona University, Flagstaff, AZ 86011, USA}

\author[0009-0009-2966-7507]{Gavin Boyle} 
\affil{El Sauce Observatory --- Obstech, Coquimbo, Chile}
\affil{Cavendish Laboratory, J. J. Thomson Avenue, Cambridge, CB3 0HE, UK}

\author[0000-0002-6477-1360]{Rodrigo Leiva} 
\affiliation{Millennium Institute for Astrophysics, Santiago, Chile}
\affiliation{Instituto de astrofísica de Andalucía, CSIC, Glorieta de la Astronomía s/n, 18008 Granada, Spain}

\author[0000-0001-7070-3842]{Vincent Suc} 
\affiliation{Millennium Institute for Astrophysics, Santiago, Chile}
\affil{Facultad de Ingenier\'ia y Ciencias, Universidad Adolfo Ib\'{a}\~{n}ez, Av. Diagonal las Torres 2640, Pe\~{n}alol\'{e}n, Santiago, Chile}
\affil{El Sauce Observatory --- Obstech, Coquimbo, Chile}

\begin{abstract}

Essential information about the formation and evolution of planetary systems can be found in their architectures---in particular, in stellar obliquity ($\psi$)---as they serve as a signature of their dynamical evolution. Here, we present ESPRESSO observations of the Rossiter-Mclaughlin (RM) effect of 8 warm gas giants, revealing that independent of the eccentricities, all of them have relatively aligned orbits. Our 5 warm Jupiters---WASP-106 b, WASP-130 b, TOI-558 b, TOI-4515 b, and TOI-5027 b---have sky-projected obliquities $|\lambda|\simeq0-10$ deg while the 2 less massive warm Saturns---K2-139 b and K2-329 A b---are slightly misaligned having $|\lambda|\simeq15-25$ deg. Furthermore, for K2-139 b, K2-329 A b, and TOI-4515 b, we also measure true 3D obliquities $\psi\simeq15-30$ deg. We also report a non-detection of the RM effect produced by TOI-2179 b. Through hierarchical Bayesian modeling of the true 3D obliquities of hot and warm Jupiters, we find that around single stars, warm Jupiters are statistically more aligned than hot Jupiters. Independent of eccentricities, 95\% of the warm Jupiters have $\psi\lesssim25$ deg with no misaligned planets, while hot Jupiters show an almost isotropic distribution of misaligned systems. This implies that around single stars, warm Jupiters form in primordially aligned protoplanetary disks and subsequently evolve in a more quiescent way than hot Jupiters. Finally, we find that Saturns may have slightly more misaligned orbits than warm Jupiters, but more obliquity measurements are necessary to be conclusive.


\end{abstract}


\section{Introduction} \label{sec:int}

Almost 6,000 exoplanets have been confirmed, revealing a large diversity of planetary systems. However, ever since the discovery of 51 Pegasi b \citep{Mayor95}, the existence of gas giants in short-period orbits has continued to elude a definitive explanation. Generally, the theoretical channels proposed to explain these types of planets fall into two categories \citep[see][for a review on the possible origins of close-in gas giants]{Dawson18}. One is high-eccentricity (high-$e$) tidal migration, in which cold Jupiters form beyond the iceline and are then launched onto highly eccentric orbits---either through planet-planet scattering \cite[e.g.,][]{Rasio96,beauge2012}, von Zeipel-Lidov-Kozai (ZLK) oscillations \citep[e.g.,][]{Wu03,Fabrycky07,Naoz11}, or secular interactions \citep[e.g.,][]{Wu11,Petrovich15}---after which tidal friction acts to circularize and shrink the orbits. The other proposed channel posits that close-in planets attained their current orbits while still embedded in their parental gas disks, either by in situ formation \citep[e.g.,][]{Batygin16,Boley16} or by formation beyond the ice line, followed by inward migration driven by nebular tides \citep[e.g.,][]{Goldreich80,Lin86}.

By characterizing the architectures of these systems, we can get information about their formation and evolution, thus testing all these possible scenarios. One of the main signatures of the dynamical evolution of planetary systems is the stellar obliquity ($\psi$)---the angle between the stellar spin axis and the planet's orbital axis. Significant progress has been made in recent years in measuring the sky-projected obliquity ($\lambda$) distribution of hot Jupiters. Measurements performed via the Rossiter-McLaughlin (RM) effect have uncovered a broad distribution of $\lambda$ for hot Jupiters, ranging from well-aligned systems to systems on highly misaligned orbits \citep[see][for a recent review on stellar obliquities]{Albrecht2022}. It has been also shown that the obliquities of the host of hot Jupiters show a clear dependence with the stellar effective temperature \citep[e.g.,][]{Winn2010}. Hot Jupiters around cooler stars below the Kraft break at $\sim6100$ K \citep{Kraft1967} tend to be well-aligned likely due to tidal realignment, while those around hot stars above the break are frequently misaligned. These results have been interpreted as evidence that hot Jupiter formation involves dynamical perturbations that excite orbital inclinations \citep{Albrecht2012}. However, since hot Jupiters experience strong tidal forces---which can change the pristine obliquity distribution---and represent an infrequent outcome of planet formation \citep{Batalha2013}, it is unclear whether the observed obliquity distribution is related to the formation of hot Jupiters specifically, or instead reflects more general aspects of star and planet formation.

One way to answer this question is by measuring the stellar obliquities of planets in longer-period orbits (i.e., warm planets), which are less affected by tides, thus serving as more pristine probes of their original formation conditions. This is not an easy task, as warm planets have less frequent transits than their hotter counterparts, making their detection and follow-up observations more challenging. Furthermore, their transits are longer, requiring more telescope time to observe them. Nevertheless, there are some warm planets with measured stellar obliquity. Recently, \citet{Rice2022} identified a potential tendency toward alignment of the sky-projected obliquity of warm Jupiters around single stars. If confirmed, this would imply different formation mechanisms for hot and warm Jupiters around single stars. However, this result was affected by small-number statistics, as the sample of WJs was composed of only 12 planets. Additionally, 8 out of those 12 planets have quasi-circular orbits ($e<0.2$), while high-$e$ mechanisms should be tested in eccentric systems. Furthermore, 10 of those 12 planets orbit stars below the Kraft break, thus the results are only valid for cool stars. Perhaps surprisingly, more recent measurements in eccentric warm Jupiters suggest that this trend toward alignment could be independent of eccentricity \citep[e.g.,][]{Sedaghati2023,Espinoza-Retamal2023b,Hu2024}. Moreover, \citet{Wang2024} revealed that the trend could be independent of the stellar effective temperature. In any case, the sample of warm Jupiters with measured obliquities is still small, and more measurements are necessary to test these emerging trends and assess the underlying mechanisms behind the formation of this population.

In this work, we present ESPRESSO observations of the RM effect of 8 warm gas giants: K2-139 b \citep{Barragan2018}, K2-329 A b \citep{Sha2021}, WASP-106 b \citep{Smith2014}, WASP-130 b \citep{Hellier2017}, TOI-558 b \citep{Ikwut-Ukwa2022}, TOI-2179 b \citep{Schlecker2020}, TOI-4515 b \citep{Carleo2024}, and TOI-5027 b \citep{Tala2024}. Four of these ESPRESSO datasets---namely the ones of K2-139 b, K2-329 A b, WASP-106 b, and WASP-130 b---were previously studied by \citet{Prinoth2024} but with a focus on atmospheric characterization. Here, we present a more detailed study of the stellar obliquity. Combining in-transit and out-of-transit radial velocities (RVs) with photometry, we put a more precise constraint on the sky-projected obliquity $\lambda$ and measure, for K2-139 b, K2-329 A b, and TOI-4515 b, the true 3D obliquity $\psi$. The precise in-transit spectroscopic observations reveal that independent of their eccentricity, all the planets have relatively small obliquities, except TOI-2179 b, where we report a non-detection of the RM effect. We also show that the less massive planets have slightly more misaligned orbits than the massive ones.

We describe our observations in Section \ref{sec:obs}. In Section \ref{sec:ana}, we present our analysis. In Section \ref{sec:results}, we present our results for the different planets. Finally, we discuss the implications of these measurements in Section \ref{sec:discussion} and summarize our findings in Section \ref{sec:conclusion}.

\newpage
\section{Observations} \label{sec:obs}

\vspace{-0.7cm}

\begin{deluxetable*}{lcccccccc}[t]
\tablecaption{Summary of ESPRESSO observations. \label{tab:espresso}}
\tablecolumns{5}
\tablewidth{0pt}
\tablehead{Planet & Date & \# of spectra & Exposure & S/N$^{a}$ & $\sigma_{\rm RV}\,^{b}$ & Seeing & Airmass & CCF\\
 & (UT) & (in-transit) & Time (s) & &(m/s) & ($^{\prime\prime}$)& & Template}
\startdata
K2-139 b   & 2022 Jul 05 & 23 (18) & 900 & 110 & 0.8 & $0.4\,-\,1.4$ & $1.33\rightarrow1.01\rightarrow1.29$&G9\\
K2-329 A b   & 2022 Aug 23 & 30 (20) & 600 & 59 & 1.0 & $0.4\,-\,0.8$ & $1.13\rightarrow1.07\rightarrow1.86$&G2\\
K2-329 A b   & 2022 Oct 12 & 29 (18) & 605 & 31 & 2.4 & $0.7\,-\,1.6$ & $1.25\rightarrow1.07\rightarrow1.46$&G2\\
WASP-130 b & 2022 Jun 08 & 47 (30) & 420 & 89 & 1.0 & $0.5\,-\,1.2$ & $1.19\rightarrow1.06\rightarrow1.45$&G8\\
WASP-106 b & 2023 Mar 17 & 38 (29) & 600 & 80 & 1.8 & $0.5\,-\,1.1$ & $1.27\rightarrow1.07\rightarrow2.22$&F9\\
TOI-558 b  & 2022 Oct 23 & 34 (19) & 405 & 43 & 4.1 & $0.5\,-\,0.9$ & $1.43\rightarrow1.20\rightarrow1.24$&G2\\
TOI-2179 b & 2021 Nov 14 & 15 (11) & 570 & 48 & 2.4 & $0.5\,-\,0.8$ & $1.24\rightarrow1.37$&G2\\
TOI-4515 b & 2024 Aug 28 & 23 (11) & 780 & 51 & 1.4 & $0.6\,-\,1.1$ & $2.16\rightarrow1.45\rightarrow1.81$&G9\\
TOI-5027 b & 2024 Jul 12 & 33 (18) & 600 & 37 & 2.8 & $1.3\,-\,3.1$ & $1.56\rightarrow1.41\rightarrow1.64$&F9\\
\enddata
\tablecomments{$^{a}$Median signal to noise ratio (S/N) per pixel of the spectral series, calculated at $\sim$\,550\,nm as the quadrature sum of the S/N values from the two slices where the central wavelength is closest to 550 nm. $^{b}$Median RV uncertainty.}
\end{deluxetable*}

\subsection{ESPRESSO Transit Spectroscopy} \label{sec:espresso}

The 8 systems presented here are part of a dedicated survey with the ESPRESSO spectrograph \citep{Pepe20}, whose main objective is to measure the stellar obliquity of systems hosting transiting warm gas giants. ESPRESSO is a highly-stabilized, fiber-fed, cross-dispersed echelle spectrograph installed at the Incoherent Combined Coudé Focus of ESO's Paranal Observatory in Chile. It covers a wavelength range from 380 to 788 nm at a resolving power of $R\approx$\,140,000 in single unit telescope high-resolution mode. We observed a single transit of each planet with ESPRESSO, except K2-329 A b, which was observed twice. Table \ref{tab:espresso} shows a summary of the ESPRESSO observations.

All datasets were reduced using the dedicated ESPRESSO data reduction pipeline \citep{Sosnowska15,Modigliani20} run on the ESO Reflex environment \citep{Freudling13}, including all standard reduction steps, which also provides RVs by fitting a Gaussian model to the calculated Cross-Correlation Function (CCF). The CCF was calculated at steps of 0.5 km/s (representing the sampling of the spectrograph) for $\pm 20$ km/s centered on the estimated systemic velocity and using the template with the closest match to the spectral type of each target (see Table \ref{tab:espresso}). The ESPRESSO RVs of the different targets, along with the best model, are shown in Figure \ref{fig:RMs}.

\subsection{Photometry}

Simultaneously with ESPRESSO, we observed the transits of WASP-106 b, WASP-130 b, TOI-558 b, TOI-2179 b, and TOI-5027 b with the station of the Observatoire Moana located in El Sauce (ES) observatory in Chile. Observatoire Moana is a global network of small-aperture robotic optical telescopes. The ES station consists of a 0.6 m corrected Dall-Kirkham robotic telescope coupled to an Andor iKon-L 936 deep depletion 2k $\times$ 2k CCD with a scale of 0.67$^{\prime\prime}$ per pixel. We observed the transits of WASP-106 b, WASP-130 b, TOI-2179 b, and TOI-5027 b using a Sloan $r'$ filter, while for TOI-558 b we used a Sloan $i'$ filter. Data reduction was done using a dedicated pipeline that automatically performs the CCD reduction steps, followed by the measurement of the aperture photometry for the brightest stars in the field. The pipeline also generates the differential light curve of the target star by identifying the optimal comparison stars based on color, brightness, and proximity to the target. The MOANA/ES light curves of the different targets, along with the best model, are shown in the different plots of Appendix \ref{app:plots}.

\subsection{Spectroscopy}\label{sec:spec}

For TOI-2179, we obtained 4 spectra with the FEROS instrument \citep{feros} installed at the 2.2m MPG telescope in the ESO La Silla Observatory in Chile. The goal of these observations was to combine the new RVs with those presented in the discovery paper \citep{Schlecker2020} to identify possible additional companions in the system. The observations were performed between January of 2022 and July of 2024. We adopted an exposure time of 1200s and used the simultaneous calibration technique. FEROS data was processed with the \texttt{ceres} \citep{ceres} pipeline, which delivers as final results precision RVs computed with the CCF technique.

\subsection{Archival Observations}

As described in Section \ref{sec:glob}, in this work we not only studied the RM effect but performed a global analysis combining transit photometry with in-transit and out-of-transit RVs in order to better constrain the parameters of the different planets and their orbits. To do that, we made use of most of the publicly available RVs and photometry of the different targets, including K2 \citep{Howell2014} and TESS \citep{Ricker2015} data. To search for and download the K2 and TESS light curves of the different targets, we used the \texttt{lightkurve} package \citep{Lightkurve}. For K2 data, we worked with the light curves generated with the EVEREST pipeline \citep{Luger2016,Luger2018}, while for TESS data, with the ones generated with the SPOC pipeline \citep{Jenkins16}. In our analysis, we combined the consecutive sectors of different TESS years and assumed different years and cadences were different instruments.

\begin{itemize}

\item For K2-139 b, we used the K2 light curve from campaign 7, which includes three transits, the CHEOPS \citep{Benz2021} light curves presented in \citet{Smith2022} including 4 transits, and a ground-based transit observation that was obtained on 2018 Aug 18 by the Evans 50 cm robotic telescope at ES Observatory with an exposure time of 60 s and $Rc$ filter, which is publicly available in ExoFOP\footnote{\url{https://exofop.ipac.caltech.edu/tess/}}. As for the RVs, we used 10 FIES, 6 HARPS, and 3 HARPS-N measurements presented in \citet{Barragan2018}. 

\item For K2-329 A b, we used the K2 light curves from campaigns 12 and 19. Also, we included one ground-based light curve obtained with the PEST telescope using an $R_c$ filter \citep[see][]{Sha2021}, and one more recent transit present in TESS sector 42 (Year 4). As for the RVs, we used 10 PFS, 13 HARPS, and 27 FEROS measurements presented in \citet{Sha2021}. 

\item For WASP-106 b, we used the 2-minute cadence light curves of TESS sectors 9 (Year 1), 36 (Year 3), and 45 and 46 (Year 4). As for the RVs, we used 19 CORALIE and 9 SOPHIE measurements presented in \citet{Smith2014}, and 11 HARPS out-of-transit measurements publicly available in the ESO archive\footnote{\url{https://archive.eso.org/eso/eso_archive_main.html}}. We did not consider the in-transit HARPS RVs presented in \citet{Harre2023} taken to constrain the stellar obliquity. 

\item For WASP-130 b, we used the 2-minute cadence TESS light curves of sectors 11 (Year 1), 37 (Year 3), and 64 (Year 5). Each sector contained two transits. As for the RVs, we used 26 CORALIE measurements presented in \citet{Hellier2017} and 18 out-of-transit HARPS measurements available in the ESO archive. In the analysis we considered the pre and post CORALIE update data as different instruments \citep[see][]{Hellier2017}. Similar to the previous case, we did not consider the in-transit HARPS RVs taken to constrain the stellar obliquity.

\item For TOI-558 b, we used the 30-minute cadence TESS light curves of sectors 2 and 3 (Year 1), the 2-minute cadence TESS light curves of sectors 29 and 30 (Year 3), and three ground-based light curves obtained with Las Cumbres Observatory Global Telescope \citep[LCOGT,][]{Brown2013} using the South African Astronomical Observatory (SAAO) station in the $z'$ band, and the Cerro Tololo Inter-American Observatory (CTIO) station in the $B$ and $i'$ bands \citep[see][]{Ikwut-Ukwa2022}. As for the RVs, we used 14 PFS measurements presented in \citet{Ikwut-Ukwa2022}.

\begin{figure*}[t!]
    \centering
    \includegraphics[width=\textwidth]{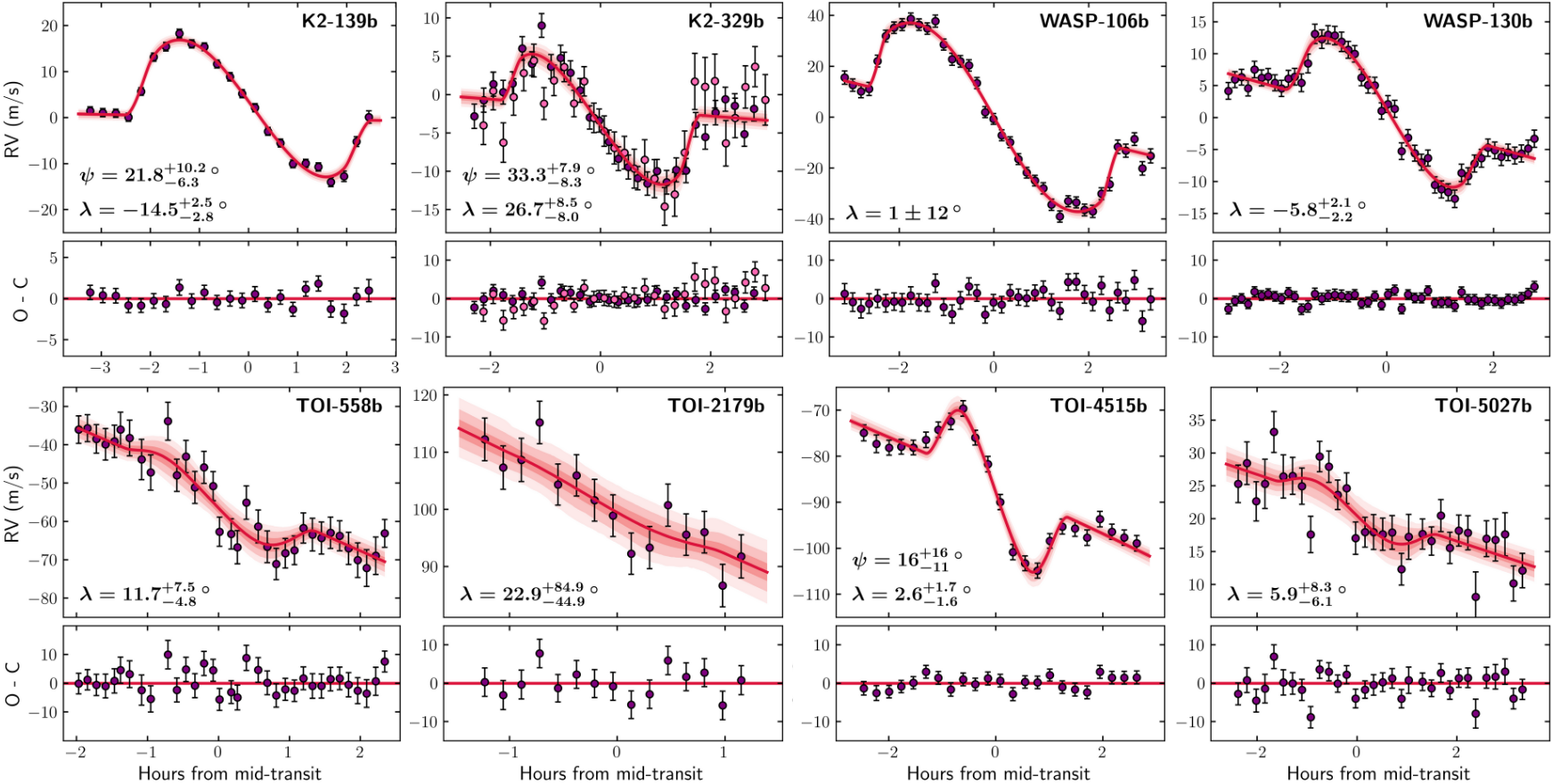}
    \caption{ESPRESSO observations of the RM effect produced by our different targets (purple---and also pink for K2-329 A b to show the second transit), along with the best model (red). Shaded red areas represent $1\sigma$, $2\sigma$, and $3\sigma$ models. All error bars include a white noise jitter term applied in quadrature to the RV data points. The different datasets are available in Appendix \ref{app:plots}.}
    \label{fig:RMs}
\end{figure*}

\item For TOI-2179 b, we used the 30-minute cadence TESS light curves of sectors 1 and 2 (Year 1), and the 2-minute cadence TESS light curves of sectors 28 and 29 (Year 3), and 68 and 69 (Year 5). Additionally, we also used the ground-based follow-up light curves presented in \citet{Schlecker2020} taken with the CHAT telescope and LCOGT, both using an $i'$ filter. As for the RVs, we used the 25 FEROS RVs presented in \citet{Schlecker2020} and 4 new FEROS observations described in Section \ref{sec:spec}.

\item For TOI-4515 b, we used the 30-minute cadence TESS light curve of sector 17 (Year 2), the 2-minute cadence TESS light curves of sectors 42 and 43 (Year 4), and 57 (Year 5). We also used the two ground-based light curves presented in \citet{Carleo2024}. The first one was taken with KeplerCam using an $i'$ filter, and the second one with LCOGT/CTIO in the $g'$ band. As for the RVs, we used 26 HARPS-N, 5 FEROS, and 18 TRES measurements presented in \citet{Carleo2024}.

\item Finally, for TOI-5027 b, we used the 30-minute cadence TESS light curve from sector 12 (Year 1), the 10-minute cadence TESS light curve from sector 39 (Year 3), the 2-minute cadence TESS light curve from sector 66 (Year 5), and one ground-based light curve obtained with LCOGT/CTIO in the $i'$ band \citep{Tala2024}. As for the RVs, we used the 27 FEROS RVs presented in \citet{Tala2024}.

\end{itemize}

\section{Analysis} \label{sec:ana}

\subsection{Stellar Parameters}\label{sec:stellar}

We performed a homogeneous analysis of the stellar properties of the 8 stars considered in this work. We followed the two-step iterative procedure presented in \citet{Brahm2019}. Briefly, in the first step, we compute the stellar atmospheric parameters using the \texttt{zaspe} package \citep{zaspe}, where a co-added high-resolution spectrum is compared with a grid of synthetic ones to determine the one that produces the best fit. The search is performed in the spectral regions that are most sensitive to changes in the stellar parameters, and reliable error bars are computed through Monte Carlo simulations. In the second step, we compute the stellar physical parameters by performing a spectral energy distribution fit. We fit public broadband photometric data to synthetic magnitudes generated from the \texttt{PARSEC} isochrones \citep{parsec} by taking into account the distance to the star using the Gaia data release (DR) 3 \citep{gaia:dr3} parallax. In this step, the stellar temperature derived with \texttt{zaspe} is used as a prior, while the metallicity is held fixed. From the stellar mass and radius obtained with the second step, we obtain a more precise \logg, which is held fixed in a new iteration of the first step. This procedure is repeated until reaching convergence in \logg. The stellar parameters of the different stars are presented in Appendix \ref{app:stellar}.

\subsection{Photometric Analysis}\label{sec:phot}

In order to update the orbital ephemeris of our different targets, and to look for Transit Timing Variations (TTVs), we performed a photometric analysis with \texttt{juliet} \citep{juliet} for each planet using all the photometric observations described in Section \ref{sec:obs}. \texttt{juliet} uses \texttt{batman} \citep{batman} for the transit model and the \texttt{dynesty} dynamic nested sampler \citep{dynesty2} to perform bayesian analysis and explore the likelihood space to obtain posterior probability distributions. For each target, we placed uninformative priors on the impact parameter $b$ and radius ratio $R_p/R_{\star}$, with an informative prior on the stellar density $\rho_\star$ that was constrained in Section \ref{sec:stellar}. We sampled the limb darkening parameters using the quadratic $q_1$ and $q_2$ from \citet{Kipping13} with uniform priors. We placed Gaussian priors for each transit mid-point based on the expected values calculated from the orbital period and time of mid-transit from the discovery paper of each planet, placing a large width of 0.1 days on the Gaussian prior standard deviation to not impact the derived transit midpoints. To account for variability and systematic noise in the TESS, K2, and CHEOPS light curves, we included a Matern-3/2 Gaussian Process (GP) as implemented in \texttt{celerite} \citep{celerite} and available in \texttt{juliet}. Each TESS year (i.e., all combined sectors of each year) and K2 campaign had its own GP kernel to account for differences in variability captured in different epochs, cadences, and instruments. For the LCOGT/CTIO light curves of TOI-558 b and TOI-4515 b in the $B$ and $g'$ filters respectively, we included a linear detrending using the airmass as input as the observed flux was seen to be correlated with the airmass of the observations. From this analysis, we ruled out the presence of TTVs larger than $\sim10$ minutes for K2-129, $\sim3$ minutes for K2-329 A b, $\sim4$ minutes for WASP-106 b, $\sim2$ minutes for WASP-130 b, $\sim10$ minutes for TOI-558 b, $\sim5$ minutes for TOI-2179 b, $\sim2$ minutes for TOI-4515 b, and $\sim3$ minutes for TOI-5027 b (at $1\sigma$). Additionally, we obtained improved orbital ephemeris for all the planets and detrended light curves. 

\begin{figure}[t]
    \centering
    \includegraphics[width=\columnwidth]{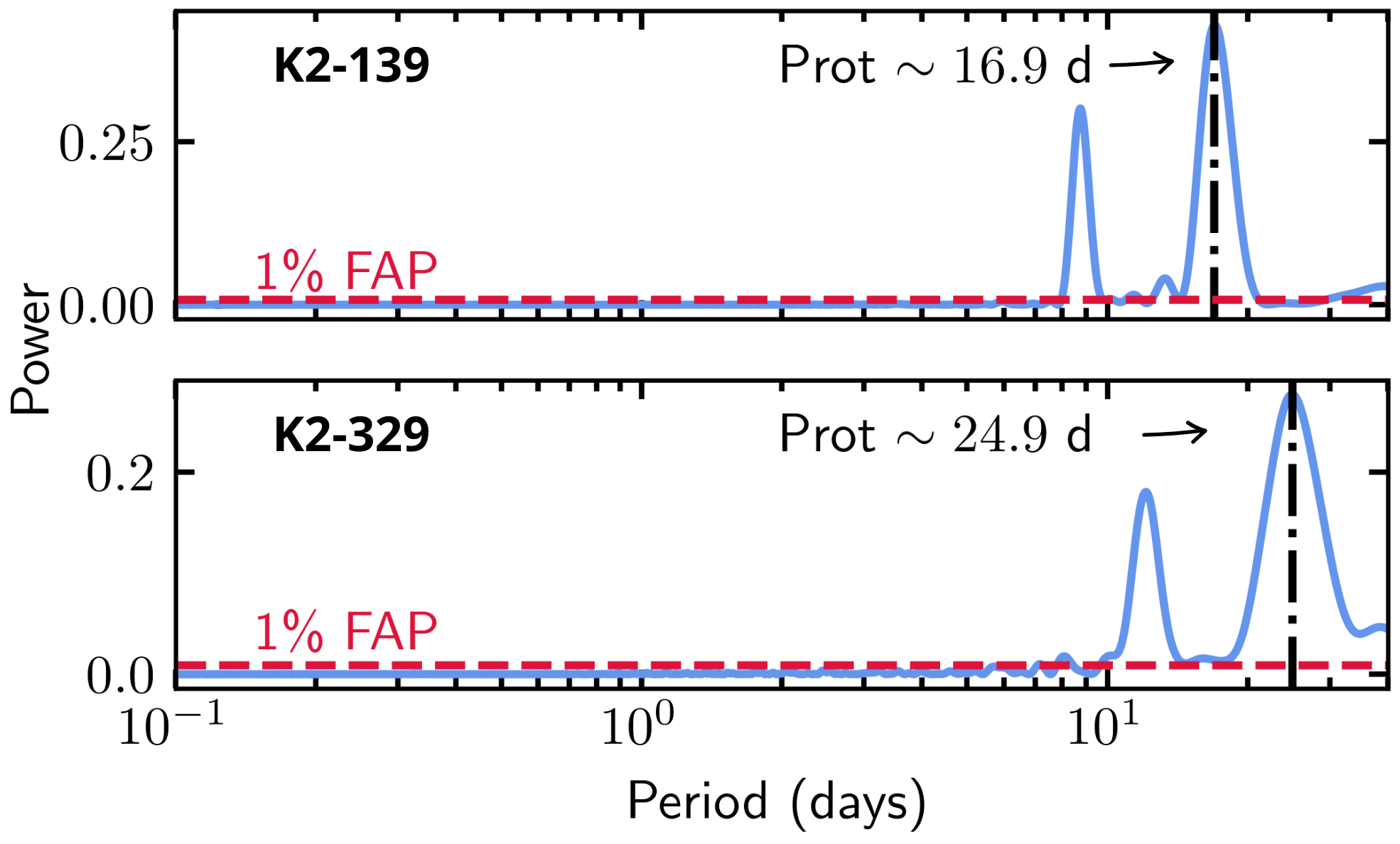}
    \caption{Lomb-Scargle periodogram of the K2 light curves of K2-139 (upper) and K2-329 A (bottom), with transit data masked out. A false alarm probability level of 1\% is marked with a dashed red line. The highest-power peaks are marked with an arrow, indicating the most likely stellar rotation period. In both cases, an alias corresponding to half the period is also clearly visible.}
    \label{fig:periodogram}
\end{figure}

For the K2 targets K2-139 and K2-329 A, we identify a modulation in the K2 light curves related to the rotational period of the stars ($P_{\rm rot}$). Figure \ref{fig:periodogram} shows the Lomb-Scargle periodograms. We tried to estimate an error bar from the periodograms themselves and using a quasi-periodic GP with \texttt{juliet}. However, because latitudinal differential rotation enforces a lower limit of 10\% to the measurement precision \citep[e.g.,][]{Epstein2014,Aigrain2015}, we ended up adopting $P_{\rm rot} = 17.0\pm1.7$ days and $24.9\pm2.5$ days for K2-139 and K2-329 A, respectively. Additionally, \citet{Carleo2024} reported a rotational period of $15.6\pm0.3$ days for TOI-4515 based on the modulation seen in the WASP \citep{Pollacco2006} light curve of this star. Here, we adopted $15.6\pm1.6$ days based on the same argument as for our K2 targets.


\subsection{Global Modeling}\label{sec:glob}

In order to constrain the parameters of the different planets and their orbits, we performed a joint fit of all the observations described in Section \ref{sec:obs} using \texttt{ironman}\footnote{\url{https://github.com/jiespinozar/ironman}} \citep{Espinoza-Retamal2024}. In brief, \texttt{ironman} is a Python code that can jointly fit in-transit and out-of-transit RVs together with transit photometry. To model the RVs \texttt{ironman} uses \texttt{rmfit} \citep{Stefansson22}, which uses \texttt{radvel} \citep{radvel} to model the Keplerian orbit and the framework from \citet{Hirano10} to model the RM effect. To model the photometry \texttt{ironman} uses \texttt{batman} \citep{batman}. Finally, to get the posteriors \texttt{ironman} uses the \texttt{dynesty} Dynamic Nested Sampler \citep{dynesty2}.

Since \texttt{ironman} has the capability of working with different parameterizations, we derived both, the sky-projected $\lambda$ and the true 3D obliquity $\psi$, for K2-129 b, K2-329 A b, and TOI-4515 b (given that we measured a rotational period for the host stars), and only the sky-projected obliquity $\lambda$ for WASP-106 b, WASP-130 b, TOI-558 b, TOI-2179 b, and TOI-5027 b. In both parametrizations, \texttt{ironman} samples $\lambda$ directly, and the difference is the $v\sin{i_\star}$ sampling. To derive $\psi$, the code follows the parametrization used in \citet{Stefansson22}, which broadly follows the methodology suggested by \citet{Masuda2020} to account for the fact that the equatorial velocity and $v\sin{i_{\star}}$ are not independent variables. The sampled parameters here are the stellar rotational period $P_{\rm rot}$, the stellar radius $R_{\star}$, and the cosine of the stellar inclination $\cos{i_{\star}}$, and $\psi$ is derived at the end as a product of the sampling. In turn, to derive only $\lambda$, the $v\sin{i_{\star}}$ is directly sampled.

In this analysis, we only considered the detrended TESS and K2 light curves close to the transits, in windows of $\sim6$ hours around the transit midpoint, to reduce the computational cost. We included independent jitter terms for each instrument to account for possible instrumental systematics. In all systems, we placed uninformative priors for almost all parameters, except for the orbital periods and times of mid-transit, which were constrained in Section \ref{sec:phot}, and for the stellar parameters that were constrained in Section \ref{sec:stellar}. When deriving only $\lambda$, we placed an informative prior on the stellar density. In turn, when also deriving $\psi$, we placed informative priors on the stellar radius, mass, and rotational period. For all systems, we also placed informative priors on $\beta_{\rm ESPRESSO}$, the intrinsic linewidth accounting for instrumental and macroturbulence broadening. We considered an instrumental broadening of 2.15 km/s because of the ESPRESSO resolution, and for the macroturbulence broadening, we used the macroturbulence law from \citet{Valenti2005} to estimate it if $\teff<6000$ K, or the macroturbulence law for hot stars from \citet{Gray84} if $\teff>6000$ K. We added the instrumental and macroturbulence broadening in quadrature to set our priors for each target, with an uncertainty of 2 km/s. All priors and resulting posteriors for each system are shown in Appendix \ref{app:posteriors}.

\section{Results}\label{sec:results}

\subsection{K2-139 b}

K2-139 b \citep{Barragan2018} is a circular warm Saturn ($M_p=0.34\pm0.04\,M_J$) in a slightly misaligned orbit with $\lambda = -14.5_{-2.8}^{+2.5}$ deg and $\psi = 21.8_{-6.3}^{+10.2}$ deg. The measured sky-projected obliquity is consistent with one reported in \citet{Prinoth2024} of $\lambda = -17\pm7$ deg using the same ESPRESSO dataset, but it is better constrained thanks to the joint fit of all the available data. Additionally, thanks to the measurement of the rotational period of the host star, here we also measured a true 3D obliquity. As for the eccentricity, \citet{Barragan2018} reported a non-zero eccentricity $e=0.12_{-0.08}^{+0.12}$, while \citet{Smith2022} adopted circular orbits given the poor constraint. In this work, we modeled the system both ways, with a circular and an eccentric orbit. The Bayesian evidence ($\log{Z}$) favored the circular model, having a Bayes factor $\Delta\log{Z} = 4.9>2$, so it is the model we adopted. Overall, the rest of the parameters of the planet are in good agreement (within $2\sigma$) with the ones reported in both \citet{Barragan2018} and \citet{Smith2022}.

\subsection{K2-329 A b}

K2-329 A b \citep{Sha2021} is a slightly eccentric ($e=0.08\pm0.03$) and slightly misaligned warm Saturn ($M_p=0.25\pm0.02\,M_J$) with $\lambda = 26.7_{-8.0}^{+8.5}$ deg and $\psi = 33.3_{-8.3}^{+7.9}$ deg. Similar to the previous case, the derived sky-projected obliquity value is consistent with the one reported in \citet{Prinoth2024} of $\lambda = 24^{+13}_{-9}$ deg using the same ESPRESSO dataset, but it is better constrained, and thanks to the additional measurement of the rotational period of the host star, here we also measured a true 3D obliquity. Since we observed two transits of this planet with ESPRESSO, in the \texttt{ironman} analysis we considered the two datasets as taken with the same instrument but allowed a different jitter term to take into account possible differences in the observing conditions each night. We also tested a fit with a different RV offset for each transit but found a negligible difference, so we assumed it was the same for both datasets. Additionally, we tested a circular versus eccentric orbit for this planet, and in this case, the Bayes factor is $\Delta\log{Z} = 0.9<2$, which suggests that there is not sufficient evidence to favor any of the models in particular. We adopted the eccentric model as the eccentricity is constrained at $>2\sigma$, and it is also consistent with the value reported in \citet{Sha2021} of $e=0.0697^{+0.040}_{-0.041}$. The rest of the parameters of the planet are also in good agreement (within $2\sigma$) with the ones reported there. Additionally, \citet{El-Badry21} reported a comoving M-dwarf companion in this system with $T_{\rm eff}\approx3000$ K and a current projected separation of $\sim8800$ au identified using data from the Gaia early DR3 \citep{Gaia_eDR3}. 

\subsection{WASP-106 b}

WASP-106 b \citep{Smith2014} is a circular and well-aligned warm Jupiter with $\lambda=1\pm12$ deg. This value is consistent with the one reported in \citet{Prinoth2024} of $\lambda=-3^{+9}_{-10}$ deg using the same ESPRESSO dataset, the one reported in \citet{Harre2023} of $\lambda =-1\pm11$ deg using two HARPS and one HARPS-N transit observations, and the one reported in \citet{Wright2023} of $\lambda=6^{+17}_{-16}$ deg using NEID observations. Although our constraint on the sky-projected obliquity appears less precise than previous results, we argue that our measurement has better-estimated error bars. The impact parameter we measure $b=0.09_{-0.07}^{+0.14}$ is quite small which produces a degeneracy between $v\sin{i_\star}$ and $\lambda$ in the RM effect \citep[see, e.g.,][]{Albrecht2011}. As we are using an uninformative prior on $v\sin{i_\star}$ in order to not affect the results of the RM analysis, the value reported here has self-consistent uncertainties that take into account possible correlation with all the other parameters in the global fit. Regarding eccentricity, both \citet{Smith2014} and \citet{Prinoth2024} worked with circular orbits, while \citet{Harre2023} and \citet{Wright2023} worked with eccentric orbits. Here, we compared both cases and found a $\Delta \log{Z} = 5.5>2$ in favor of the circular model, so it is the one we adopted. The rest of the parameters of the planet are in good agreement (within $2\sigma$) with the ones reported in these previous works.

\newpage
\subsection{WASP-130 b}

WASP-130 b \citep{Hellier2017} is a circular and aligned warm Jupiter with $\lambda=-5.8_{-2.2}^{+2.1}$ deg. This value is consistent with the one reported in \citet{Prinoth2024} of $\lambda=-2\pm2$ deg. Here we also tested circular versus eccentric orbits and found a $\Delta\log{Z}=3.8>2$ in favor of the circular model, so it is the one we adopted. All parameters of the planet are in good agreement with the ones reported in \citet{Hellier2017}.

\subsection{TOI-558 b}

TOI-558 b \citep{Ikwut-Ukwa2022} is an eccentric $(e=0.31\pm0.02)$ warm Jupiter in an aligned orbit with $\lambda=11.7_{-4.8}^{+7.5}$ deg. Given the clearly eccentric orbit, here we only considered the eccentric case and did not compare it with a circular fit. An important point to consider in this system is that because of the high impact parameter $b=0.92\pm0.01$, the RM effect has a low amplitude. Nevertheless, the sky-projected obliquity is well-constrained due to the high sensitivity of the RM effect to small changes in $\lambda$ at the current orbital configuration of the system (see Figure \ref{fig:RM_b}). This system demonstrates the capability of ESPRESSO in detecting such low-amplitude signals, and the utility of doing global fits of in-transit and out-of-transit RVs with photometry using \texttt{ironman}. All parameters of the planet are in good agreement with the ones reported in \citet{Ikwut-Ukwa2022}.

\subsection{TOI-2179 b}

TOI-2179 b \citep{Schlecker2020} is an eccentric ($e=0.57\pm0.01$) warm Jupiter. Unfortunately, for this planet, we were unable to constrain the obliquity given the low amplitude of the RM effect due to the high impact parameter value $b=0.91\pm0.01$ and the small $v\sin{i_\star}$ of $0.7^{+0.9}_{-0.5}$ km/s. Even with the ESPRESSO precision, we were unable to detect the RM signal. We decided to make public these results to show that even with ESPRESSO, it would be difficult to constrain the stellar obliquity of this planet. Additionally, we detect the presence of an RV trend in the data, suggesting the presence of a longer-period outer companion in the system. We modeled the trend using a quadratic model of the form $\ddot{\gamma}(t-t_a)^2+\dot{\gamma}(t-t_a)$, where $t_a$ is an arbitrary time set to the BJD of the first RV observation of this star. The parameters of the detected trend are, in this case, $\ddot{\gamma}=0.00008\pm0.00002$ m/s/day$^2$ and $\dot{\gamma}=-0.09\pm0.04$ m/s/day. Future RV monitoring and/or Gaia DR4 astrometry will be required to confirm this companion and to possibly measure the mutual inclination between the orbits of the two planets, which, in the absence of an obliquity measurement, will help constrain the dynamical history of this system \citep[e.g.,][]{Espinoza-Retamal2023a}.

\subsection{TOI-4515 b}

TOI-4515 b \citep{Carleo2024} is an eccentric ($e=0.46\pm0.01$) and aligned warm Jupiter with $\lambda=2.6^{+1.7}_{-1.6}$ deg and $\psi=16^{+16}_{-11}$ deg. Given the clearly eccentric orbit of the planet, here we only considered the eccentric case and did not compare it with a circular fit. This is the most eccentric planet for which we have measured the obliquity in this work, and thanks to the measured rotational period of the star, here we also constrain the true 3D obliquity. All parameters are in good agreement with those reported in \citet{Carleo2024}.

\begin{figure}[t!]
    \centering
    \includegraphics[width=\columnwidth]{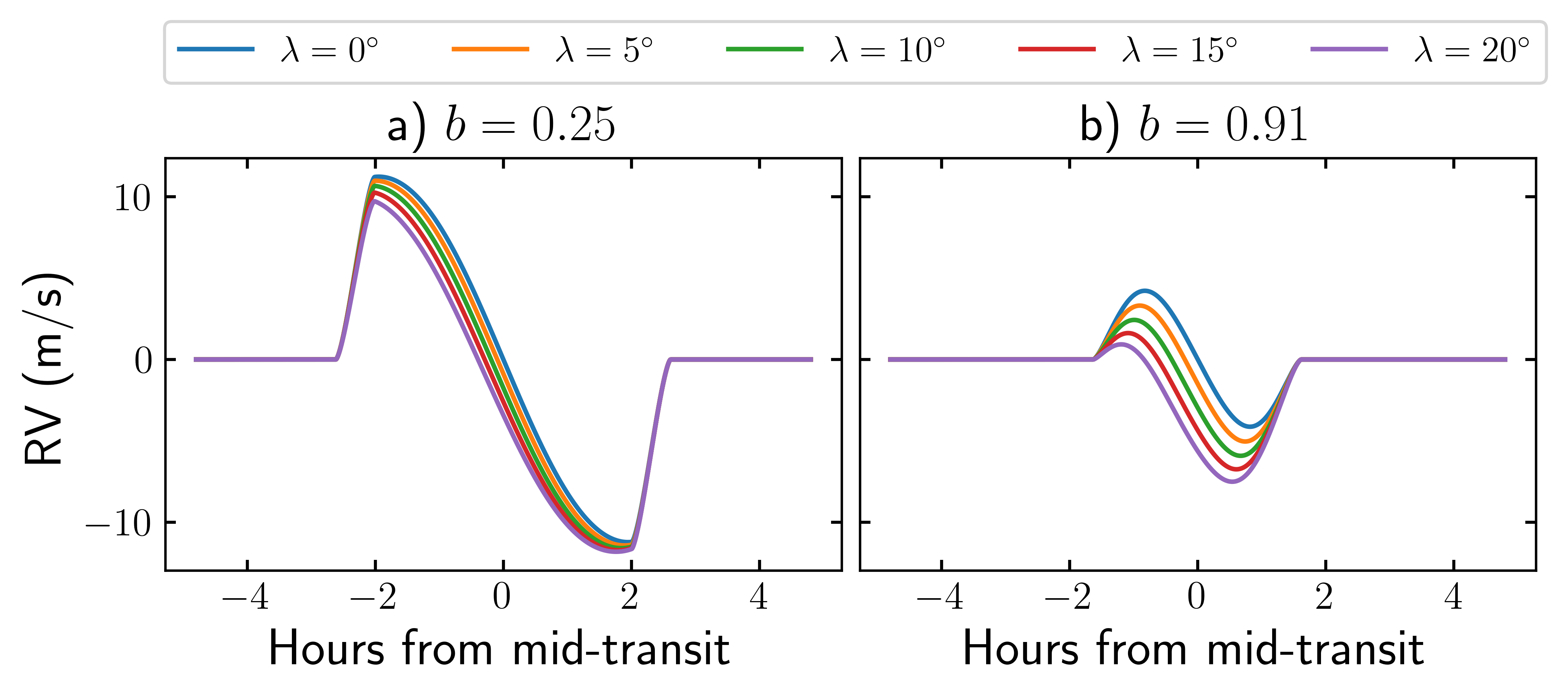}
    \caption{Simulated RM effect for a system with the same configuration as TOI-5027 b but assuming an impact parameter of a) $b=0.25$ and b) $b=0.91$. Different colors represent different sky-projected obliquities to show that for high impact parameter values, although the amplitude of the RM effect is lower, the effect becomes highly sensitive to small changes in $\lambda$. In particular, the shape becomes asymmetric faster than for the smaller impact parameter case. Thanks to the ESPRESSO capability of detecting such low-amplitude signals, we were able to measure the sky-projected obliquity for TOI-558 b and TOI-5027 b with a relatively good precision.}
    \label{fig:RM_b}
\end{figure}

\subsection{TOI-5027 b}

TOI-5027 b \citep{Tala2024} is an eccentric and aligned warm Jupiter with $e=0.35\pm0.03$ and $\lambda=5.9^{+8.3}_{-6.1}$ deg. Similar to TOI-558 b and TOI-4515 b, given the clearly eccentric orbit, here we only considered the eccentric case and did not compare it with a circular fit. Here, we also measured a high impact parameter $b=0.91\pm0.02$, which produces a low amplitude of the RM effect, but we were able to constrain the sky-projected obliquity thanks to the ESPRESSO precision and the global fit (see Figure \ref{fig:RM_b}). Additionally, in this system, we detect a linear trend in the RVs with $\dot{\gamma} = -0.12\pm0.04$ m/s/day, which might suggest the presence of a longer-period companion. Future RV monitoring and/or Gaia DR4 astrometry will be necessary to confirm or rule out this possible outer companion. The rest of the parameters of the planet are in good agreement (within $2\sigma$) with the ones reported in \citet{Tala2024}.

\begin{figure*}[t!]
    \centering
    \includegraphics[width=\textwidth]{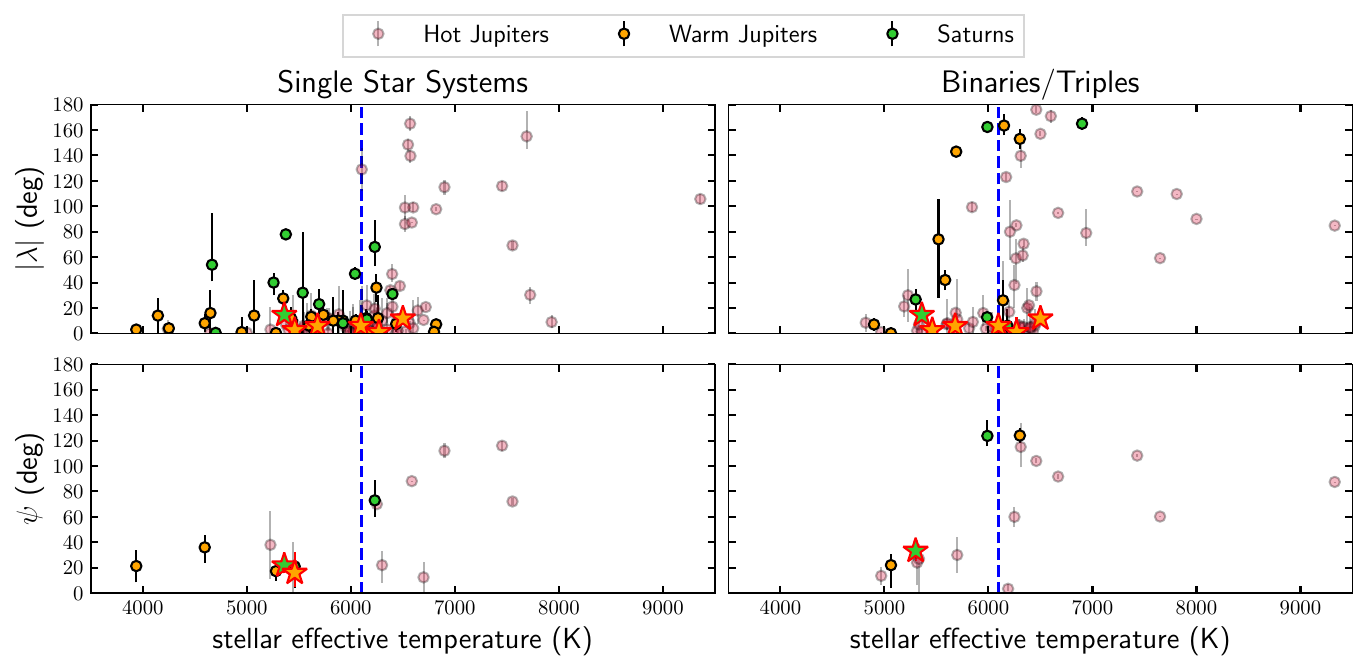}
    \caption{Sky-projected obliquity $\lambda$ (upper panels) and true 3D obliquity $\psi$ (bottom panels) measurements as a function of stellar effective temperature for the sample of planets around single star (left panels) and known binary/triple systems (right panels). The dashed blue line demarks the Kraft break at $\teff\approx6100$ K. Hot Jupiters ($0.4<M_p/M_{\rm J}<13$ and $a/R_\star<11$) are shown in red, Warm Jupiters ($0.4<M_p/M_{\rm J}<13$ and $a/R_\star>11$) in orange, and Saturns ($0.2<M_p/M_{\rm J}<0.4$) in green. Measurements reported in this work are shown as stars, while circles are the data from TEPCat, \citet{Wang2024}, and \citet{Knudstrup2024}.}
    \label{fig:obl_vs_teff}
\end{figure*}

\section{Discussion}\label{sec:discussion}

In this work, we have measured the stellar obliquity of 7 warm gas giants---K2-139 b, K2-329 A b, WASP-106 b, WASP-130 b, TOI-558 b, TOI-4515 b, and TOI-5027 b---and found that they all have relatively aligned orbits thus continuing with the observed trends for warm planets. For consistency with previous works, and following \citet{Rice2022}, for the statistical analysis we define hot Jupiters as $0.4\leq M_p/M_{\rm J}\leq13$ and $a/R_\star\leq11$, warm Jupiters as $0.4\leq M_p/M_{\rm J}\leq13$ and $a/R_\star>11$, and Saturns as $0.2\leq M_p/M_{\rm J}<0.4$. Figure \ref{fig:obl_vs_teff} shows the sky-projected and true 3D obliquity measurements of the different types of gas giants as a function of the stellar effective temperature. These measurements come from TEPCat\footnote{\url{https://www.astro.keele.ac.uk/jkt/tepcat/tepcat.html}} \citep{Southworth2011} as of September 2024, with complementary recent measurements from \citet{Wang2024} and \citet{Knudstrup2024}. As suggested by \citet{Albrecht2022}, we removed controversial measurements from the sample and did not consider them in the analysis. The removed systems include CoRoT-1, CoRoT-19, HAT-P-27, HATS-14, LTT 1445, TOI-451, WASP-1, WASP-2, WASP-23, WASP-49, and WASP-134. We also separated the sample into two to show the different properties of planets orbiting single stars and known binary/triple systems. We considered as binaries/triples, systems with a reported comoving companion in the catalog of binaries from \citet{El-Badry21}, or systems flagged as having more than one star in the NASA Exoplanet Archive\footnote{\url{https://exoplanetarchive.ipac.caltech.edu/}} \citep{Akeson2013}.

\begin{figure*}[t!]
    \centering
    \includegraphics[width=\textwidth]{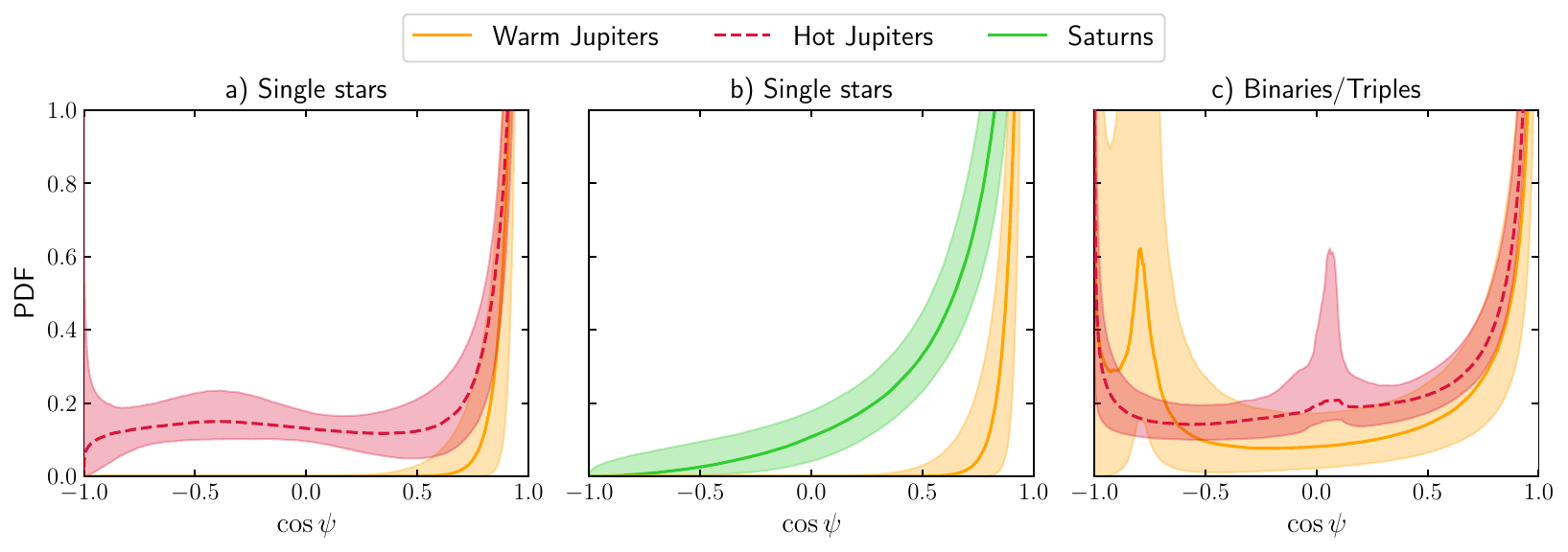}
    \caption{Inferred stellar obliquity distribution of hot Jupiters (red), warm Jupiters (orange), and Saturns (green). This inference was done from the sky-projected obliquity measurements following the methodology from \citet{Dong23}, without including information about the stellar inclination. Shaded regions show $1\sigma$ models. a) Sample of Jovian planets around single stars. b) Sample of warm planets around single stars. c) Sample of Jovian planets around binaries/triples.}
    \label{fig:psi_dist}
\end{figure*}

For both single stars and binaries/triples, we can see in Figure \ref{fig:obl_vs_teff} that hot Jupiters above the Kraft break at $\teff\approx6100$ K \citep{Kraft1967} appear to be misaligned, while below the break they are typically well aligned likely due to strong tidal forces that realigned the orbits. For warm Jupiters, however, the picture appears to be different. Around single stars, all 30 warm Jupiters are relatively well aligned with the star having $|\lambda|\lesssim20$ deg. The only exceptions are HAT-P-17 \citep[$\lambda=-27.5\pm6.7$ deg,][]{Mancini2022} and KELT-6 b \citep[$\lambda=-36\pm11$ deg,][]{Damasso2015} having slightly misaligned orbits. There is no apparent dependence of the stellar obliquity on the effective temperature of the host star. This was recently discussed by \citet{Wang2024}. In turn, although the sample is composed of 8 systems, there are warm Jupiters in highly misaligned orbits around binaries/triples, including K2-290 c \citep{Hjorth2021}, TIC 241249530 b \citep{Gupta2024}, and WASP-8 b \citep{Bourrier2017}. This could suggest a different formation mechanism for warm Jupiters around single stars and binaries/triples, but more obliquity measurements are necessary to be conclusive, especially around multi-star systems.

As for the Saturns, we observe that they are slightly more misaligned than warm Jupiters around single stars, while around binaries/triples, 2 out of the 5 Saturns are highly misaligned: WASP-131 b \citep{Doyle2023} and WASP-167 b \citep{Temple2017}. We do not separate between hot and warm Saturns as i) the numbers of obliquity measurements are small, with 5 hot and 10 warm Saturns in single stars and 4 hot and 1 warm Saturn in binaries/triples, and ii) their observed obliquity distributions are similar.

\subsection{The true 3D obliquity distribution}

To better quantify the possible difference in the obliquity distributions of the different populations, using the new measurements we have provided here, we used a Hierarchical Bayesian model \citep[HBM; e.g.,][]{Hogg2010} for the underlying obliquity distribution \citep[e.g.,][]{Morton2014,Munoz2018}. We worked with the framework presented by \citet{Dong23}\footnote{\url{https://github.com/jiayindong/obliquity}}, which models the underlying distribution of $\cos{\psi}$ across an exoplanet population using a mixture model of Beta distributions \citep[e.g.,][]{Gelman14}. This HBM framework allows the user to derive the underlying true 3D obliquity distribution from observed sky-projected obliquities and has the capacity to capture anything from an isotropic distribution to a strongly multi-modal population, depending on the number of Beta components assumed. Although the framework can work with different observables to derive the $\cos{\psi}$ distribution, here we derived it only from the $\lambda$ measurements and did not include information about the stellar inclination. All Beta distribution components are modeled using 3 parameters each: $w$, $\mu$, and $\kappa$. The parameter $w$ describes the weight of the component, while $\mu$ and $1/\kappa$ correspond to the mean and variance of each beta distribution component. The greater the value of $\kappa$, the smaller the variance, i.e., the distribution is more concentrated. The $\mu$ and $\kappa$ parameters can be related to the typical $\alpha$ and $\beta$ parameters of a beta distribution: $\alpha = \mu \kappa$ and $\beta=(1-\mu)\kappa$. In all cases, we assumed uninformative priors between 0 and 1 for $\mu$ and between -4 and 10 for $\log\kappa$. All derived distributions are shown in Figure \ref{fig:psi_dist}.

We have found that the underlying obliquity distributions of hot and warm Jupiters around single stars are different (Figure \ref{fig:psi_dist}). On the one hand, the $\cos{\psi}$ distribution of warm Jupiters can be modeled with only 1 beta component with $\mu=0.99\pm0.01$ and $\log{\kappa}=3.35^{+1.03}_{-0.95}$. This implies that warm Jupiters are typically well-aligned with no misaligned systems. Further, direct integration shows 95\% of the warm Jupiters have $\psi\lesssim25$ deg. We also tested modeling warm Jupiters with two beta components but found a weight of the second component consistent with 0. On the other hand, the $\psi$ distribution of hot Jupiters can be modeled with a two-component beta distribution with $w_0=0.22\pm0.11$, $\mu_0=0.39^{+0.18}_{-0.10}$, and $\log{\kappa_0}=1.4\pm1.1$ for the misaligned component and $w_1=0.78\pm0.11$, $\mu_1=0.98^{+0.01}_{-0.07}$, and $\log{\kappa_1}=2.7^{+1.1}_{-2.2}$ for the well-aligned component. This means that hot Jupiters are typically well-aligned, having an almost isotropic tail of misaligned systems. Direct integration shows 95\% of the hot Jupiters have $\psi\lesssim120$ deg. The difference in the obliquity distributions of hot and warm Jupiters around single stars suggests that they must form and evolve in a different way. The lack of misaligned warm Jupiters implies they must form more quiescently than hot Jupiters.

\begin{figure}
    \centering
    \includegraphics[width=\columnwidth]{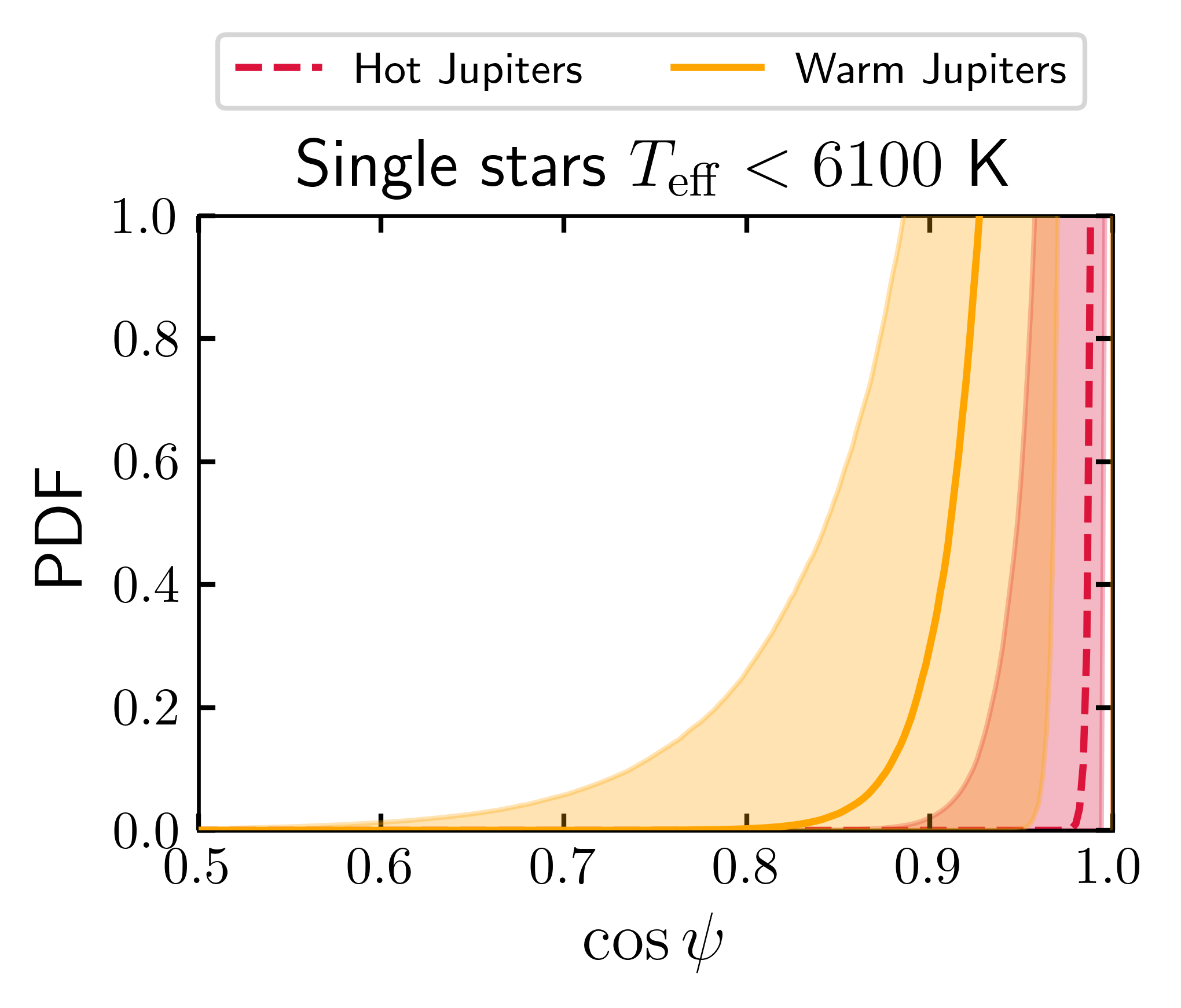}
    \caption{Inferred stellar obliquity distribution of hot Jupiters (red) and warm Jupiters (orange) around cool stars below the Kraft break. This inference was done from the sky-projected obliquity measurements following the methodology from \citet{Dong23}, without including information about the stellar inclination. Shaded regions show $1\sigma$ models.}
    \label{fig:cool_stars}
\end{figure}

Given that the sample of warm Jupiter systems with measured obliquity is mostly composed of cool single stars, we repeated the analysis to have a fairer comparison. In this second analysis, we only considered single stars below the break. We have found that the $\cos{\psi}$ distribution of the cool hosts of warm Jupiters can be modeled with one beta component with $\mu=0.99\pm0.01$ and $\log{\kappa}=4.48^{+2.15}_{-1.34}$, in good agreement with the previous result for the whole population with no cut in $T_{\rm eff}$. Similarly, we have found that the $\cos{\psi}$ distribution of the cool hosts of hot Jupiters can be modeled with one beta component with $\mu=0.99\pm0.01$ and $\log{\kappa}=7.03^{+1.99}_{-2.30}$. Similar results were obtained by \citet{Morgan2024} using the $v\sin{i_\star}$ method. The distributions of both hot and warm Jupiters show a preference for alignment, although using more direct measurements from the RM effect, we can see the distributions with more detail. As shown in Figure \ref{fig:cool_stars}, the distribution for the sample of hot Jupiters is more concentrated around $\sim0$ deg than that of warm Jupiters. As suggested by \citet{Zanazzi2024}, the small misalignment seen in the sample of warm Jupiters is likely a primordial feature, while the higher concentration of extremely aligned hot Jupiters suggests that tidal realignment is more effective and more important in the evolution of these systems.

We also repeated the analysis for the sample of 15 Saturns around single stars. We found that the $\cos{\psi}$ distribution of Saturns can be modeled with only 1 beta component with $\mu=0.89^{+0.04}_{-0.05}$ and $\log{\kappa}=1.12\pm0.6$. This implies that Saturns around single stars are typically well aligned but not as much as warm Jupiters. Direct integration shows 95\% of the Saturns around single stars have $\psi\lesssim80$ deg.

For completeness, we repeated the analysis for the sample of hot and warm Jupiters around binaries/triples (Figure \ref{fig:psi_dist}b). For warm Jupiters, we have found a bimodal $\cos{\psi}$ distribution peaked at $\cos{\psi}\sim0$ and $\cos{\psi}\sim-0.76$. The second peak is produced by the 3 systems in retrograde orbits in Figure \ref{fig:obl_vs_teff}. However, we acknowledge that the error bars here are large due to the small size of the sample. Therefore, more measurements are necessary to be conclusive. For hot Jupiters, we have found similar results to those for the sample around single stars. The $\cos{\psi}$ distribution can be modeled with 2 beta components with $w_0=0.12^{+0.31}_{-0.07}$, $\mu_0=0.52^{+0.07}_{-0.21}$, and $\log{\kappa_0}=2.1^{+4.4}_{-2.3}$ for the misaligned component and $w_1=0.88^{+0.07}_{-0.31}$, $\mu_1=0.84^{+0.14}_{-0.05}$, and $\log{\kappa_1}=-0.15^{+3.17}_{-0.25}$ for the well-aligned component. As shown in Figure \ref{fig:psi_dist}, we found little evidence ($\sim1.5\sigma$) for a preponderance of polar hot Jupiters around binaries/triples. A similar result was found by \citet{Albrecht2021} but using the entire sample of systems with measured obliquity and a frequentist analysis. Nonetheless, more recent studies by \citet{Dong23} and \citet{Siegel2023}, using Bayesian approaches, did not find evidence for a clustering around $\sim90$ deg. However, when limiting the sample to less massive planets, and now to binaries, the preponderance of polar orbits is detected (with different confidence levels) even with a Bayesian analysis \citep{Espinoza-Retamal2024, Knudstrup2024}.

\begin{figure*}[t!]
    \centering
    \includegraphics[width=\textwidth]{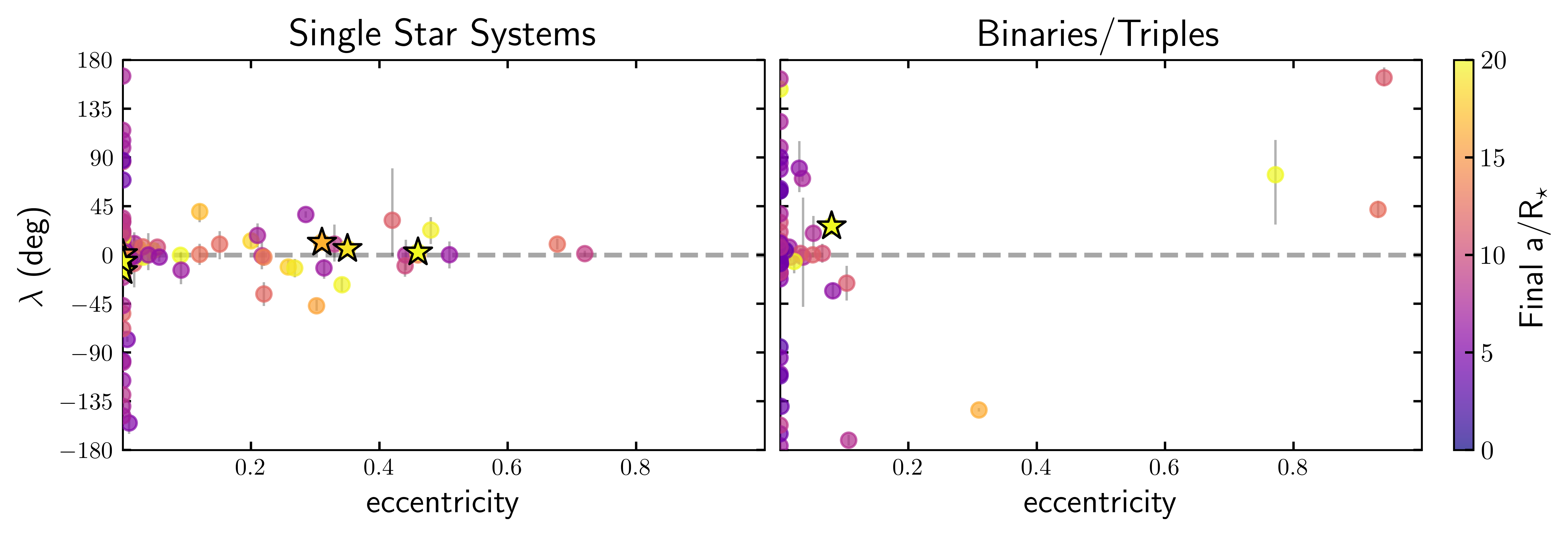}
    \caption{Sky-projected obliquity as a function of eccentricity for all giant planets ($M_p>0.2 M_J$) colored by their final scaled semimajor axis $(a/R_\star)_{\rm final} = (a/R_\star)(1-e^2)$ to show which gas giants are or will become hot planets with enough time. The left panel shows single stars, while the right panel shows known binary/triple systems. Measurements reported in this work are shown as stars, while circles are the data from TEPCat, \citet{Wang2024}, and \citet{Knudstrup2024}.}
    \label{fig:obl_vs_ecc}
\end{figure*}

\subsection{A trend with eccentricity}\label{sec:lambda_vs_ecc}

The work by \citet{Espinoza-Retamal2023b} noted a relation between the sky-projected obliquity and the eccentricity of giant exoplanets in single-star systems. Figure \ref{fig:obl_vs_ecc} shows an updated version of the $\lambda-e$ diagram for planets with $M_p>0.2\,M_J$, including the eccentric TOI-558 b, TOI-4515 b, and TOI-5027 b. Circular systems span the full range of obliquities, but they decrease with increasing eccentricity. Systems with $0.1\lesssim e\lesssim0.4$ are spreading from well-aligned to moderately misaligned ($|\lambda| \sim 45$ deg). All systems with $e\gtrsim0.4$ are well aligned. To test the statistical significance of this trend, we performed Kolmogorov-Smirnov (KS) tests on the distribution of obliquities, accounting for asymmetric uncertainties via a Monte Carlo bootstrap approach. We find that the obliquity distribution of eccentric ($e > 0.1$) gas giants around single stars is inconsistent with an isotropic distribution at the $8.6\sigma$ level. This result is somewhat unexpected given that large eccentricities are likely relics of a dynamically active history that could have excited large stellar obliquities. Possible resolutions to this problem include excitation of eccentricities by disk-planet interactions, especially if the warm Jupiters resided inside gas-depleted cavities \citep[e.g.,][]{Debras2021,Romanova2023}. Alternatively, planet-planet interactions may have excited the eccentricities without exciting the inclinations either by coplanar high-$e$ migration driven by unseen massive long-period outer companion \citep{Petrovich15} or an ejected planet following scattering events \citep{ford_rasio2008,ford_2011_scattering}. Future RV and/or Gaia astrometric observations will be crucial to detect or rule out companions, thus distinguishing between these two possible explanations for the eccentric warm Jupiters. 

This $\lambda-e$ trend may also extend to more massive brown dwarfs and less massive Neptunes/sub-Saturns, as shown by \citet{Doyle2024} and \citet{Rubenzahl2024}, respectively. In contrast, the highly eccentric and misaligned warm Jupiters in multiple-star systems suggest that the stellar companions in those systems have played a role in shaping the planetary orbits, possibly through ZLK oscillations \citep[as originally proposed for HD 80606 b by][]{Wu03}. Consistently, our bootstrap simulations show that the obliquity distribution for eccentric ($e > 0.1$) gas giants in binaries/triples is consistent with an isotropic distribution at the $1.3\sigma$ level. Future obliquity measurements of warm Jupiters in binaries will be important to further test this emerging trend.

\subsection{A potential trend with mass}\label{sec:mass_vs_obliquity}

We observe a potential trend with mass in the obliquity distribution of gas giants around single stars. Excluding hot Jupiters, the distribution of close-in planets has fewer misaligned planets with increasing mass. Our 5 warm Jupiters---WASP-106 b, WASP-130 b, TOI-558 b, TOI-4515 b, and TOI-5027 b---have sky-projected obliquities $|\lambda|\simeq0-10$ deg while the 2 less massive warm Saturns---K2-139 b and K2-329 A b---are slightly misaligned having $|\lambda|\simeq15-25$ deg. Furthermore, this holds at the population level where warm Jupiters around single stars are all well aligned with the typical warm Jupiter having $\psi\sim5$ deg while less massive Saturns are slightly more misaligned with the typical Saturn having $\psi\sim20$ deg. 

This trend may be a consequence of an underlying mechanism that excites larger obliquities as the planetary masses decrease. Natural candidates may rely on the secular excitation due to distant planetary companions. For instance, the disk-driven resonance mechanism proposed by \citet{Petrovich2020} and further extended by \citet{louden} and \citet{jj_disk}. This mechanism relies on the interaction between the close-in planet, an outer companion, and an evaporating protoplanetary disk. For systems with close-in Neptunes and outer Jovian companions, this mechanism can produce polar orbits. In turn, when applied to systems with planets with similar masses, like a close-in Jupiter and an outer Jupiter companion, the mechanism is unable of producing significant misalignments unless the outer planet is already largely inclined relative to the disk. As Saturns lie in the middle of Neptunes and Jupiters, the mechanism is expected to produce modest misalignments, just as observed. Another possible explanation is secular chaos \citep[e.g.,][]{Wu11} operating on later stages. In this case, because of the equipartition of angular momentum deficit, less massive planets will gradually become eccentric and/or inclined. Similarly to the disk-driven resonance, if cold Jupiters are present in the system, inner hot/warm Saturns are expected to acquire higher inclinations than warm Jupiters. In any case, more obliquity measurements of Saturns will be necessary to draw more conclusive results. 

This trend with mass may actually extend to a wider range of planet masses. The recent studies by \citet{Espinoza-Retamal2024} and \citet{Knudstrup2024} have found a prevalence of polar orbits for even less massive Neptunes/sub-Saturns---although they considered the whole sample, including systems in binaries/triples. Furthermore, recent obliquity measurements of aligned brown dwarfs \citep[e.g.,][]{Giacalone2024,dosSantos2024,Brady2024,Doyle2024} suggest that the trend might continue for even larger masses. This whole picture is also consistent with the break in the obliquity distribution at $M_p/M_\star=2\times10^{-3}$ recently found by \citet{Rusznak2024}, and with the fact that more massive hot Jupiters are also more closely aligned than less massive ones \citep[e.g.,][]{Hebrard2010}. More obliquity measurements across a wide range of masses will be necessary to test this interesting emerging trend further.

\subsection{On the formation of the different populations}

In our analysis, we show that the different populations of gas giants have different $\psi$ distributions (Figure \ref{fig:psi_dist}). One explanation for these different distributions is that the populations might originate from different formation processes. 

\paragraph{Hot Jupiters} The distribution of hot Jupiters around both single stars and binaries/triples---which is roughly the same---is consistent with formation through high-$e$ migration with subsequent tidal realignment in systems around cooler stars \citep[e.g.,][]{Albrecht2012, Rice2022a, Zanazzi2024}. This is also supported by the fact that $\sim75\%$ of the hot Jupiters have longer-period outer companions \citep{Bryan2016} capable of, e.g., driving the ZLK oscillations. However, it is also possible that hot Jupiters around cool stars form in a more quiescent way than around hot stars, as they tend to be more aligned.

\paragraph{Warm Jupiters} The overall sample of warm Jupiters is unlikely to follow the formation pathway of high-$e$ migration of the hot Jupiters. For one, the lack of misaligned warm Jupiters around single stars (95\% of them have $\psi\lesssim25$ deg) points toward formation in a well-aligned protoplanetary disk, possibly followed by inward migration driven by nebular tides \citep[e.g.,][]{Goldreich80,Lin86} hinted by the growing sample of resonant warm Jupiters \citep[e.g.,][]{Bozhilov2023,Vitkova2024}. This is also supported by observations of primordial alignment of directly imaged young systems \citep[e.g.,][]{Kraus2020,Sepulveda2024}, protoplanetary disks \citep[e.g.,][]{Davies2019}, and debris disks \citep[e.g.,][]{Hurt2023}.

As discussed in Section \ref{sec:lambda_vs_ecc}, the trend of spin-orbit alignment around single stars extends even to the sample of eccentric warm Jupiters, indicating that the excitation of eccentricities in these systems is not due to the ZLK mechanism \citep{Espinoza-Retamal2023b}. Instead, processes that excite the eccentricities retaining the primordial alignment may be at play. In contrast, eccentric warm Jupiters in multi-star systems present significant obliquities, consistent with the expectations from ZLK migration driven by the stellar companions \citep[e.g.,][]{Wu03,Fabrycky07}.

\paragraph{Saturns} There is a hint that Saturns around single stars are slightly more misaligned than warm Jupiters as shown in Figure \ref{fig:psi_dist} and discussed in Section \ref{sec:mass_vs_obliquity}. 
A possible explanation is that the level of (modest) obliquity excitation is a function of the planetary mass. Alternatives include secular excitation processes like secular chaos \citep[e.g.,][]{Wu11} or disk-driven resonance \citet{Petrovich2020}, that transfer angular momentum deficit from wider-orbit Jovian companions to the orbits of these inner planets. Naturally, this transfer becomes more efficient at driving the inclinations as the planetary masses decrease. 


\section{Conclusions and Summary}\label{sec:conclusion}

In this work, we have presented new observations of 8 warm gas giant systems. We have jointly analyzed ESPRESSO observations of the RM effect produced by the transiting planets, with transit photometry and out-of-transit RVs using the publicly available code \texttt{ironman}. We have concluded that:

\begin{itemize}
    \item Independent of the orbital eccentricity and host star \teff, the 5 warm Jupiters studied here---WASP-106 b, WASP-130 b, TOI-558 b, TOI-4515 b, and TOI-5027 b---are all consistent with being well aligned with the star, having $|\lambda|\simeq0-10$ deg.
    
    \item The 2 warm Saturns studied here---K2-139 b and K2-329 A b---have slightly more misaligned orbits than the more massive warm Jupiters, having $|\lambda|\simeq15-25$ deg, and $\psi\simeq20-30$ deg.
    
    \item  We report a non-detection of the RM effect produced by the transit of TOI-2179 b. This is a product of the small projected velocity $v\sin{i_\star}=0.7^{+0.9}_{-0.5}$ km/s and the high value of the impact parameter $b=0.91\pm0.01$ affecting the amplitude of the RM effect.
    
    \item The stellar obliquity of close-in gas giants around single stars seems to be correlated with their orbital eccentricity. In particular, obliquity decreases with increasing eccentricity, implying that processes that excite eccentricities while retaining the primordial alignment may be at play.
    
    \item Bayesian modeling of sky-projected obliquity measurements of gas giants around single stars shows that warm Jupiters are statistically more aligned than hot Jupiters. Hot Jupiters have an almost isotropic tail of misaligned systems where 95\% of them have $\psi\lesssim120$ deg. In turn, there are no misaligned warm Jupiters, with 95\% of them having $\psi\lesssim25$ deg. This implies that around single stars, warm Jupiters form in primordially aligned protoplanetary disks and subsequently evolve in a more quiescent way than hot Jupiters.

    \item The Bayesian modeling also showed that Saturns are slightly more misaligned than warm Jupiters around single stars. The typical Saturn has $\psi\sim20$ deg while the typical Jupiter has $\psi\sim5$ deg.
    
    \item The obliquity distribution of hot Jupiters around single stars and binaries/triples is roughly the same, pointing toward a similar violent and chaotic formation mechanism for these populations. Warm Jupiters around binaries might also be formed in a similar way, but more obliquity measurements are necessary to be conclusive.
\end{itemize}

Future RV and/or Gaia astrometric observations will be crucial in detecting and characterizing longer-period outer companions in these systems. Moreover, astrometric observations will have the capability of measuring mutual inclinations between the orbits of the transiting planet and the possible astrometric companion \citep{Espinoza-Retamal2023a}, thus allowing for detailed characterization of the architectures of these systems, which will constrain their histories further.

\section*{acknowledgments}
We would like to thank Songhu Wang, Xian-Yu Wang, Brandon Radzom, and the exoplanet group at Indiana University in general for their useful discussions. We also thank the anonymous referee for their thoughtful comments and suggestions that strengthened the manuscript.
J.I.E.-R. and C.P. gratefully acknowledge support from the John and A-Lan Reynolds Faculty Research Fund and support from ANID BASAL project FB210003.
J.I.E.-R. acknowledges support from ANID Doctorado Nacional grant 2021-21212378.
A.J., C.P., R.B., and M.H. acknowledge support from ANID Millennium  Science  Initiative ICN12\_009, IM23-0001. A.J. acknowledges additional support from FONDECYT project 1210718.
R.B. acknowledges additional support from FONDECYT Project 1241963. 
C.P. acknowledges support from CASSACA grant CCJRF2105 and FONDECYT Regular grant 1210425.
This research has made use of the Exoplanet Follow-up Observation Program (ExoFOP) website, which is operated by the California Institute of Technology, under contract with NASA under the Exoplanet Exploration Program. This paper includes data collected by the TESS and K2 missions. Funding for the TESS and K2 missions is provided by NASA's Science Mission Directorate.

\facilities{VLT/ESPRESSO, Observatoire Moana, TESS, K2, LCOGT, CHEOPS, CHAT, PEST, MPG/FEROS, ESO 3.6m/HARPS, TNG/HARPS-N, Magellan Clay/PFS, NOT/FIES, Euler/CORALIE, OHP/SOPHIE, FLWO/TRES.
}

\software{
\texttt{astropy}~\citep{astropy},
\texttt{batman}~\citep{batman},
\texttt{ceres}~\citep{ceres},
\texttt{celerite}~\citep{celerite},
\texttt{dynesty}~\citep{dynesty2},
\texttt{ironman}~\citep{Espinoza-Retamal2024},
\texttt{juliet}~\citep{juliet},
\texttt{lightkurve}~\citep{Lightkurve}
\texttt{radvel}~\citep{radvel},
\texttt{rmfit}~\citep{Stefansson22},
\texttt{zaspe}~\citep{zaspe}.
}

\bibliography{sample631}{}

\begin{thebibliography}{}
\expandafter\ifx\csname natexlab\endcsname\relax\def\natexlab#1{#1}\fi
\providecommand{\url}[1]{\href{#1}{#1}}
\providecommand{\dodoi}[1]{doi:~\href{http://doi.org/#1}{\nolinkurl{#1}}}
\providecommand{\doeprint}[1]{\href{http://ascl.net/#1}{\nolinkurl{http://ascl.net/#1}}}
\providecommand{\doarXiv}[1]{\href{https://arxiv.org/abs/#1}{\nolinkurl{https://arxiv.org/abs/#1}}}

\bibitem[{{Aigrain} {et~al.}(2015){Aigrain}, {Llama}, {Ceillier}, {Chagas}, {Davenport}, {Garc{\'\i}a}, {Hay}, {Lanza}, {McQuillan}, {Mazeh}, {de Medeiros}, {Nielsen}, \& {Reinhold}}]{Aigrain2015}
{Aigrain}, S., {Llama}, J., {Ceillier}, T., {et~al.} 2015, \mnras, 450, 3211, \dodoi{10.1093/mnras/stv853}

\bibitem[{{Akeson} {et~al.}(2013){Akeson}, {Chen}, {Ciardi}, {Crane}, {Good}, {Harbut}, {Jackson}, {Kane}, {Laity}, {Leifer}, {Lynn}, {McElroy}, {Papin}, {Plavchan}, {Ram{\'\i}rez}, {Rey}, {von Braun}, {Wittman}, {Abajian}, {Ali}, {Beichman}, {Beekley}, {Berriman}, {Berukoff}, {Bryden}, {Chan}, {Groom}, {Lau}, {Payne}, {Regelson}, {Saucedo}, {Schmitz}, {Stauffer}, {Wyatt}, \& {Zhang}}]{Akeson2013}
{Akeson}, R.~L., {Chen}, X., {Ciardi}, D., {et~al.} 2013, \pasp, 125, 989, \dodoi{10.1086/672273}

\bibitem[{{Albrecht} {et~al.}(2011){Albrecht}, {Winn}, {Johnson}, {Butler}, {Crane}, {Shectman}, {Thompson}, {Narita}, {Sato}, {Hirano}, {Enya}, \& {Fischer}}]{Albrecht2011}
{Albrecht}, S., {Winn}, J.~N., {Johnson}, J.~A., {et~al.} 2011, \apj, 738, 50, \dodoi{10.1088/0004-637X/738/1/50}

\bibitem[{{Albrecht} {et~al.}(2012){Albrecht}, {Winn}, {Johnson}, {Howard}, {Marcy}, {Butler}, {Arriagada}, {Crane}, {Shectman}, {Thompson}, {Hirano}, {Bakos}, \& {Hartman}}]{Albrecht2012}
---. 2012, \apj, 757, 18, \dodoi{10.1088/0004-637X/757/1/18}

\bibitem[{{Albrecht} {et~al.}(2022){Albrecht}, {Dawson}, \& {Winn}}]{Albrecht2022}
{Albrecht}, S.~H., {Dawson}, R.~I., \& {Winn}, J.~N. 2022, \pasp, 134, 082001, \dodoi{10.1088/1538-3873/ac6c09}

\bibitem[{{Albrecht} {et~al.}(2021){Albrecht}, {Marcussen}, {Winn}, {Dawson}, \& {Knudstrup}}]{Albrecht2021}
{Albrecht}, S.~H., {Marcussen}, M.~L., {Winn}, J.~N., {Dawson}, R.~I., \& {Knudstrup}, E. 2021, \apjl, 916, L1, \dodoi{10.3847/2041-8213/ac0f03}

\bibitem[{{Astropy Collaboration} {et~al.}(2013){Astropy Collaboration}, {Robitaille}, {Tollerud}, {Greenfield}, {Droettboom}, {Bray}, {Aldcroft}, {Davis}, {Ginsburg}, {Price-Whelan}, {Kerzendorf}, {Conley}, {Crighton}, {Barbary}, {Muna}, {Ferguson}, {Grollier}, {Parikh}, {Nair}, {Unther}, {Deil}, {Woillez}, {Conseil}, {Kramer}, {Turner}, {Singer}, {Fox}, {Weaver}, {Zabalza}, {Edwards}, {Azalee Bostroem}, {Burke}, {Casey}, {Crawford}, {Dencheva}, {Ely}, {Jenness}, {Labrie}, {Lim}, {Pierfederici}, {Pontzen}, {Ptak}, {Refsdal}, {Servillat}, \& {Streicher}}]{astropy}
{Astropy Collaboration}, {Robitaille}, T.~P., {Tollerud}, E.~J., {et~al.} 2013, \aap, 558, A33, \dodoi{10.1051/0004-6361/201322068}

\bibitem[{{Barrag{\'a}n} {et~al.}(2018){Barrag{\'a}n}, {Gandolfi}, {Smith}, {Deeg}, {Fridlund}, {Persson}, {Donati}, {Endl}, {Csizmadia}, {Grziwa}, {Nespral}, {Hatzes}, {Cochran}, {Fossati}, {Brems}, {Cabrera}, {Cusano}, {Eigm{\"u}ller}, {Eiroa}, {Erikson}, {Guenther}, {Korth}, {Lorenzo-Oliveira}, {Mancini}, {P{\"a}tzold}, {Prieto-Arranz}, {Rauer}, {Rebollido}, {Saario}, \& {Zakhozhay}}]{Barragan2018}
{Barrag{\'a}n}, O., {Gandolfi}, D., {Smith}, A.~M.~S., {et~al.} 2018, \mnras, 475, 1765, \dodoi{10.1093/mnras/stx3207}

\bibitem[{{Batalha} {et~al.}(2013){Batalha}, {Rowe}, {Bryson}, {Barclay}, {Burke}, {Caldwell}, {Christiansen}, {Mullally}, {Thompson}, {Brown}, {Dupree}, {Fabrycky}, {Ford}, {Fortney}, {Gilliland}, {Isaacson}, {Latham}, {Marcy}, {Quinn}, {Ragozzine}, {Shporer}, {Borucki}, {Ciardi}, {Gautier}, {Haas}, {Jenkins}, {Koch}, {Lissauer}, {Rapin}, {Basri}, {Boss}, {Buchhave}, {Carter}, {Charbonneau}, {Christensen-Dalsgaard}, {Clarke}, {Cochran}, {Demory}, {Desert}, {Devore}, {Doyle}, {Esquerdo}, {Everett}, {Fressin}, {Geary}, {Girouard}, {Gould}, {Hall}, {Holman}, {Howard}, {Howell}, {Ibrahim}, {Kinemuchi}, {Kjeldsen}, {Klaus}, {Li}, {Lucas}, {Meibom}, {Morris}, {Pr{\v{s}}a}, {Quintana}, {Sanderfer}, {Sasselov}, {Seader}, {Smith}, {Steffen}, {Still}, {Stumpe}, {Tarter}, {Tenenbaum}, {Torres}, {Twicken}, {Uddin}, {Van Cleve}, {Walkowicz}, \& {Welsh}}]{Batalha2013}
{Batalha}, N.~M., {Rowe}, J.~F., {Bryson}, S.~T., {et~al.} 2013, \apjs, 204, 24, \dodoi{10.1088/0067-0049/204/2/24}

\bibitem[{{Batygin} {et~al.}(2016){Batygin}, {Bodenheimer}, \& {Laughlin}}]{Batygin16}
{Batygin}, K., {Bodenheimer}, P.~H., \& {Laughlin}, G.~P. 2016, \apj, 829, 114, \dodoi{10.3847/0004-637X/829/2/114}

\bibitem[{{Beaug{\'e}} \& {Nesvorn{\'y}}(2012)}]{beauge2012}
{Beaug{\'e}}, C., \& {Nesvorn{\'y}}, D. 2012, \apj, 751, 119, \dodoi{10.1088/0004-637X/751/2/119}

\bibitem[{{Benz} {et~al.}(2021){Benz}, {Broeg}, {Fortier}, {Rando}, {Beck}, {Beck}, {Queloz}, {Ehrenreich}, {Maxted}, {Isaak}, {Billot}, {Alibert}, {Alonso}, {Ant{\'o}nio}, {Asquier}, {Bandy}, {B{\'a}rczy}, {Barrado}, {Barros}, {Baumjohann}, {Bekkelien}, {Bergomi}, {Biondi}, {Bonfils}, {Borsato}, {Brandeker}, {Busch}, {Cabrera}, {Cessa}, {Charnoz}, {Chazelas}, {Collier Cameron}, {Corral Van Damme}, {Cortes}, {Davies}, {Deleuil}, {Deline}, {Delrez}, {Demangeon}, {Demory}, {Erikson}, {Farinato}, {Fossati}, {Fridlund}, {Futyan}, {Gandolfi}, {Garcia Munoz}, {Gillon}, {Guterman}, {Gutierrez}, {Hasiba}, {Heng}, {Hernandez}, {Hoyer}, {Kiss}, {Kovacs}, {Kuntzer}, {Laskar}, {Lecavelier des Etangs}, {Lendl}, {L{\'o}pez}, {Lora}, {Lovis}, {L{\"u}ftinger}, {Magrin}, {Malvasio}, {Marafatto}, {Michaelis}, {de Miguel}, {Modrego}, {Munari}, {Nascimbeni}, {Olofsson}, {Ottacher}, {Ottensamer}, {Pagano}, {Palacios}, {Pall{\'e}}, {Peter}, {Piazza}, {Piotto}, {Pizarro}, {Pollaco}, {Ragazzoni}, {Ratti}, {Rauer}, {Ribas}, {Rieder},
  {Rohlfs}, {Safa}, {Salatti}, {Santos}, {Scandariato}, {S{\'e}gransan}, {Simon}, {Smith}, {Sordet}, {Sousa}, {Steller}, {Szab{\'o}}, {Szoke}, {Thomas}, {Tschentscher}, {Udry}, {Van Grootel}, {Viotto}, {Walter}, {Walton}, {Wildi}, \& {Wolter}}]{Benz2021}
{Benz}, W., {Broeg}, C., {Fortier}, A., {et~al.} 2021, Experimental Astronomy, 51, 109, \dodoi{10.1007/s10686-020-09679-4}

\bibitem[{{Boley} {et~al.}(2016){Boley}, {Granados Contreras}, \& {Gladman}}]{Boley16}
{Boley}, A.~C., {Granados Contreras}, A.~P., \& {Gladman}, B. 2016, \apjl, 817, L17, \dodoi{10.3847/2041-8205/817/2/L17}

\bibitem[{{Bourrier} {et~al.}(2017){Bourrier}, {Cegla}, {Lovis}, \& {Wyttenbach}}]{Bourrier2017}
{Bourrier}, V., {Cegla}, H.~M., {Lovis}, C., \& {Wyttenbach}, A. 2017, \aap, 599, A33, \dodoi{10.1051/0004-6361/201629973}

\bibitem[{{Bozhilov} {et~al.}(2023){Bozhilov}, {Antonova}, {Hobson}, {Brahm}, {Jord{\'a}n}, {Henning}, {Eberhardt}, {Rojas}, {Batygin}, {Torres-Miranda}, {Stassun}, {Millholland}, {Stoeva}, {Minev}, {Espinoza}, {Ricker}, {Latham}, {Dragomir}, {Kunimoto}, {Jenkins}, {Ting}, {Seager}, {Winn}, {Villasenor}, {Bouma}, {Medina}, \& {Trifonov}}]{Bozhilov2023}
{Bozhilov}, V., {Antonova}, D., {Hobson}, M.~J., {et~al.} 2023, \apjl, 946, L36, \dodoi{10.3847/2041-8213/acbd4f}

\bibitem[{{Brady} {et~al.}(2024){Brady}, {Bean}, {Stef{\'a}nsson}, {Brown}, {Seifahrt}, {Basant}, {Das}, {Luque}, \& {St{\"u}rmer}}]{Brady2024}
{Brady}, M., {Bean}, J., {Stef{\'a}nsson}, G., {et~al.} 2024, arXiv e-prints, arXiv:2411.10402, \dodoi{10.48550/arXiv.2411.10402}

\bibitem[{{Brahm} {et~al.}(2017{\natexlab{a}}){Brahm}, {Jord{\'a}n}, \& {Espinoza}}]{ceres}
{Brahm}, R., {Jord{\'a}n}, A., \& {Espinoza}, N. 2017{\natexlab{a}}, \pasp, 129, 034002, \dodoi{10.1088/1538-3873/aa5455}

\bibitem[{{Brahm} {et~al.}(2017{\natexlab{b}}){Brahm}, {Jord{\'a}n}, {Hartman}, \& {Bakos}}]{zaspe}
{Brahm}, R., {Jord{\'a}n}, A., {Hartman}, J., \& {Bakos}, G. 2017{\natexlab{b}}, \mnras, 467, 971, \dodoi{10.1093/mnras/stx144}

\bibitem[{{Brahm} {et~al.}(2019){Brahm}, {Espinoza}, {Jord{\'a}n}, {Henning}, {Sarkis}, {Jones}, {D{\'\i}az}, {Jenkins}, {Vanzi}, {Zapata}, {Petrovich}, {Kossakowski}, {Rabus}, {Rojas}, \& {Torres}}]{Brahm2019}
{Brahm}, R., {Espinoza}, N., {Jord{\'a}n}, A., {et~al.} 2019, \aj, 158, 45, \dodoi{10.3847/1538-3881/ab279a}

\bibitem[{{Bressan} {et~al.}(2012){Bressan}, {Marigo}, {Girardi}, {Salasnich}, {Dal Cero}, {Rubele}, \& {Nanni}}]{parsec}
{Bressan}, A., {Marigo}, P., {Girardi}, L., {et~al.} 2012, \mnras, 427, 127, \dodoi{10.1111/j.1365-2966.2012.21948.x}

\bibitem[{{Brown} {et~al.}(2013){Brown}, {Baliber}, {Bianco}, {Bowman}, {Burleson}, {Conway}, {Crellin}, {Depagne}, {De Vera}, {Dilday}, {Dragomir}, {Dubberley}, {Eastman}, {Elphick}, {Falarski}, {Foale}, {Ford}, {Fulton}, {Garza}, {Gomez}, {Graham}, {Greene}, {Haldeman}, {Hawkins}, {Haworth}, {Haynes}, {Hidas}, {Hjelstrom}, {Howell}, {Hygelund}, {Lister}, {Lobdill}, {Martinez}, {Mullins}, {Norbury}, {Parrent}, {Paulson}, {Petry}, {Pickles}, {Posner}, {Rosing}, {Ross}, {Sand}, {Saunders}, {Shobbrook}, {Shporer}, {Street}, {Thomas}, {Tsapras}, {Tufts}, {Valenti}, {Vander Horst}, {Walker}, {White}, \& {Willis}}]{Brown2013}
{Brown}, T.~M., {Baliber}, N., {Bianco}, F.~B., {et~al.} 2013, \pasp, 125, 1031, \dodoi{10.1086/673168}

\bibitem[{{Bryan} {et~al.}(2016){Bryan}, {Knutson}, {Howard}, {Ngo}, {Batygin}, {Crepp}, {Fulton}, {Hinkley}, {Isaacson}, {Johnson}, {Marcy}, \& {Wright}}]{Bryan2016}
{Bryan}, M.~L., {Knutson}, H.~A., {Howard}, A.~W., {et~al.} 2016, \apj, 821, 89, \dodoi{10.3847/0004-637X/821/2/89}

\bibitem[{{Carleo} {et~al.}(2024){Carleo}, {Malavolta}, {Desidera}, {Nardiello}, {Wang}, {Turrini}, {Lanza}, {Baratella}, {Marzari}, {Benatti}, {Biazzo}, {Bieryla}, {Brahm}, {Bonavita}, {Collins}, {Hellier}, {Locci}, {Hobson}, {Maggio}, {Mantovan}, {Messina}, {Pinamonti}, {Rodriguez}, {Sozzetti}, {Stassun}, {Wang}, {Ziegler}, {Damasso}, {Giacobbe}, {Murgas}, {Parviainen}, {Andreuzzi}, {Barkaoui}, {Berlind}, {Bignamini}, {Borsa}, {Brice{\~n}o}, {Brogi}, {Cabona}, {Calkins}, {Capuzzo-Dolcetta}, {Cecconi}, {Colon}, {Cosentino}, {Dragomir}, {Esquerdo}, {Henning}, {Ghedina}, {Goeke}, {Gratton}, {Horta}, {Gupta}, {Jenkins}, {Jord{\'a}n}, {Knapic}, {Latham}, {Mireles}, {Law}, {Lorenzi}, {Lund}, {Maldonado}, {Mann}, {Molinari}, {Pall{\'e}}, {Paegert}, {Pedani}, {Quinn}, {Scandariato}, {Seager}, {Winn}, {Wohler}, \& {Zingales}}]{Carleo2024}
{Carleo}, I., {Malavolta}, L., {Desidera}, S., {et~al.} 2024, \aap, 682, A135, \dodoi{10.1051/0004-6361/202348207}

\bibitem[{{Damasso} {et~al.}(2015){Damasso}, {Esposito}, {Nascimbeni}, {Desidera}, {Bonomo}, {Bieryla}, {Malavolta}, {Biazzo}, {Sozzetti}, {Covino}, {Latham}, {Gandolfi}, {Rainer}, {Petrovich}, {Collins}, {Boccato}, {Claudi}, {Cosentino}, {Gratton}, {Lanza}, {Maggio}, {Micela}, {Molinari}, {Pagano}, {Piotto}, {Poretti}, {Smareglia}, {Di Fabrizio}, {Giacobbe}, {Gomez-Jimenez}, {Murabito}, {Molinaro}, {Affer}, {Barbieri}, {Bedin}, {Benatti}, {Borsa}, {Maldonado}, {Mancini}, {Scandariato}, {Southworth}, \& {Zanmar Sanchez}}]{Damasso2015}
{Damasso}, M., {Esposito}, M., {Nascimbeni}, V., {et~al.} 2015, \aap, 581, L6, \dodoi{10.1051/0004-6361/201526995}

\bibitem[{{Davies}(2019)}]{Davies2019}
{Davies}, C.~L. 2019, \mnras, 484, 1926, \dodoi{10.1093/mnras/stz086}

\bibitem[{{Dawson} \& {Johnson}(2018)}]{Dawson18}
{Dawson}, R.~I., \& {Johnson}, J.~A. 2018, \araa, 56, 175, \dodoi{10.1146/annurev-astro-081817-051853}

\bibitem[{{Debras} {et~al.}(2021){Debras}, {Baruteau}, \& {Donati}}]{Debras2021}
{Debras}, F., {Baruteau}, C., \& {Donati}, J.-F. 2021, \mnras, 500, 1621, \dodoi{10.1093/mnras/staa3397}

\bibitem[{{Dong} \& {Foreman-Mackey}(2023)}]{Dong23}
{Dong}, J., \& {Foreman-Mackey}, D. 2023, \aj, 166, 112, \dodoi{10.3847/1538-3881/ace105}

\bibitem[{{Doyle} {et~al.}(2023){Doyle}, {Cegla}, {Anderson}, {Lendl}, {Bourrier}, {Bryant}, {Vines}, {Allart}, {Bayliss}, {Burleigh}, {Buchschacher}, {Casewell}, {Hawthorn}, {Jenkins}, {Lafarga}, {Moyano}, {Psaridi}, {Roguet-Kern}, {Sosnowska}, \& {Wheatley}}]{Doyle2023}
{Doyle}, L., {Cegla}, H.~M., {Anderson}, D.~R., {et~al.} 2023, \mnras, 522, 4499, \dodoi{10.1093/mnras/stad1240}

\bibitem[{{Doyle} {et~al.}(2024){Doyle}, {Ca{\~n}as}, {Libby-Roberts}, {Cegla}, {Stef{\'a}nsson}, {Anderson}, {Armstrong}, {Bender}, {Bayliss}, {Carmichael}, {Casewell}, {Kanodia}, {Lafarga}, {Lin}, {Mahadevan}, {Monson}, {Robertson}, \& {Veras}}]{Doyle2024}
{Doyle}, L., {Ca{\~n}as}, C.~I., {Libby-Roberts}, J.~E., {et~al.} 2024, arXiv e-prints, arXiv:2411.18567, \dodoi{10.48550/arXiv.2411.18567}

\bibitem[{{El-Badry} {et~al.}(2021){El-Badry}, {Rix}, \& {Heintz}}]{El-Badry21}
{El-Badry}, K., {Rix}, H.-W., \& {Heintz}, T.~M. 2021, \mnras, 506, 2269, \dodoi{10.1093/mnras/stab323}

\bibitem[{{Epstein} \& {Pinsonneault}(2014)}]{Epstein2014}
{Epstein}, C.~R., \& {Pinsonneault}, M.~H. 2014, \apj, 780, 159, \dodoi{10.1088/0004-637X/780/2/159}

\bibitem[{{Espinoza} {et~al.}(2019){Espinoza}, {Kossakowski}, \& {Brahm}}]{juliet}
{Espinoza}, N., {Kossakowski}, D., \& {Brahm}, R. 2019, \mnras, 490, 2262, \dodoi{10.1093/mnras/stz2688}

\bibitem[{{Espinoza-Retamal} {et~al.}(2023{\natexlab{a}}){Espinoza-Retamal}, {Zhu}, \& {Petrovich}}]{Espinoza-Retamal2023a}
{Espinoza-Retamal}, J.~I., {Zhu}, W., \& {Petrovich}, C. 2023{\natexlab{a}}, \aj, 166, 231, \dodoi{10.3847/1538-3881/ad00b9}

\bibitem[{{Espinoza-Retamal} {et~al.}(2023{\natexlab{b}}){Espinoza-Retamal}, {Brahm}, {Petrovich}, {Jord{\'a}n}, {Stef{\'a}nsson}, {Sedaghati}, {Hobson}, {Mu{\~n}oz}, {Boyle}, {Leiva}, \& {Suc}}]{Espinoza-Retamal2023b}
{Espinoza-Retamal}, J.~I., {Brahm}, R., {Petrovich}, C., {et~al.} 2023{\natexlab{b}}, \apjl, 958, L20, \dodoi{10.3847/2041-8213/ad096d}

\bibitem[{{Espinoza-Retamal} {et~al.}(2024){Espinoza-Retamal}, {Stef{\'a}nsson}, {Petrovich}, {Brahm}, {Jord{\'a}n}, {Sedaghati}, {Lucero}, {Pinto}, {Mu{\~n}oz}, {Boyle}, {Leiva}, \& {Suc}}]{Espinoza-Retamal2024}
{Espinoza-Retamal}, J.~I., {Stef{\'a}nsson}, G., {Petrovich}, C., {et~al.} 2024, \aj, 168, 185, \dodoi{10.3847/1538-3881/ad70b8}

\bibitem[{{Fabrycky} \& {Tremaine}(2007)}]{Fabrycky07}
{Fabrycky}, D., \& {Tremaine}, S. 2007, \apj, 669, 1298, \dodoi{10.1086/521702}

\bibitem[{{Ferreira dos Santos} {et~al.}(2024){Ferreira dos Santos}, {Rice}, {Wang}, \& {Wang}}]{dosSantos2024}
{Ferreira dos Santos}, T., {Rice}, M., {Wang}, X.-Y., \& {Wang}, S. 2024, \aj, 168, 145, \dodoi{10.3847/1538-3881/ad6b7f}

\bibitem[{{Ford} {et~al.}(2001){Ford}, {Havlickova}, \& {Rasio}}]{ford_2011_scattering}
{Ford}, E.~B., {Havlickova}, M., \& {Rasio}, F.~A. 2001, \icarus, 150, 303, \dodoi{10.1006/icar.2001.6588}

\bibitem[{{Ford} \& {Rasio}(2008)}]{ford_rasio2008}
{Ford}, E.~B., \& {Rasio}, F.~A. 2008, \apj, 686, 621, \dodoi{10.1086/590926}

\bibitem[{{Foreman-Mackey} {et~al.}(2017){Foreman-Mackey}, {Agol}, {Angus}, \& {Ambikasaran}}]{celerite}
{Foreman-Mackey}, D., {Agol}, E., {Angus}, R., \& {Ambikasaran}, S. 2017, AJ, 154, 220, \dodoi{10.3847/1538-3881/aa9332}

\bibitem[{{Freudling} {et~al.}(2013){Freudling}, {Romaniello}, {Bramich}, {Ballester}, {Forchi}, {Garc{\'\i}a-Dabl{\'o}}, {Moehler}, \& {Neeser}}]{Freudling13}
{Freudling}, W., {Romaniello}, M., {Bramich}, D.~M., {et~al.} 2013, \aap, 559, A96, \dodoi{10.1051/0004-6361/201322494}

\bibitem[{{Fulton} {et~al.}(2018){Fulton}, {Petigura}, {Blunt}, \& {Sinukoff}}]{radvel}
{Fulton}, B.~J., {Petigura}, E.~A., {Blunt}, S., \& {Sinukoff}, E. 2018, \pasp, 130, 044504, \dodoi{10.1088/1538-3873/aaaaa8}

\bibitem[{{Gaia Collaboration} {et~al.}(2021){Gaia Collaboration}, {Brown}, {Vallenari}, {Prusti}, {de Bruijne}, {Babusiaux}, {Biermann}, {Creevey}, {Evans}, {Eyer}, {Hutton}, {Jansen}, {Jordi}, {Klioner}, {Lammers}, {Lindegren}, {Luri}, {Mignard}, {Panem}, {Pourbaix}, {Randich}, {Sartoretti}, {Soubiran}, {Walton}, {Arenou}, {Bailer-Jones}, {Bastian}, {Cropper}, {Drimmel}, {Katz}, {Lattanzi}, {van Leeuwen}, {Bakker}, {Cacciari}, {Casta{\~n}eda}, {De Angeli}, {Ducourant}, {Fabricius}, {Fouesneau}, {Fr{\'e}mat}, {Guerra}, {Guerrier}, {Guiraud}, {Jean-Antoine Piccolo}, {Masana}, {Messineo}, {Mowlavi}, {Nicolas}, {Nienartowicz}, {Pailler}, {Panuzzo}, {Riclet}, {Roux}, {Seabroke}, {Sordo}, {Tanga}, {Th{\'e}venin}, {Gracia-Abril}, {Portell}, {Teyssier}, {Altmann}, {Andrae}, {Bellas-Velidis}, {Benson}, {Berthier}, {Blomme}, {Brugaletta}, {Burgess}, {Busso}, {Carry}, {Cellino}, {Cheek}, {Clementini}, {Damerdji}, {Davidson}, {Delchambre}, {Dell'Oro}, {Fern{\'a}ndez-Hern{\'a}ndez}, {Galluccio}, {Garc{\'\i}a-Lario},
  {Garcia-Reinaldos}, {Gonz{\'a}lez-N{\'u}{\~n}ez}, {Gosset}, {Haigron}, {Halbwachs}, {Hambly}, {Harrison}, {Hatzidimitriou}, {Heiter}, {Hern{\'a}ndez}, {Hestroffer}, {Hodgkin}, {Holl}, {Jan{\ss}en}, {Jevardat de Fombelle}, {Jordan}, {Krone-Martins}, {Lanzafame}, {L{\"o}ffler}, {Lorca}, {Manteiga}, {Marchal}, {Marrese}, {Moitinho}, {Mora}, {Muinonen}, {Osborne}, {Pancino}, {Pauwels}, {Petit}, {Recio-Blanco}, {Richards}, {Riello}, {Rimoldini}, {Robin}, {Roegiers}, {Rybizki}, {Sarro}, {Siopis}, {Smith}, {Sozzetti}, {Ulla}, {Utrilla}, {van Leeuwen}, {van Reeven}, {Abbas}, {Abreu Aramburu}, {Accart}, {Aerts}, {Aguado}, {Ajaj}, {Altavilla}, {{\'A}lvarez}, {{\'A}lvarez Cid-Fuentes}, {Alves}, {Anderson}, {Anglada Varela}, {Antoja}, {Audard}, {Baines}, {Baker}, {Balaguer-N{\'u}{\~n}ez}, {Balbinot}, {Balog}, {Barache}, {Barbato}, {Barros}, {Barstow}, {Bartolom{\'e}}, {Bassilana}, {Bauchet}, {Baudesson-Stella}, {Becciani}, {Bellazzini}, {Bernet}, {Bertone}, {Bianchi}, {Blanco-Cuaresma}, {Boch}, {Bombrun}, {Bossini},
  {Bouquillon}, {Bragaglia}, {Bramante}, {Breedt}, {Bressan}, {Brouillet}, {Bucciarelli}, {Burlacu}, {Busonero}, {Butkevich}, {Buzzi}, {Caffau}, {Cancelliere}, {C{\'a}novas}, {Cantat-Gaudin}, {Carballo}, {Carlucci}, {Carnerero}, {Carrasco}, {Casamiquela}, {Castellani}, {Castro-Ginard}, {Castro Sampol}, {Chaoul}, {Charlot}, {Chemin}, {Chiavassa}, {Cioni}, {Comoretto}, {Cooper}, {Cornez}, {Cowell}, {Crifo}, {Crosta}, {Crowley}, {Dafonte}, {Dapergolas}, {David}, {David}, {de Laverny}, {De Luise}, {De March}, {De Ridder}, {de Souza}, {de Teodoro}, {de Torres}, {del Peloso}, {del Pozo}, {Delbo}, {Delgado}, {Delgado}, {Delisle}, {Di Matteo}, {Diakite}, {Diener}, {Distefano}, {Dolding}, {Eappachen}, {Edvardsson}, {Enke}, {Esquej}, {Fabre}, {Fabrizio}, {Faigler}, {Fedorets}, {Fernique}, {Fienga}, {Figueras}, {Fouron}, {Fragkoudi}, {Fraile}, {Franke}, {Gai}, {Garabato}, {Garcia-Gutierrez}, {Garc{\'\i}a-Torres}, {Garofalo}, {Gavras}, {Gerlach}, {Geyer}, {Giacobbe}, {Gilmore}, {Girona}, {Giuffrida}, {Gomel}, {Gomez},
  {Gonzalez-Santamaria}, {Gonz{\'a}lez-Vidal}, {Granvik}, {Guti{\'e}rrez-S{\'a}nchez}, {Guy}, {Hauser}, {Haywood}, {Helmi}, {Hidalgo}, {Hilger}, {H{\l}adczuk}, {Hobbs}, {Holland}, {Huckle}, {Jasniewicz}, {Jonker}, {Juaristi Campillo}, {Julbe}, {Karbevska}, {Kervella}, {Khanna}, {Kochoska}, {Kontizas}, {Kordopatis}, {Korn}, {Kostrzewa-Rutkowska}, {Kruszy{\'n}ska}, {Lambert}, {Lanza}, {Lasne}, {Le Campion}, {Le Fustec}, {Lebreton}, {Lebzelter}, {Leccia}, {Leclerc}, {Lecoeur-Taibi}, {Liao}, {Licata}, {Lindstr{\o}m}, {Lister}, {Livanou}, {Lobel}, {Madrero Pardo}, {Managau}, {Mann}, {Marchant}, {Marconi}, {Marcos Santos}, {Marinoni}, {Marocco}, {Marshall}, {Martin Polo}, {Mart{\'\i}n-Fleitas}, {Masip}, {Massari}, {Mastrobuono-Battisti}, {Mazeh}, {McMillan}, {Messina}, {Michalik}, {Millar}, {Mints}, {Molina}, {Molinaro}, {Moln{\'a}r}, {Montegriffo}, {Mor}, {Morbidelli}, {Morel}, {Morris}, {Mulone}, {Munoz}, {Muraveva}, {Murphy}, {Musella}, {Noval}, {Ord{\'e}novic}, {Orr{\`u}}, {Osinde}, {Pagani}, {Pagano},
  {Palaversa}, {Palicio}, {Panahi}, {Pawlak}, {Pe{\~n}alosa Esteller}, {Penttil{\"a}}, {Piersimoni}, {Pineau}, {Plachy}, {Plum}, {Poggio}, {Poretti}, {Poujoulet}, {Pr{\v{s}}a}, {Pulone}, {Racero}, {Ragaini}, {Rainer}, {Raiteri}, {Rambaux}, {Ramos}, {Ramos-Lerate}, {Re Fiorentin}, {Regibo}, {Reyl{\'e}}, {Ripepi}, {Riva}, {Rixon}, {Robichon}, {Robin}, {Roelens}, {Rohrbasser}, {Romero-G{\'o}mez}, {Rowell}, {Royer}, {Rybicki}, {Sadowski}, {Sagrist{\`a} Sell{\'e}s}, {Sahlmann}, {Salgado}, {Salguero}, {Samaras}, {Sanchez Gimenez}, {Sanna}, {Santove{\~n}a}, {Sarasso}, {Schultheis}, {Sciacca}, {Segol}, {Segovia}, {S{\'e}gransan}, {Semeux}, {Shahaf}, {Siddiqui}, {Siebert}, {Siltala}, {Slezak}, {Smart}, {Solano}, {Solitro}, {Souami}, {Souchay}, {Spagna}, {Spoto}, {Steele}, {Steidelm{\"u}ller}, {Stephenson}, {S{\"u}veges}, {Szabados}, {Szegedi-Elek}, {Taris}, {Tauran}, {Taylor}, {Teixeira}, {Thuillot}, {Tonello}, {Torra}, {Torra}, {Turon}, {Unger}, {Vaillant}, {van Dillen}, {Vanel}, {Vecchiato}, {Viala}, {Vicente},
  {Voutsinas}, {Weiler}, {Wevers}, {Wyrzykowski}, {Yoldas}, {Yvard}, {Zhao}, {Zorec}, {Zucker}, {Zurbach}, \& {Zwitter}}]{Gaia_eDR3}
{Gaia Collaboration}, {Brown}, A.~G.~A., {Vallenari}, A., {et~al.} 2021, \aap, 649, A1, \dodoi{10.1051/0004-6361/202039657}

\bibitem[{{Gaia Collaboration} {et~al.}(2023){Gaia Collaboration}, {Vallenari}, {Brown}, {Prusti}, {de Bruijne}, {Arenou}, {Babusiaux}, {Biermann}, {Creevey}, {Ducourant}, {Evans}, {Eyer}, {Guerra}, {Hutton}, {Jordi}, {Klioner}, {Lammers}, {Lindegren}, {Luri}, {Mignard}, {Panem}, {Pourbaix}, {Randich}, {Sartoretti}, {Soubiran}, {Tanga}, {Walton}, {Bailer-Jones}, {Bastian}, {Drimmel}, {Jansen}, {Katz}, {Lattanzi}, {van Leeuwen}, {Bakker}, {Cacciari}, {Casta{\~n}eda}, {De Angeli}, {Fabricius}, {Fouesneau}, {Fr{\'e}mat}, {Galluccio}, {Guerrier}, {Heiter}, {Masana}, {Messineo}, {Mowlavi}, {Nicolas}, {Nienartowicz}, {Pailler}, {Panuzzo}, {Riclet}, {Roux}, {Seabroke}, {Sordo}, {Th{\'e}venin}, {Gracia-Abril}, {Portell}, {Teyssier}, {Altmann}, {Andrae}, {Audard}, {Bellas-Velidis}, {Benson}, {Berthier}, {Blomme}, {Burgess}, {Busonero}, {Busso}, {C{\'a}novas}, {Carry}, {Cellino}, {Cheek}, {Clementini}, {Damerdji}, {Davidson}, {de Teodoro}, {Nu{\~n}ez Campos}, {Delchambre}, {Dell'Oro}, {Esquej},
  {Fern{\'a}ndez-Hern{\'a}ndez}, {Fraile}, {Garabato}, {Garc{\'\i}a-Lario}, {Gosset}, {Haigron}, {Halbwachs}, {Hambly}, {Harrison}, {Hern{\'a}ndez}, {Hestroffer}, {Hodgkin}, {Holl}, {Jan{\ss}en}, {Jevardat de Fombelle}, {Jordan}, {Krone-Martins}, {Lanzafame}, {L{\"o}ffler}, {Marchal}, {Marrese}, {Moitinho}, {Muinonen}, {Osborne}, {Pancino}, {Pauwels}, {Recio-Blanco}, {Reyl{\'e}}, {Riello}, {Rimoldini}, {Roegiers}, {Rybizki}, {Sarro}, {Siopis}, {Smith}, {Sozzetti}, {Utrilla}, {van Leeuwen}, {Abbas}, {{\'A}brah{\'a}m}, {Abreu Aramburu}, {Aerts}, {Aguado}, {Ajaj}, {Aldea-Montero}, {Altavilla}, {{\'A}lvarez}, {Alves}, {Anders}, {Anderson}, {Anglada Varela}, {Antoja}, {Baines}, {Baker}, {Balaguer-N{\'u}{\~n}ez}, {Balbinot}, {Balog}, {Barache}, {Barbato}, {Barros}, {Barstow}, {Bartolom{\'e}}, {Bassilana}, {Bauchet}, {Becciani}, {Bellazzini}, {Berihuete}, {Bernet}, {Bertone}, {Bianchi}, {Binnenfeld}, {Blanco-Cuaresma}, {Blazere}, {Boch}, {Bombrun}, {Bossini}, {Bouquillon}, {Bragaglia}, {Bramante}, {Breedt},
  {Bressan}, {Brouillet}, {Brugaletta}, {Bucciarelli}, {Burlacu}, {Butkevich}, {Buzzi}, {Caffau}, {Cancelliere}, {Cantat-Gaudin}, {Carballo}, {Carlucci}, {Carnerero}, {Carrasco}, {Casamiquela}, {Castellani}, {Castro-Ginard}, {Chaoul}, {Charlot}, {Chemin}, {Chiaramida}, {Chiavassa}, {Chornay}, {Comoretto}, {Contursi}, {Cooper}, {Cornez}, {Cowell}, {Crifo}, {Cropper}, {Crosta}, {Crowley}, {Dafonte}, {Dapergolas}, {David}, {David}, {de Laverny}, {De Luise}, \& {De March}}]{gaia:dr3}
{Gaia Collaboration}, {Vallenari}, A., {Brown}, A.~G.~A., {et~al.} 2023, \aap, 674, A1, \dodoi{10.1051/0004-6361/202243940}

\bibitem[{{Gelman} {et~al.}(2014){Gelman}, {Carlin}, {Stern}, {Dunson}, {Vehtari}, \& {Rubin}}]{Gelman14}
{Gelman}, A., {Carlin}, J.~B., {Stern}, H.~S., {et~al.} 2014, {Bayesian Data Analysis}

\bibitem[{{Giacalone} {et~al.}(2024){Giacalone}, {Dai}, {Zanazzi}, {Howard}, {Dressing}, {Winn}, {Rubenzahl}, {Carmichael}, {Vowell}, {Kesseli}, {Halverson}, {Isaacson}, {Brodheim}, {Deich}, {Fulton}, {Gibson}, {Hill}, {Holden}, {Householder}, {Kaye}, {Laher}, {Lanclos}, {Payne}, {Petigura}, {Roy}, {Schwab}, {Shaum}, {Sirk}, {Smith}, {Stef{\'a}nsson}, {Walawender}, {Wang}, {Weiss}, \& {Yeh}}]{Giacalone2024}
{Giacalone}, S., {Dai}, F., {Zanazzi}, J.~J., {et~al.} 2024, \aj, 168, 189, \dodoi{10.3847/1538-3881/ad785a}

\bibitem[{{Goldreich} \& {Tremaine}(1980)}]{Goldreich80}
{Goldreich}, P., \& {Tremaine}, S. 1980, \apj, 241, 425, \dodoi{10.1086/158356}

\bibitem[{{Gray}(1984)}]{Gray84}
{Gray}, D.~F. 1984, \apj, 281, 719, \dodoi{10.1086/162149}

\bibitem[{{Gupta} {et~al.}(2024){Gupta}, {Millholland}, {Im}, {Dong}, {Jackson}, {Carleo}, {Libby-Roberts}, {Delamer}, {Giovinazzi}, {Lin}, {Kanodia}, {Wang}, {Stassun}, {Masseron}, {Dragomir}, {Mahadevan}, {Wright}, {Alvarado-Montes}, {Bender}, {Blake}, {Caldwell}, {Ca{\~n}as}, {Cochran}, {Dalba}, {Everett}, {Fernandez}, {Golub}, {Guillet}, {Halverson}, {Hebb}, {Higuera}, {Huang}, {Klusmeyer}, {Knight}, {Leroux}, {Logsdon}, {Loose}, {McElwain}, {Monson}, {Ninan}, {Nowak}, {Palle}, {Patel}, {Pepper}, {Primm}, {Rajagopal}, {Robertson}, {Roy}, {Schneider}, {Schwab}, {Schweiker}, {Sgro}, {Shimizu}, {Simard}, {Stef{\'a}nsson}, {Stevens}, {Villanueva}, {Wisniewski}, {Will}, \& {Ziegler}}]{Gupta2024}
{Gupta}, A.~F., {Millholland}, S.~C., {Im}, H., {et~al.} 2024, \nat, 632, 50, \dodoi{10.1038/s41586-024-07688-3}

\bibitem[{{Harre} {et~al.}(2023){Harre}, {Smith}, {Hirano}, {Csizmadia}, {J. Triaud}, \& {Anderson}}]{Harre2023}
{Harre}, J.-V., {Smith}, A. M.~S., {Hirano}, T., {et~al.} 2023, \aj, 166, 159, \dodoi{10.3847/1538-3881/acf46d}

\bibitem[{{H{\'e}brard} {et~al.}(2010){H{\'e}brard}, {D{\'e}sert}, {D{\'\i}az}, {Boisse}, {Bouchy}, {Lecavelier Des Etangs}, {Moutou}, {Ehrenreich}, {Arnold}, {Bonfils}, {Delfosse}, {Desort}, {Eggenberger}, {Forveille}, {Gregorio}, {Lagrange}, {Lovis}, {Pepe}, {Perrier}, {Pont}, {Queloz}, {Santerne}, {Santos}, {S{\'e}gransan}, {Sing}, {Udry}, \& {Vidal-Madjar}}]{Hebrard2010}
{H{\'e}brard}, G., {D{\'e}sert}, J.~M., {D{\'\i}az}, R.~F., {et~al.} 2010, \aap, 516, A95, \dodoi{10.1051/0004-6361/201014327}

\bibitem[{{Hellier} {et~al.}(2017){Hellier}, {Anderson}, {Collier Cameron}, {Delrez}, {Gillon}, {Jehin}, {Lendl}, {Maxted}, {Neveu-VanMalle}, {Pepe}, {Pollacco}, {Queloz}, {S{\'e}gransan}, {Smalley}, {Southworth}, {Triaud}, {Udry}, {Wagg}, \& {West}}]{Hellier2017}
{Hellier}, C., {Anderson}, D.~R., {Collier Cameron}, A., {et~al.} 2017, \mnras, 465, 3693, \dodoi{10.1093/mnras/stw3005}

\bibitem[{{Hirano} {et~al.}(2010){Hirano}, {Suto}, {Taruya}, {Narita}, {Sato}, {Johnson}, \& {Winn}}]{Hirano10}
{Hirano}, T., {Suto}, Y., {Taruya}, A., {et~al.} 2010, \apj, 709, 458, \dodoi{10.1088/0004-637X/709/1/458}

\bibitem[{{Hjorth} {et~al.}(2021){Hjorth}, {Albrecht}, {Hirano}, {Winn}, {Dawson}, {Zanazzi}, {Knudstrup}, \& {Sato}}]{Hjorth2021}
{Hjorth}, M., {Albrecht}, S., {Hirano}, T., {et~al.} 2021, Proceedings of the National Academy of Science, 118, e2017418118, \dodoi{10.1073/pnas.2017418118}

\bibitem[{{Hogg} {et~al.}(2010){Hogg}, {Myers}, \& {Bovy}}]{Hogg2010}
{Hogg}, D.~W., {Myers}, A.~D., \& {Bovy}, J. 2010, \apj, 725, 2166, \dodoi{10.1088/0004-637X/725/2/2166}

\bibitem[{{Howell} {et~al.}(2014){Howell}, {Sobeck}, {Haas}, {Still}, {Barclay}, {Mullally}, {Troeltzsch}, {Aigrain}, {Bryson}, {Caldwell}, {Chaplin}, {Cochran}, {Huber}, {Marcy}, {Miglio}, {Najita}, {Smith}, {Twicken}, \& {Fortney}}]{Howell2014}
{Howell}, S.~B., {Sobeck}, C., {Haas}, M., {et~al.} 2014, \pasp, 126, 398, \dodoi{10.1086/676406}

\bibitem[{{Hu} {et~al.}(2024){Hu}, {Rice}, {Wang}, {Wang}, {Shporer}, {Teske}, {Yee}, {Butler}, {Shectman}, {Crane}, {Collins}, \& {Collins}}]{Hu2024}
{Hu}, Q., {Rice}, M., {Wang}, X.-Y., {et~al.} 2024, \aj, 167, 175, \dodoi{10.3847/1538-3881/ad2855}

\bibitem[{{Hurt} \& {MacGregor}(2023)}]{Hurt2023}
{Hurt}, S.~A., \& {MacGregor}, M.~A. 2023, \apj, 954, 10, \dodoi{10.3847/1538-4357/accf9d}

\bibitem[{{Ikwut-Ukwa} {et~al.}(2022){Ikwut-Ukwa}, {Rodriguez}, {Quinn}, {Zhou}, {Vanderburg}, {Ali}, {Bunten}, {Gaudi}, {Latham}, {Howell}, {Huang}, {Bieryla}, {Collins}, {Carmichael}, {Rabus}, {Eastman}, {Collins}, {Tan}, {Schwarz}, {Myers}, {Stockdale}, {Kielkopf}, {Radford}, {Oelkers}, {Jenkins}, {Ricker}, {Seager}, {Vanderspek}, {Winn}, {Burt}, {Butler}, {Calkins}, {Crane}, {Gnilka}, {Esquerdo}, {Fong}, {Kreidberg}, {Mink}, {Rodriguez}, {Schlieder}, {Shectman}, {Shporer}, {Teske}, {Ting}, {Villase{\~n}or}, \& {Yahalomi}}]{Ikwut-Ukwa2022}
{Ikwut-Ukwa}, M., {Rodriguez}, J.~E., {Quinn}, S.~N., {et~al.} 2022, \aj, 163, 9, \dodoi{10.3847/1538-3881/ac2ee1}

\bibitem[{{Jenkins} {et~al.}(2016){Jenkins}, {Twicken}, {McCauliff}, {Campbell}, {Sanderfer}, {Lung}, {Mansouri-Samani}, {Girouard}, {Tenenbaum}, {Klaus}, {Smith}, {Caldwell}, {Chacon}, {Henze}, {Heiges}, {Latham}, {Morgan}, {Swade}, {Rinehart}, \& {Vanderspek}}]{Jenkins16}
{Jenkins}, J.~M., {Twicken}, J.~D., {McCauliff}, S., {et~al.} 2016, in Society of Photo-Optical Instrumentation Engineers (SPIE) Conference Series, Vol. 9913, Software and Cyberinfrastructure for Astronomy IV, ed. G.~{Chiozzi} \& J.~C. {Guzman}, 99133E, \dodoi{10.1117/12.2233418}

\bibitem[{{Kaufer} {et~al.}(1999){Kaufer}, {Stahl}, {Tubbesing}, {N{\o}rregaard}, {Avila}, {Francois}, {Pasquini}, \& {Pizzella}}]{feros}
{Kaufer}, A., {Stahl}, O., {Tubbesing}, S., {et~al.} 1999, The Messenger, 95, 8

\bibitem[{{Kipping}(2013)}]{Kipping13}
{Kipping}, D.~M. 2013, \mnras, 435, 2152, \dodoi{10.1093/mnras/stt1435}

\bibitem[{{Knudstrup} {et~al.}(2024){Knudstrup}, {Albrecht}, {Winn}, {Gandolfi}, {Zanazzi}, {Persson}, {Fridlund}, {Marcussen}, {Chontos}, {Keniger}, {Eisner}, {Bieryla}, {Isaacson}, {Howard}, {Hirsch}, {Murgas}, {Narita}, {Palle}, {Kawai}, \& {Baker}}]{Knudstrup2024}
{Knudstrup}, E., {Albrecht}, S.~H., {Winn}, J.~N., {et~al.} 2024, \aap, 690, A379, \dodoi{10.1051/0004-6361/202450627}

\bibitem[{{Kraft}(1967)}]{Kraft1967}
{Kraft}, R.~P. 1967, \apj, 150, 551, \dodoi{10.1086/149359}

\bibitem[{{Kraus} {et~al.}(2020){Kraus}, {Le Bouquin}, {Kreplin}, {Davies}, {Hone}, {Monnier}, {Gardner}, {Kennedy}, \& {Hinkley}}]{Kraus2020}
{Kraus}, S., {Le Bouquin}, J.-B., {Kreplin}, A., {et~al.} 2020, \apjl, 897, L8, \dodoi{10.3847/2041-8213/ab9d27}

\bibitem[{{Kreidberg}(2015)}]{batman}
{Kreidberg}, L. 2015, \pasp, 127, 1161, \dodoi{10.1086/683602}

\bibitem[{{Lightkurve Collaboration} {et~al.}(2018){Lightkurve Collaboration}, {Cardoso}, {Hedges}, {Gully-Santiago}, {Saunders}, {Cody}, {Barclay}, {Hall}, {Sagear}, {Turtelboom}, {Zhang}, {Tzanidakis}, {Mighell}, {Coughlin}, {Bell}, {Berta-Thompson}, {Williams}, {Dotson}, \& {Barentsen}}]{Lightkurve}
{Lightkurve Collaboration}, {Cardoso}, J.~V.~d.~M., {Hedges}, C., {et~al.} 2018, {Lightkurve: Kepler and TESS time series analysis in Python}, Astrophysics Source Code Library.
\newblock \doeprint{1812.013}

\bibitem[{{Lin} \& {Papaloizou}(1986)}]{Lin86}
{Lin}, D.~N.~C., \& {Papaloizou}, J. 1986, \apj, 309, 846, \dodoi{10.1086/164653}

\bibitem[{{Louden} \& {Millholland}(2024)}]{louden}
{Louden}, E.~M., \& {Millholland}, S.~C. 2024, \apj, 974, 304, \dodoi{10.3847/1538-4357/ad74ff}

\bibitem[{{Luger} {et~al.}(2016){Luger}, {Agol}, {Kruse}, {Barnes}, {Becker}, {Foreman-Mackey}, \& {Deming}}]{Luger2016}
{Luger}, R., {Agol}, E., {Kruse}, E., {et~al.} 2016, \aj, 152, 100, \dodoi{10.3847/0004-6256/152/4/100}

\bibitem[{{Luger} {et~al.}(2018){Luger}, {Kruse}, {Foreman-Mackey}, {Agol}, \& {Saunders}}]{Luger2018}
{Luger}, R., {Kruse}, E., {Foreman-Mackey}, D., {Agol}, E., \& {Saunders}, N. 2018, \aj, 156, 99, \dodoi{10.3847/1538-3881/aad230}

\bibitem[{{Mancini} {et~al.}(2022){Mancini}, {Esposito}, {Covino}, {Southworth}, {Poretti}, {Andreuzzi}, {Barbato}, {Biazzo}, {Borsato}, {Bruni}, {Damasso}, {Di Fabrizio}, {Evans}, {Granata}, {Lanza}, {Naponiello}, {Nascimbeni}, {Pinamonti}, {Sozzetti}, {Tregloan-Reed}, {Basilicata}, {Bignamini}, {Bonomo}, {Claudi}, {Cosentino}, {Desidera}, {Fiorenzano}, {Giacobbe}, {Harutyunyan}, {Henning}, {Knapic}, {Maggio}, {Micela}, {Molinari}, {Pagano}, {Pedani}, \& {Piotto}}]{Mancini2022}
{Mancini}, L., {Esposito}, M., {Covino}, E., {et~al.} 2022, \aap, 664, A162, \dodoi{10.1051/0004-6361/202243742}

\bibitem[{{Masuda} \& {Winn}(2020)}]{Masuda2020}
{Masuda}, K., \& {Winn}, J.~N. 2020, \aj, 159, 81, \dodoi{10.3847/1538-3881/ab65be}

\bibitem[{{Mayor} \& {Queloz}(1995)}]{Mayor95}
{Mayor}, M., \& {Queloz}, D. 1995, \nat, 378, 355, \dodoi{10.1038/378355a0}

\bibitem[{{Modigliani} {et~al.}(2020){Modigliani}, {Freudling}, {Anderson}, {Lovis}, {Sosnowska}, \& {Segovia}}]{Modigliani20}
{Modigliani}, A., {Freudling}, W., {Anderson}, R.~I., {et~al.} 2020, in Astronomical Society of the Pacific Conference Series, Vol. 527, Astronomical Data Analysis Software and Systems XXIX, ed. R.~{Pizzo}, E.~R. {Deul}, J.~D. {Mol}, J.~{de Plaa}, \& H.~{Verkouter}, 667

\bibitem[{{Morgan} {et~al.}(2024){Morgan}, {Bowler}, {Tran}, {Petigura}, {Nagpal}, \& {Blunt}}]{Morgan2024}
{Morgan}, M., {Bowler}, B.~P., {Tran}, Q.~H., {et~al.} 2024, \aj, 167, 48, \dodoi{10.3847/1538-3881/ad0728}

\bibitem[{{Morton} \& {Winn}(2014)}]{Morton2014}
{Morton}, T.~D., \& {Winn}, J.~N. 2014, \apj, 796, 47, \dodoi{10.1088/0004-637X/796/1/47}

\bibitem[{{Mu{\~n}oz} \& {Perets}(2018)}]{Munoz2018}
{Mu{\~n}oz}, D.~J., \& {Perets}, H.~B. 2018, \aj, 156, 253, \dodoi{10.3847/1538-3881/aae7d0}

\bibitem[{{Naoz} {et~al.}(2011){Naoz}, {Farr}, {Lithwick}, {Rasio}, \& {Teyssandier}}]{Naoz11}
{Naoz}, S., {Farr}, W.~M., {Lithwick}, Y., {Rasio}, F.~A., \& {Teyssandier}, J. 2011, \nat, 473, 187, \dodoi{10.1038/nature10076}

\bibitem[{{Pepe} {et~al.}(2021){Pepe}, {Cristiani}, {Rebolo}, {Santos}, {Dekker}, {Cabral}, {Di Marcantonio}, {Figueira}, {Lo Curto}, {Lovis}, {Mayor}, {M{\'e}gevand}, {Molaro}, {Riva}, {Zapatero Osorio}, {Amate}, {Manescau}, {Pasquini}, {Zerbi}, {Adibekyan}, {Abreu}, {Affolter}, {Alibert}, {Aliverti}, {Allart}, {Allende Prieto}, {{\'A}lvarez}, {Alves}, {Avila}, {Baldini}, {Bandy}, {Barros}, {Benz}, {Bianco}, {Borsa}, {Bourrier}, {Bouchy}, {Broeg}, {Calderone}, {Cirami}, {Coelho}, {Conconi}, {Coretti}, {Cumani}, {Cupani}, {D'Odorico}, {Damasso}, {Deiries}, {Delabre}, {Demangeon}, {Dumusque}, {Ehrenreich}, {Faria}, {Fragoso}, {Genolet}, {Genoni}, {G{\'e}nova Santos}, {Gonz{\'a}lez Hern{\'a}ndez}, {Hughes}, {Iwert}, {Kerber}, {Knudstrup}, {Landoni}, {Lavie}, {Lillo-Box}, {Lizon}, {Maire}, {Martins}, {Mehner}, {Micela}, {Modigliani}, {Monteiro}, {Monteiro}, {Moschetti}, {Murphy}, {Nunes}, {Oggioni}, {Oliveira}, {Oshagh}, {Pall{\'e}}, {Pariani}, {Poretti}, {Rasilla}, {Rebord{\~a}o}, {Redaelli}, {Santana Tschudi},
  {Santin}, {Santos}, {S{\'e}gransan}, {Schmidt}, {Segovia}, {Sosnowska}, {Sozzetti}, {Sousa}, {Span{\`o}}, {Su{\'a}rez Mascare{\~n}o}, {Tabernero}, {Tenegi}, {Udry}, \& {Zanutta}}]{Pepe20}
{Pepe}, F., {Cristiani}, S., {Rebolo}, R., {et~al.} 2021, \aap, 645, A96, \dodoi{10.1051/0004-6361/202038306}

\bibitem[{{Petrovich}(2015)}]{Petrovich15}
{Petrovich}, C. 2015, \apj, 805, 75, \dodoi{10.1088/0004-637X/805/1/75}

\bibitem[{{Petrovich} {et~al.}(2020){Petrovich}, {Mu{\~n}oz}, {Kratter}, \& {Malhotra}}]{Petrovich2020}
{Petrovich}, C., {Mu{\~n}oz}, D.~J., {Kratter}, K.~M., \& {Malhotra}, R. 2020, \apjl, 902, L5, \dodoi{10.3847/2041-8213/abb952}

\bibitem[{{Pollacco} {et~al.}(2006){Pollacco}, {Skillen}, {Collier Cameron}, {Christian}, {Hellier}, {Irwin}, {Lister}, {Street}, {West}, {Anderson}, {Clarkson}, {Deeg}, {Enoch}, {Evans}, {Fitzsimmons}, {Haswell}, {Hodgkin}, {Horne}, {Kane}, {Keenan}, {Maxted}, {Norton}, {Osborne}, {Parley}, {Ryans}, {Smalley}, {Wheatley}, \& {Wilson}}]{Pollacco2006}
{Pollacco}, D.~L., {Skillen}, I., {Collier Cameron}, A., {et~al.} 2006, \pasp, 118, 1407, \dodoi{10.1086/508556}

\bibitem[{{Prinoth} {et~al.}(2024){Prinoth}, {Sedaghati}, {Seidel}, {Hoeijmakers}, {Brahm}, {Thorsbro}, \& {Jord{\'a}n}}]{Prinoth2024}
{Prinoth}, B., {Sedaghati}, E., {Seidel}, J.~V., {et~al.} 2024, \aj, 168, 133, \dodoi{10.3847/1538-3881/ad5a7f}

\bibitem[{{Rasio} \& {Ford}(1996)}]{Rasio96}
{Rasio}, F.~A., \& {Ford}, E.~B. 1996, Science, 274, 954, \dodoi{10.1126/science.274.5289.954}

\bibitem[{{Rice} {et~al.}(2022{\natexlab{a}}){Rice}, {Wang}, \& {Laughlin}}]{Rice2022a}
{Rice}, M., {Wang}, S., \& {Laughlin}, G. 2022{\natexlab{a}}, \apjl, 926, L17, \dodoi{10.3847/2041-8213/ac502d}

\bibitem[{{Rice} {et~al.}(2022{\natexlab{b}}){Rice}, {Wang}, {Wang}, {Stef{\'a}nsson}, {Isaacson}, {Howard}, {Logsdon}, {Schweiker}, {Dai}, {Brinkman}, {Giacalone}, \& {Holcomb}}]{Rice2022}
{Rice}, M., {Wang}, S., {Wang}, X.-Y., {et~al.} 2022{\natexlab{b}}, \aj, 164, 104, \dodoi{10.3847/1538-3881/ac8153}

\bibitem[{{Ricker} {et~al.}(2015){Ricker}, {Winn}, {Vanderspek}, {Latham}, {Bakos}, {Bean}, {Berta-Thompson}, {Brown}, {Buchhave}, {Butler}, {Butler}, {Chaplin}, {Charbonneau}, {Christensen-Dalsgaard}, {Clampin}, {Deming}, {Doty}, {De Lee}, {Dressing}, {Dunham}, {Endl}, {Fressin}, {Ge}, {Henning}, {Holman}, {Howard}, {Ida}, {Jenkins}, {Jernigan}, {Johnson}, {Kaltenegger}, {Kawai}, {Kjeldsen}, {Laughlin}, {Levine}, {Lin}, {Lissauer}, {MacQueen}, {Marcy}, {McCullough}, {Morton}, {Narita}, {Paegert}, {Palle}, {Pepe}, {Pepper}, {Quirrenbach}, {Rinehart}, {Sasselov}, {Sato}, {Seager}, {Sozzetti}, {Stassun}, {Sullivan}, {Szentgyorgyi}, {Torres}, {Udry}, \& {Villasenor}}]{Ricker2015}
{Ricker}, G.~R., {Winn}, J.~N., {Vanderspek}, R., {et~al.} 2015, Journal of Astronomical Telescopes, Instruments, and Systems, 1, 014003, \dodoi{10.1117/1.JATIS.1.1.014003}

\bibitem[{{Romanova} {et~al.}(2023){Romanova}, {Koldoba}, {Ustyugova}, {Lai}, \& {Lovelace}}]{Romanova2023}
{Romanova}, M.~M., {Koldoba}, A.~V., {Ustyugova}, G.~V., {Lai}, D., \& {Lovelace}, R.~V.~E. 2023, \mnras, 523, 2832, \dodoi{10.1093/mnras/stad987}

\bibitem[{{Rubenzahl} {et~al.}(2024){Rubenzahl}, {Howard}, {Halverson}, {Petrovich}, {Angelo}, {Stef{\'a}nsson}, {Dai}, {Householder}, {Fulton}, {Gibson}, {Roy}, {Shaum}, {Isaacson}, {Brodheim}, {Deich}, {Hill}, {Holden}, {Huber}, {Laher}, {Lanclos}, {Payne}, {Petigura}, {Schwab}, {Walawender}, {Wang}, {Weiss}, {Winn}, \& {Wright}}]{Rubenzahl2024}
{Rubenzahl}, R.~A., {Howard}, A.~W., {Halverson}, S., {et~al.} 2024, \apjl, 971, L40, \dodoi{10.3847/2041-8213/ad6985}

\bibitem[{{Rusznak} {et~al.}(2024){Rusznak}, {Wang}, {Rice}, \& {Wang}}]{Rusznak2024}
{Rusznak}, J., {Wang}, X.-Y., {Rice}, M., \& {Wang}, S. 2024, arXiv e-prints, arXiv:2412.04438, \dodoi{10.48550/arXiv.2412.04438}

\bibitem[{{Schlecker} {et~al.}(2020){Schlecker}, {Kossakowski}, {Brahm}, {Espinoza}, {Henning}, {Carone}, {Molaverdikhani}, {Trifonov}, {Molli{\`e}re}, {Hobson}, {Jord{\'a}n}, {Rojas}, {Klahr}, {Sarkis}, {Bakos}, {Bhatti}, {Osip}, {Suc}, {Ricker}, {Vanderspek}, {Latham}, {Seager}, {Winn}, {Jenkins}, {Vezie}, {Villase{\~n}or}, {Rose}, {Rodriguez}, {Rodriguez}, {Quinn}, \& {Shporer}}]{Schlecker2020}
{Schlecker}, M., {Kossakowski}, D., {Brahm}, R., {et~al.} 2020, \aj, 160, 275, \dodoi{10.3847/1538-3881/abbe03}

\bibitem[{{Sedaghati} {et~al.}(2023){Sedaghati}, {Jord{\'a}n}, {Brahm}, {Mu{\~n}oz}, {Petrovich}, \& {Hobson}}]{Sedaghati2023}
{Sedaghati}, E., {Jord{\'a}n}, A., {Brahm}, R., {et~al.} 2023, \aj, 166, 130, \dodoi{10.3847/1538-3881/acea84}

\bibitem[{{Sepulveda} {et~al.}(2024){Sepulveda}, {Huber}, {Bedding}, {Hey}, {Murphy}, {Zhang}, \& {Liu}}]{Sepulveda2024}
{Sepulveda}, A.~G., {Huber}, D., {Bedding}, T.~R., {et~al.} 2024, \aj, 168, 13, \dodoi{10.3847/1538-3881/ad4964}

\bibitem[{{Sha} {et~al.}(2021){Sha}, {Huang}, {Shporer}, {Rodriguez}, {Vanderburg}, {Brahm}, {Hagelberg}, {Matthews}, {Ziegler}, {Livingston}, {Stassun}, {Wright}, {Crane}, {Espinoza}, {Bouchy}, {Bakos}, {Collins}, {Zhou}, {Bieryla}, {Hartman}, {Wittenmyer}, {Nielsen}, {Plavchan}, {Bayliss}, {Sarkis}, {Tan}, {Cloutier}, {Mancini}, {Jord{\'a}n}, {Wang}, {Henning}, {Narita}, {Penev}, {Teske}, {Kane}, {Mann}, {Addison}, {Tamura}, {Horner}, {Barbieri}, {Burt}, {D{\'\i}az}, {Crossfield}, {Dragomir}, {Drass}, {Feinstein}, {Zhang}, {Hart}, {Kielkopf}, {Jensen}, {Montet}, {Ottoni}, {Schwarz}, {Rojas}, {Nespral}, {Torres}, {Mengel}, {Udry}, {Zapata}, {Snoddy}, {Okumura}, {Ricker}, {Vanderspek}, {Latham}, {Winn}, {Seager}, {Jenkins}, {Col{\'o}n}, {Henze}, {Krishnamurthy}, {Ting}, {Vezie}, \& {Villanueva}}]{Sha2021}
{Sha}, L., {Huang}, C.~X., {Shporer}, A., {et~al.} 2021, \aj, 161, 82, \dodoi{10.3847/1538-3881/abd187}

\bibitem[{{Siegel} {et~al.}(2023){Siegel}, {Winn}, \& {Albrecht}}]{Siegel2023}
{Siegel}, J.~C., {Winn}, J.~N., \& {Albrecht}, S.~H. 2023, \apjl, 950, L2, \dodoi{10.3847/2041-8213/acd62f}

\bibitem[{{Skrutskie} {et~al.}(2006){Skrutskie}, {Cutri}, {Stiening}, {Weinberg}, {Schneider}, {Carpenter}, {Beichman}, {Capps}, {Chester}, {Elias}, {Huchra}, {Liebert}, {Lonsdale}, {Monet}, {Price}, {Seitzer}, {Jarrett}, {Kirkpatrick}, {Gizis}, {Howard}, {Evans}, {Fowler}, {Fullmer}, {Hurt}, {Light}, {Kopan}, {Marsh}, {McCallon}, {Tam}, {Van Dyk}, \& {Wheelock}}]{2mass}
{Skrutskie}, M.~F., {Cutri}, R.~M., {Stiening}, R., {et~al.} 2006, \aj, 131, 1163, \dodoi{10.1086/498708}

\bibitem[{{Smith} \& {Csizmadia}(2022)}]{Smith2022}
{Smith}, A. M.~S., \& {Csizmadia}, S. 2022, \aj, 164, 21, \dodoi{10.3847/1538-3881/ac704c}

\bibitem[{{Smith} {et~al.}(2014){Smith}, {Anderson}, {Armstrong}, {Barros}, {Bonomo}, {Bouchy}, {Brown}, {Collier Cameron}, {Delrez}, {Faedi}, {Gillon}, {G{\'o}mez Maqueo Chew}, {H{\'e}brard}, {Jehin}, {Lendl}, {Louden}, {Maxted}, {Montagnier}, {Neveu-VanMalle}, {Osborn}, {Pepe}, {Pollacco}, {Queloz}, {Rostron}, {Segransan}, {Smalley}, {Triaud}, {Turner}, {Udry}, {Walker}, {West}, \& {Wheatley}}]{Smith2014}
{Smith}, A.~M.~S., {Anderson}, D.~R., {Armstrong}, D.~J., {et~al.} 2014, \aap, 570, A64, \dodoi{10.1051/0004-6361/201424752}

\bibitem[{{Sosnowska} {et~al.}(2015){Sosnowska}, {Lovis}, {Figueira}, {Modigliani}, {Marcantonio}, {Megevand}, \& {Pepe}}]{Sosnowska15}
{Sosnowska}, D., {Lovis}, C., {Figueira}, P., {et~al.} 2015, in Astronomical Society of the Pacific Conference Series, Vol. 495, Astronomical Data Analysis Software an Systems XXIV (ADASS XXIV), ed. A.~R. {Taylor} \& E.~{Rosolowsky}, 285, \dodoi{10.48550/arXiv.1509.05584}

\bibitem[{{Southworth}(2011)}]{Southworth2011}
{Southworth}, J. 2011, \mnras, 417, 2166, \dodoi{10.1111/j.1365-2966.2011.19399.x}

\bibitem[{{Speagle}(2020)}]{dynesty2}
{Speagle}, J.~S. 2020, \mnras, 493, 3132, \dodoi{10.1093/mnras/staa278}

\bibitem[{{Stefansson} {et~al.}(2022){Stefansson}, {Mahadevan}, {Petrovich}, {Winn}, {Kanodia}, {Millholland}, {Maney}, {Ca{\~n}as}, {Wisniewski}, {Robertson}, {Ninan}, {Ford}, {Bender}, {Blake}, {Cegla}, {Cochran}, {Diddams}, {Dong}, {Endl}, {Fredrick}, {Halverson}, {Hearty}, {Hebb}, {Hirano}, {Lin}, {Logsdon}, {Lubar}, {McElwain}, {Metcalf}, {Monson}, {Rajagopal}, {Ramsey}, {Roy}, {Schwab}, {Schweiker}, {Terrien}, \& {Wright}}]{Stefansson22}
{Stefansson}, G., {Mahadevan}, S., {Petrovich}, C., {et~al.} 2022, \apjl, 931, L15, \dodoi{10.3847/2041-8213/ac6e3c}

\bibitem[{{Tala Pinto} {et~al.}(2025){Tala Pinto}, {Jord{\'a}n}, {Acu{\~n}a}, {Jones}, {Brahm}, {Reinarz}, {Eberhardt}, {Espinoza}, {Henning}, {Hobson}, {Rojas}, {Schlecker}, {Trifonov}, {Bakos}, {Boyle}, {Csubry}, {Hartmann}, {Knepper}, {Kreidberg}, {Suc}, {Teske}, {Butler}, {Crane}, {Schectman}, {Thompson}, {Osip}, {Ricker}, {Collins}, {Watkins}, {Bieryla}, {Stockdale}, {Wang}, {Zambelli}, {Seager}, {Winn}, {Rose}, {Rice}, \& {Essack}}]{Tala2024}
{Tala Pinto}, M., {Jord{\'a}n}, A., {Acu{\~n}a}, L., {et~al.} 2025, \aap, 694, A268, \dodoi{10.1051/0004-6361/202452517}

\bibitem[{{Tayar} {et~al.}(2022){Tayar}, {Claytor}, {Huber}, \& {van Saders}}]{tayar2022}
{Tayar}, J., {Claytor}, Z.~R., {Huber}, D., \& {van Saders}, J. 2022, \apj, 927, 31, \dodoi{10.3847/1538-4357/ac4bbc}

\bibitem[{{Temple} {et~al.}(2017){Temple}, {Hellier}, {Albrow}, {Anderson}, {Bayliss}, {Beatty}, {Bieryla}, {Brown}, {Cargile}, {Collier Cameron}, {Collins}, {Col{\'o}n}, {Curtis}, {D'Ago}, {Delrez}, {Eastman}, {Gaudi}, {Gillon}, {Gregorio}, {James}, {Jehin}, {Joner}, {Kielkopf}, {Kuhn}, {Labadie-Bartz}, {Latham}, {Lendl}, {Lund}, {Malpas}, {Maxted}, {Myers}, {Oberst}, {Pepe}, {Pepper}, {Pollacco}, {Queloz}, {Rodriguez}, {S{\'e}gransan}, {Siverd}, {Smalley}, {Stassun}, {Stevens}, {Stockdale}, {Tan}, {Triaud}, {Udry}, {Villanueva}, {West}, \& {Zhou}}]{Temple2017}
{Temple}, L.~Y., {Hellier}, C., {Albrow}, M.~D., {et~al.} 2017, \mnras, 471, 2743, \dodoi{10.1093/mnras/stx1729}

\bibitem[{{Valenti} \& {Fischer}(2005)}]{Valenti2005}
{Valenti}, J.~A., \& {Fischer}, D.~A. 2005, \apjs, 159, 141, \dodoi{10.1086/430500}

\bibitem[{{V{\'\i}tkov{\'a}} {et~al.}(2024){V{\'\i}tkov{\'a}}, {Brahm}, {Trifonov}, {Kab{\'a}th}, {Jord{\'a}n}, {Henning}, {Hobson}, {Eberhardt}, {Tala Pinto}, {Rojas}, {Espinoza}, {Schlecker}, {Jones}, {Moyano}, {Eyheramendy}, {Ziegler}, {Lissauer}, {Vanderburg}, {Collins}, {Wohler}, {Watanabe}, {Ricker}, {Vanderspek}, {Seager}, {Winn}, {Jenkins}, \& {Skarka}}]{Vitkova2024}
{V{\'\i}tkov{\'a}}, M., {Brahm}, R., {Trifonov}, T., {et~al.} 2024, arXiv e-prints, arXiv:2412.05609.
\newblock \doarXiv{2412.05609}

\bibitem[{{Wang} {et~al.}(2024){Wang}, {Rice}, {Wang}, {Kanodia}, {Dai}, {Logsdon}, {Schweiker}, {Teske}, {Butler}, {Crane}, {Shectman}, {Quinn}, {Kostov}, {Osborn}, {Goeke}, {Eastman}, {Shporer}, {Rapetti}, {Collins}, {Watkins}, {Relles}, {Ricker}, {Seager}, {Winn}, \& {Jenkins}}]{Wang2024}
{Wang}, X.-Y., {Rice}, M., {Wang}, S., {et~al.} 2024, arXiv e-prints, arXiv:2408.10038, \dodoi{10.48550/arXiv.2408.10038}

\bibitem[{{Winn} {et~al.}(2010){Winn}, {Fabrycky}, {Albrecht}, \& {Johnson}}]{Winn2010}
{Winn}, J.~N., {Fabrycky}, D., {Albrecht}, S., \& {Johnson}, J.~A. 2010, \apjl, 718, L145, \dodoi{10.1088/2041-8205/718/2/L145}

\bibitem[{{Wright} {et~al.}(2023){Wright}, {Rice}, {Wang}, {Hixenbaugh}, \& {Wang}}]{Wright2023}
{Wright}, J., {Rice}, M., {Wang}, X.-Y., {Hixenbaugh}, K., \& {Wang}, S. 2023, \aj, 166, 217, \dodoi{10.3847/1538-3881/ad0131}

\bibitem[{{Wu} \& {Lithwick}(2011)}]{Wu11}
{Wu}, Y., \& {Lithwick}, Y. 2011, \apj, 735, 109, \dodoi{10.1088/0004-637X/735/2/109}

\bibitem[{{Wu} \& {Murray}(2003)}]{Wu03}
{Wu}, Y., \& {Murray}, N. 2003, \apj, 589, 605, \dodoi{10.1086/374598}

\bibitem[{{Zanazzi} \& {Chiang}(2024)}]{jj_disk}
{Zanazzi}, J.~J., \& {Chiang}, E. 2024, \mnras, 527, 7203, \dodoi{10.1093/mnras/stad3066}

\bibitem[{{Zanazzi} {et~al.}(2024){Zanazzi}, {Dewberry}, \& {Chiang}}]{Zanazzi2024}
{Zanazzi}, J.~J., {Dewberry}, J., \& {Chiang}, E. 2024, \apjl, 967, L29, \dodoi{10.3847/2041-8213/ad4644}

\end{thebibliography}
\bibliographystyle{aasjournal}

\appendix

\renewcommand{\thefigure}{A\arabic{figure}}
\setcounter{figure}{0}

\section{Plots}\label{app:plots}


\begin{figure*}[h!]
    \centering
    \includegraphics[scale=0.8]{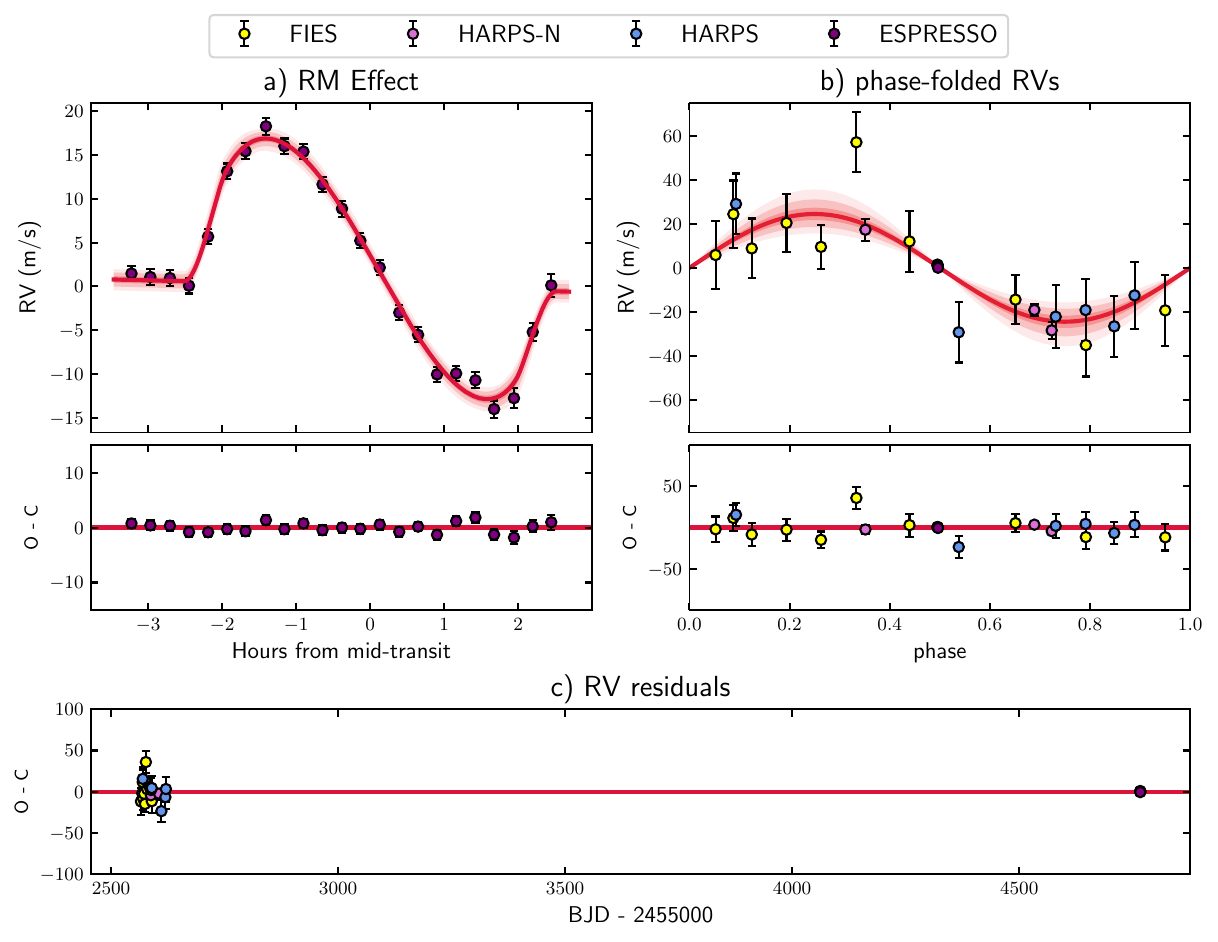}
    \includegraphics[scale=0.8]{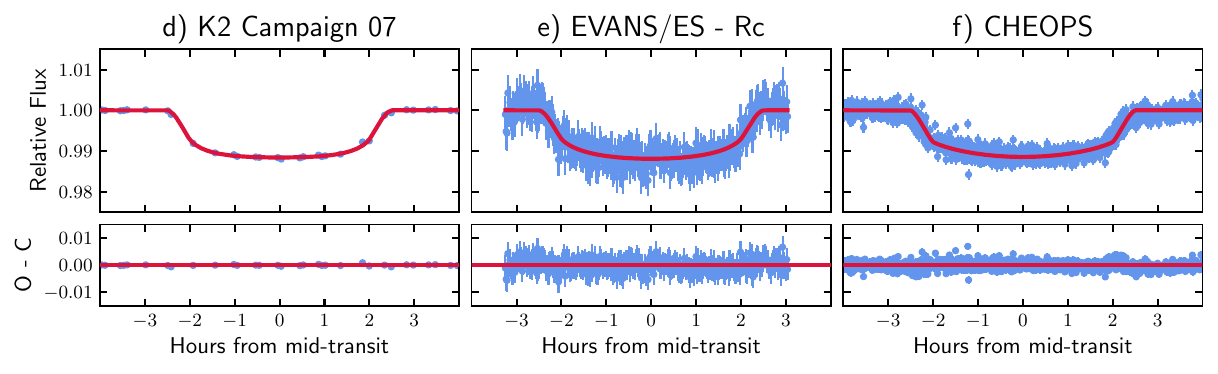}
    \caption{Observations of K2-139 along with the best model. All error bars include a white noise jitter term applied in quadrature to the RV and photometric data points. a) ESPRESSO observations of the RM effect. We show the best model as the red line, and the $1\sigma$, $2\sigma$, and $3\sigma$ models as the shaded areas b) Phase-folded out-of-transit RVs along with the best, $1\sigma$, $2\sigma$, and $3\sigma$ models. c) RV residuals as a function of time. d-f) Different light curves along with the best model. Data used to create this figure is available as the data behind the figure.}
    \label{fig:K2-139}
\end{figure*}

\clearpage

\begin{figure*}[h!]
    \centering
    \includegraphics[scale=0.8]{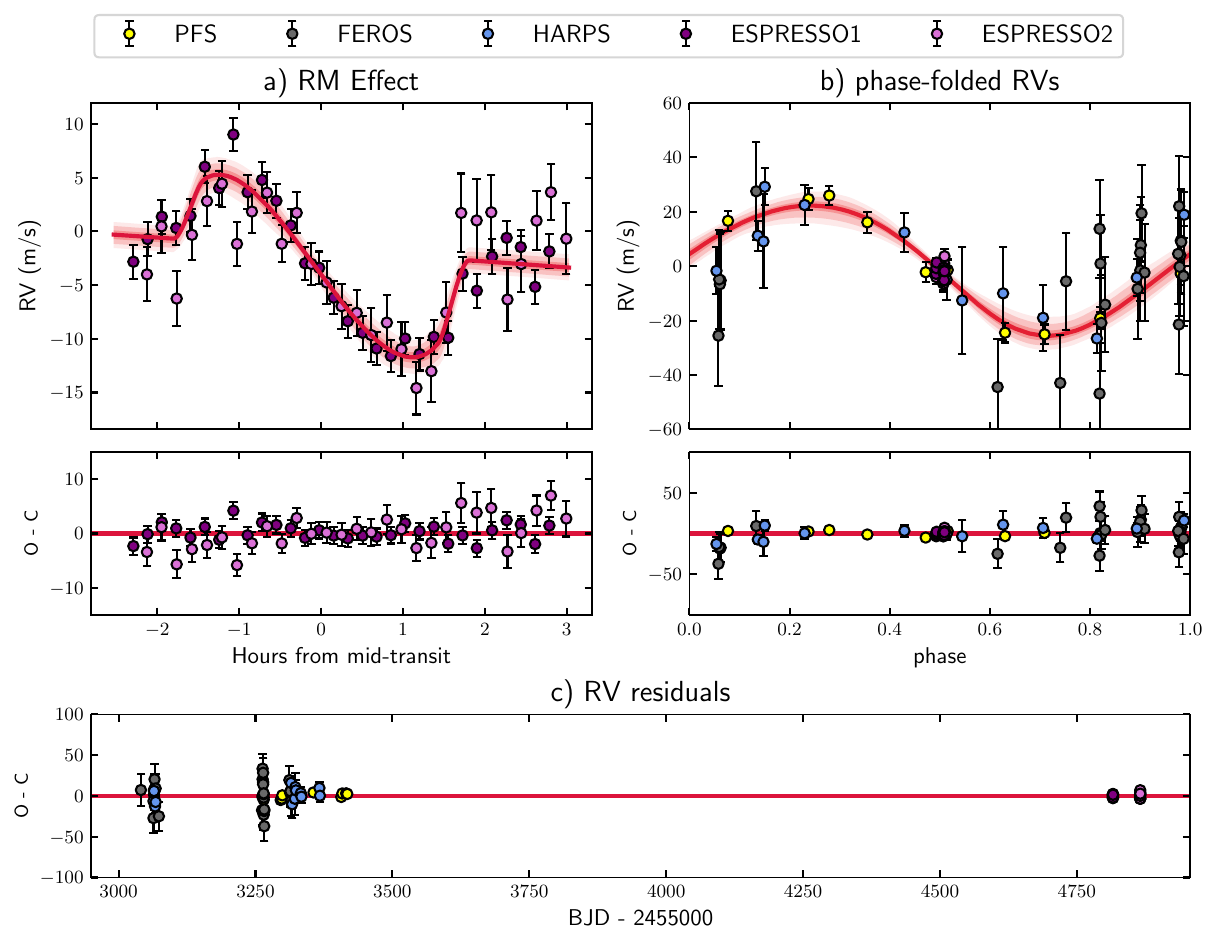}
    \includegraphics[scale=0.8]{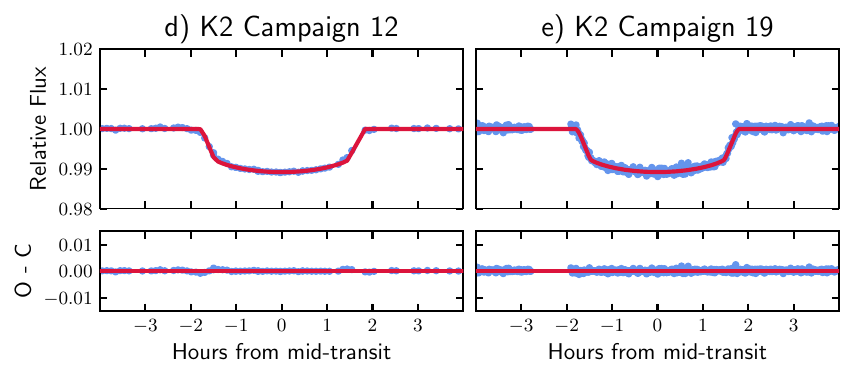}
    \includegraphics[scale=0.8]{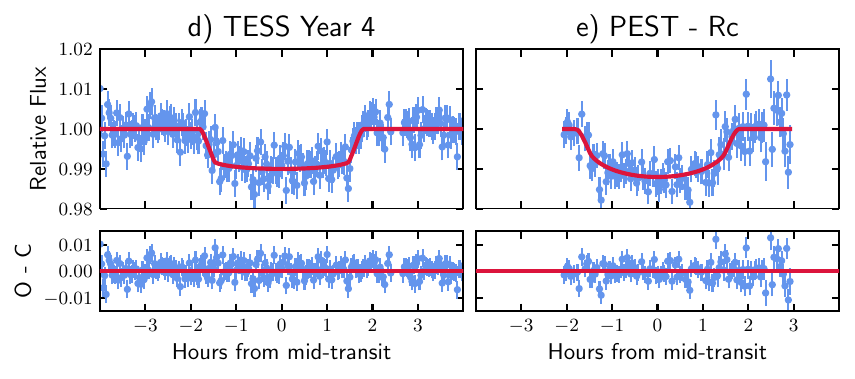}
    \caption{Same as Figure \ref{fig:K2-139} but for K2-329 A. Data used to create this figure is available as the data behind the figure.}
    \label{fig:K2-329}
\end{figure*}

\clearpage

\begin{figure*}[h!]
    \centering
    \includegraphics[scale=0.8]{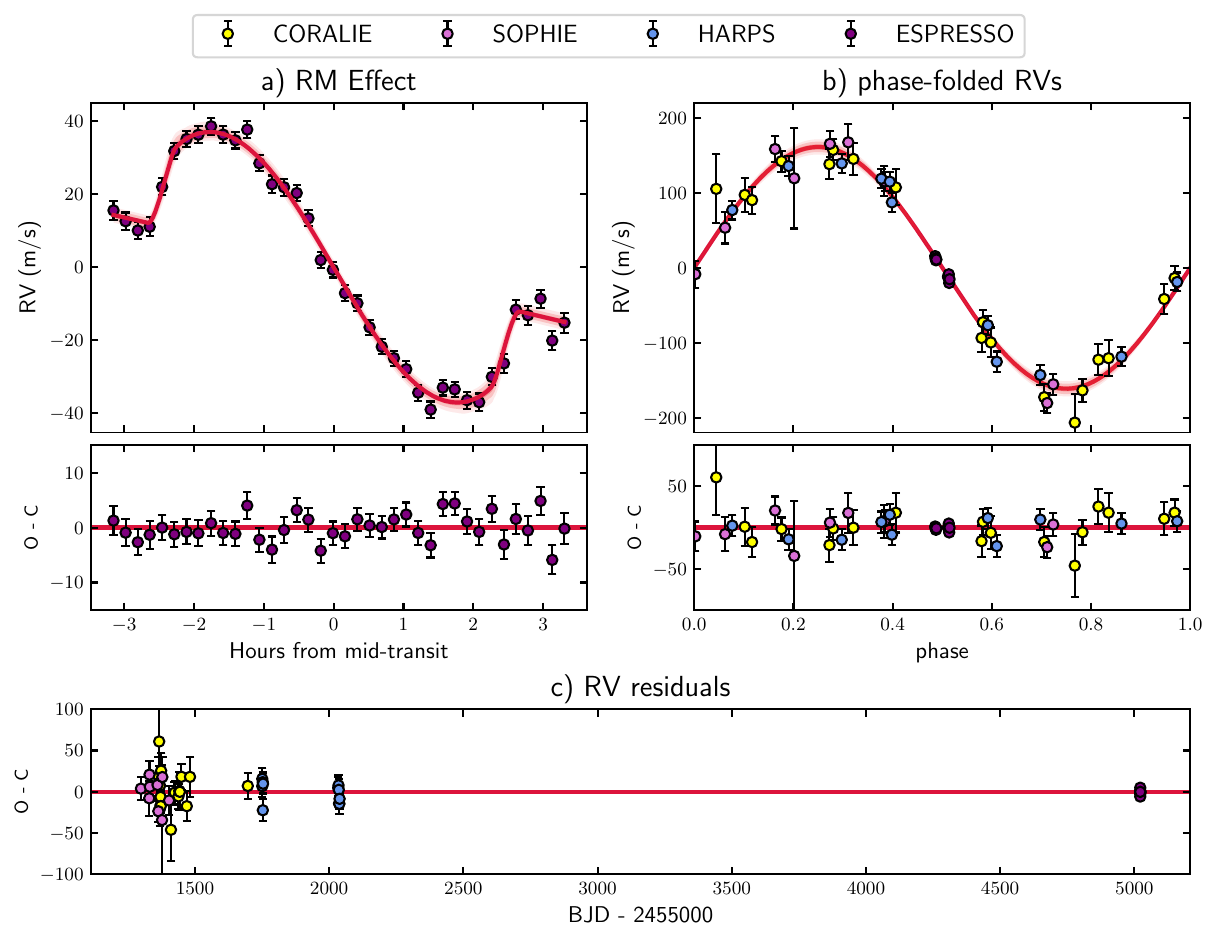}
    \includegraphics[scale=0.8]{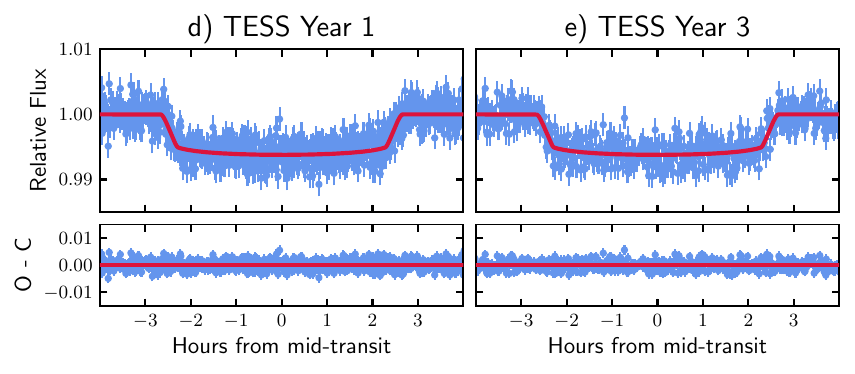}
    \includegraphics[scale=0.8]{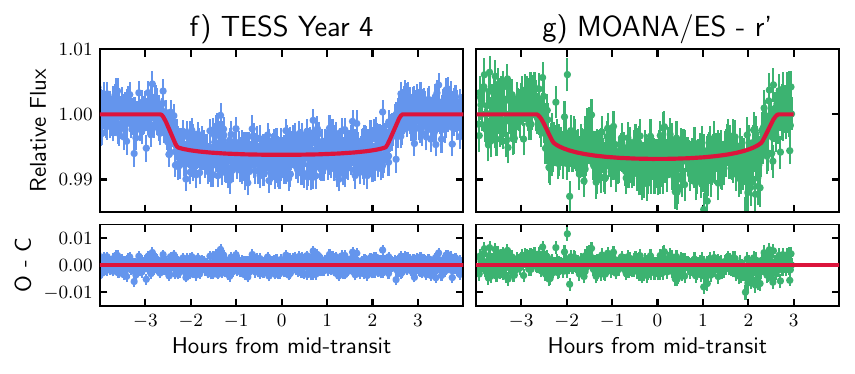}
    \caption{Same as Figure \ref{fig:K2-139} but for WASP-106. The simultaneous MOANA/ES light curve is shown in green. Data used to create this figure is available as the data behind the figure.}
    \label{fig:WASP-106}
\end{figure*}

\begin{figure*}[h!]
    \centering
    \includegraphics[scale=0.8]{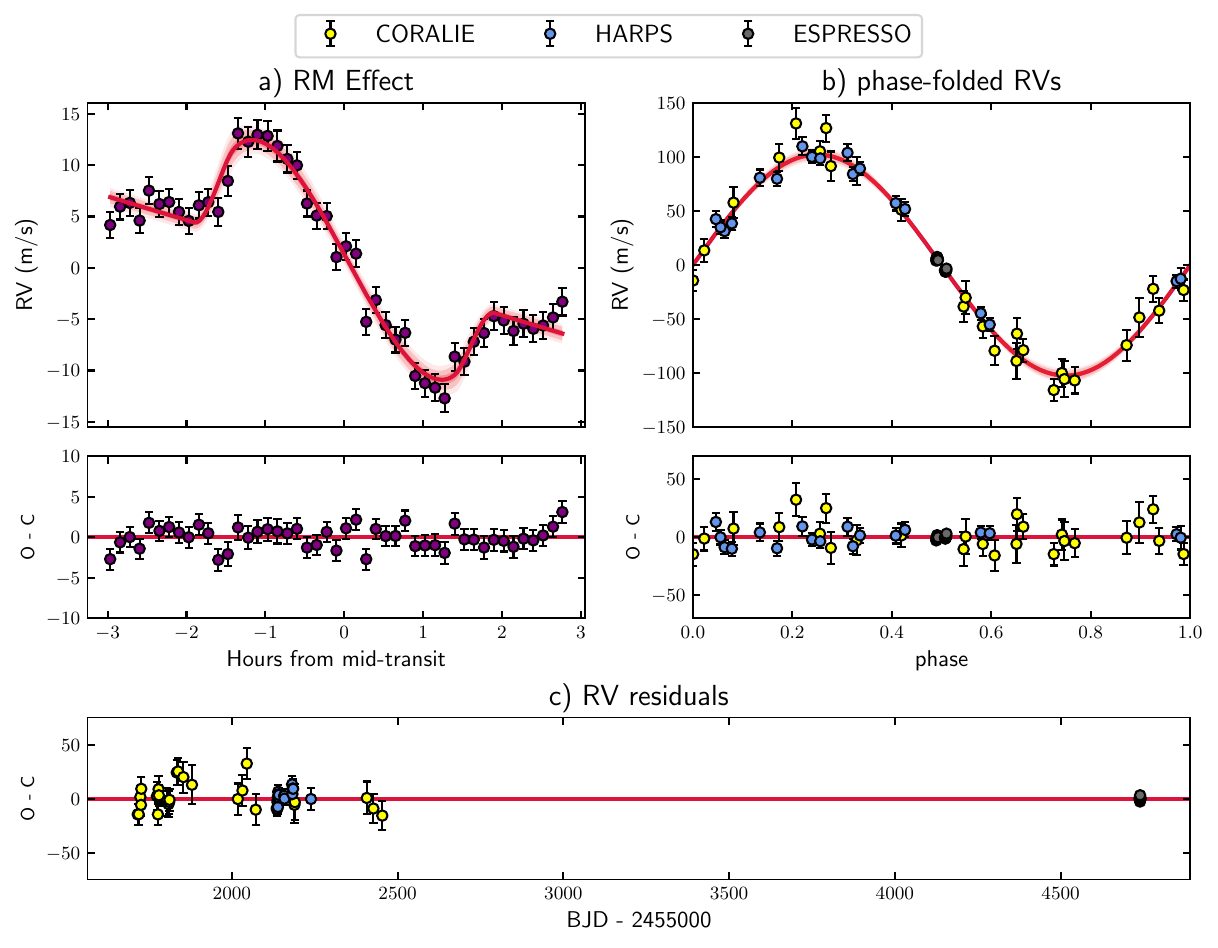}
    \includegraphics[scale=0.8]{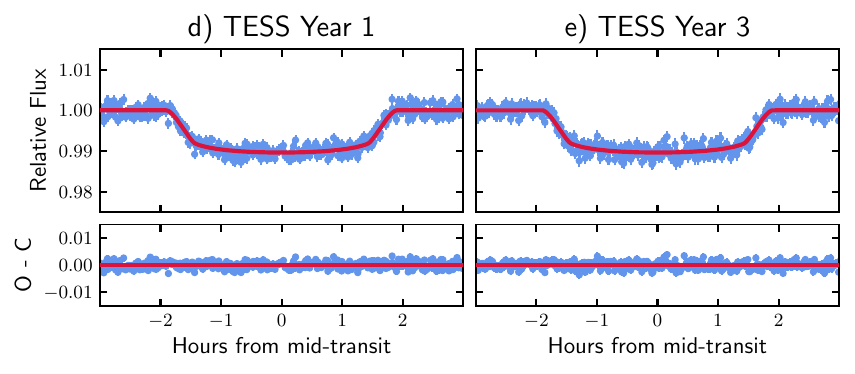}
    \includegraphics[scale=0.8]{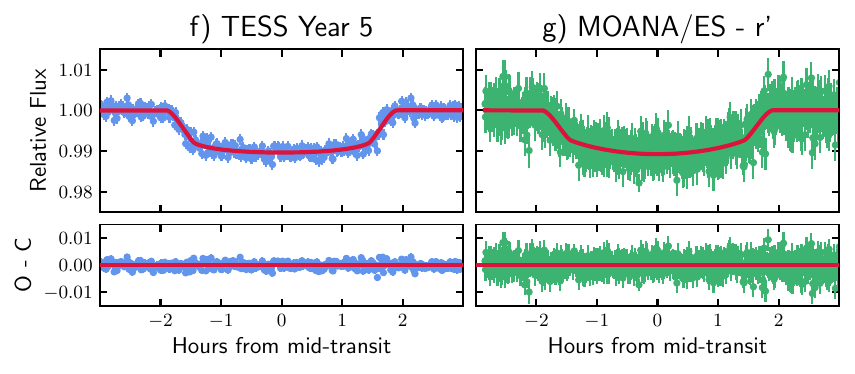}
    \caption{Same as Figure \ref{fig:K2-139} but for WASP-130. The simultaneous MOANA/ES light curve is shown in green. Data used to create this figure is available as the data behind the figure.}
    \label{fig:WASP-130}
\end{figure*}

\begin{figure*}[h!]
    \centering
    \includegraphics[scale=0.8]{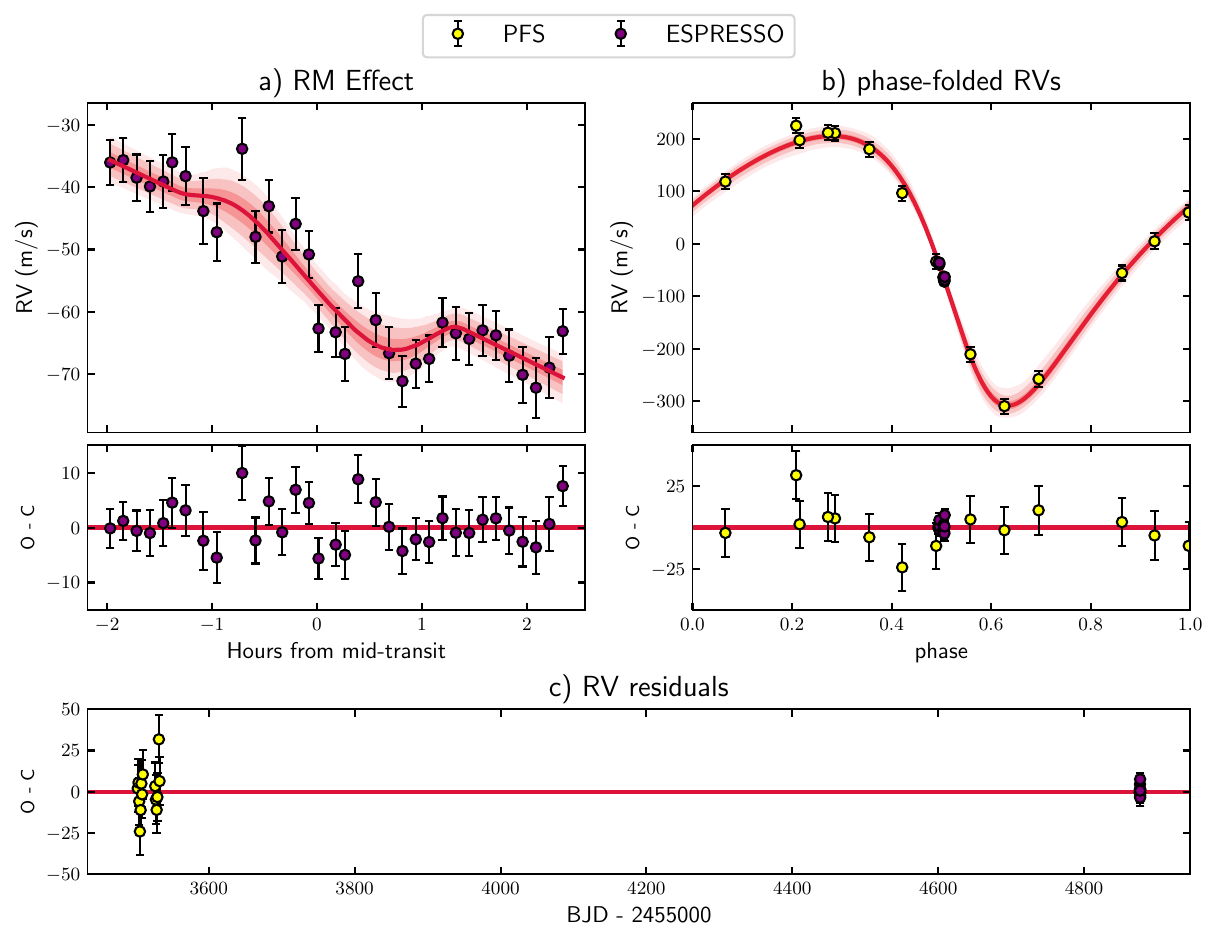}
    \includegraphics[scale=0.8]{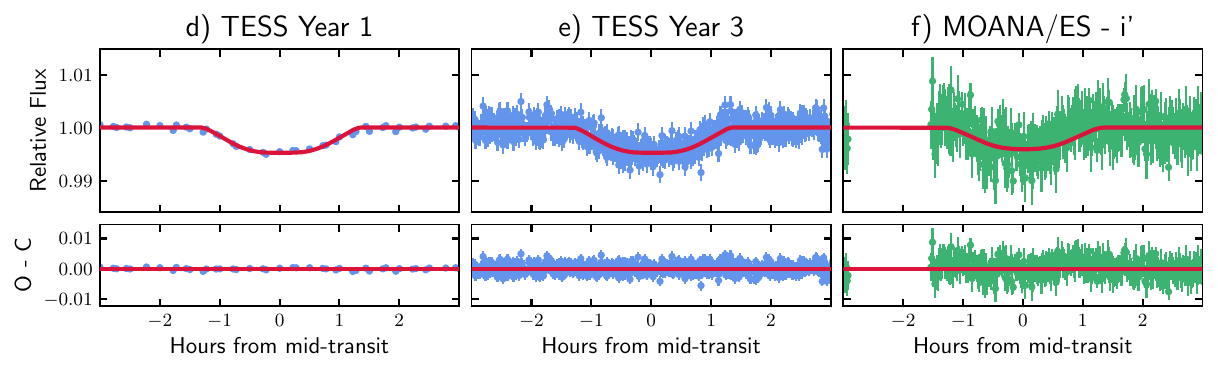}
    \includegraphics[scale=0.8]{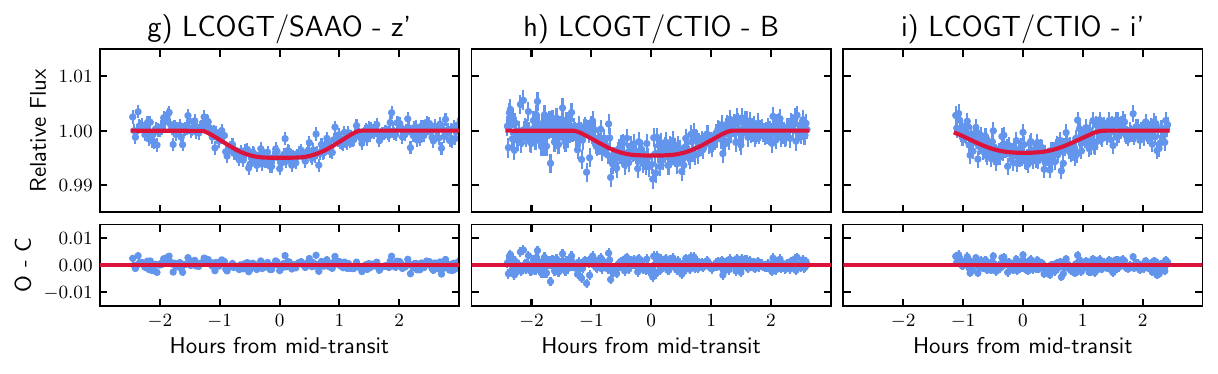}
    \caption{Same as Figure \ref{fig:K2-139} but for TOI-558. The simultaneous MOANA/ES light curve is shown in green. Data used to create this figure is available as the data behind the figure.}
    \label{fig:TOI-558}
\end{figure*}

\begin{figure*}[h!]
    \centering
    \vspace{-1cm}
    \includegraphics[scale=0.8]{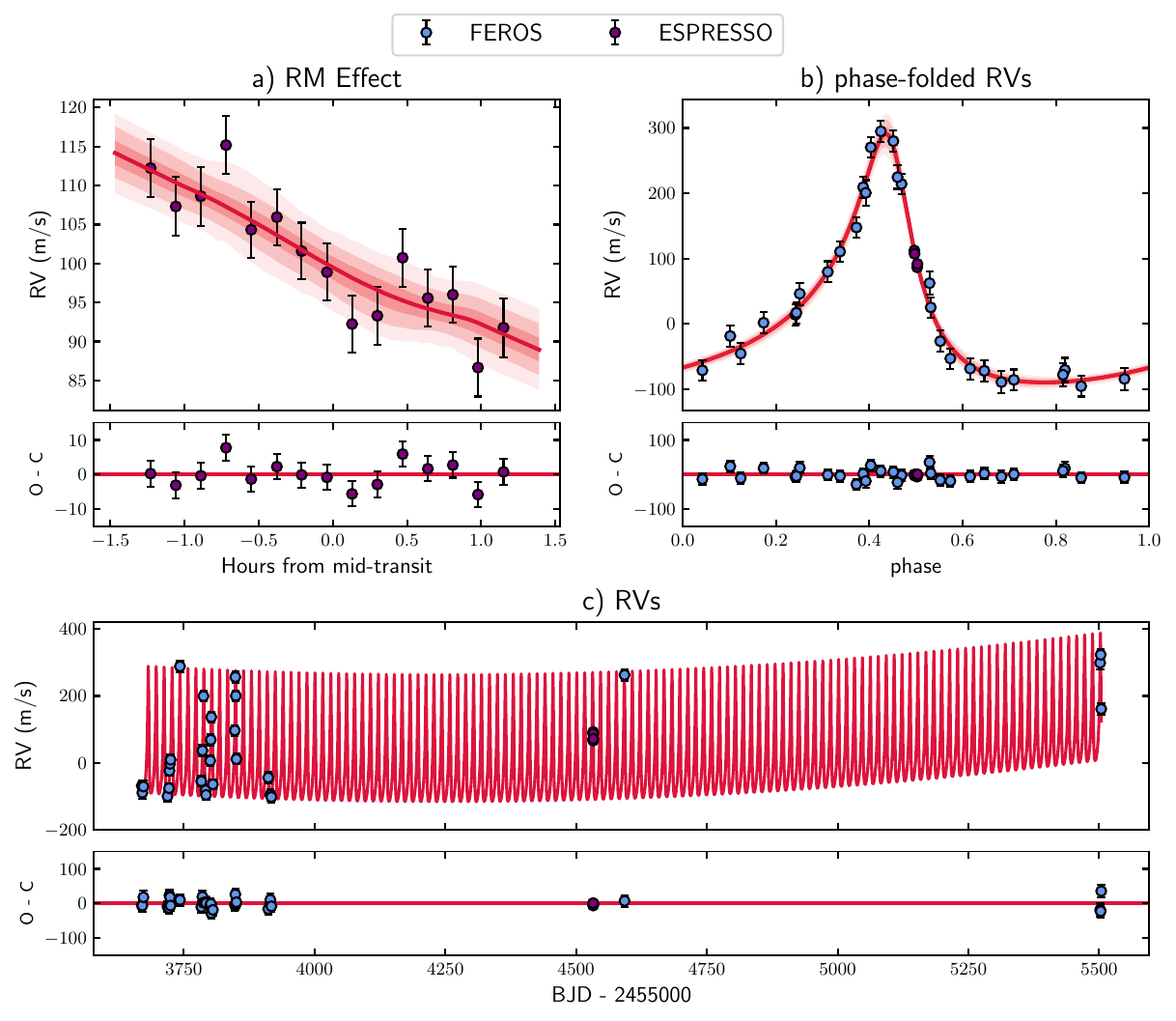}
    \includegraphics[scale=0.8]{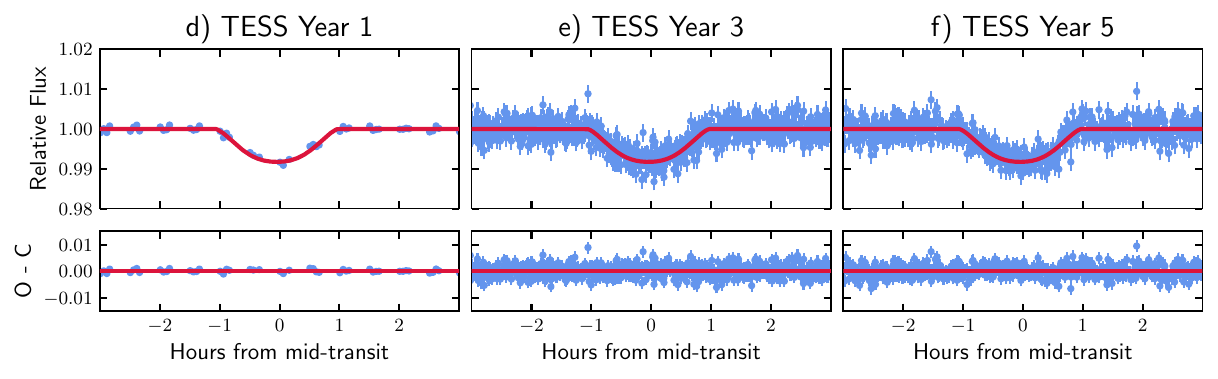}
    \includegraphics[scale=0.8]{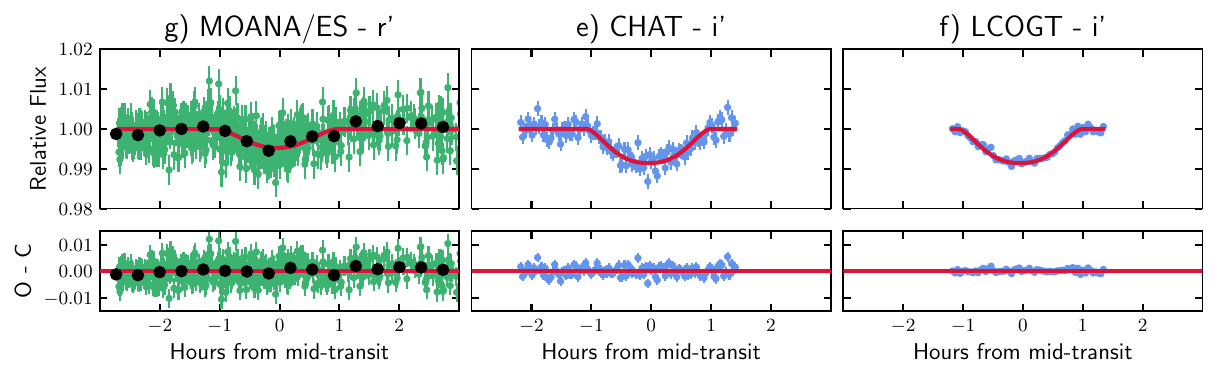}
    \caption{Same as Figure \ref{fig:K2-139} but for TOI-2179. Panel c here also shows the RVs as a function of time to show the quadratic RV trend detected at $\sim4\sigma$. The simultaneous MOANA/ES light curve is shown in green, with binned data in black. Data used to create this figure is available as the data behind the figure.}
    \label{fig:TOI-2179}
\end{figure*}

\begin{figure*}[h!]
    \centering
    \includegraphics[scale=0.8]{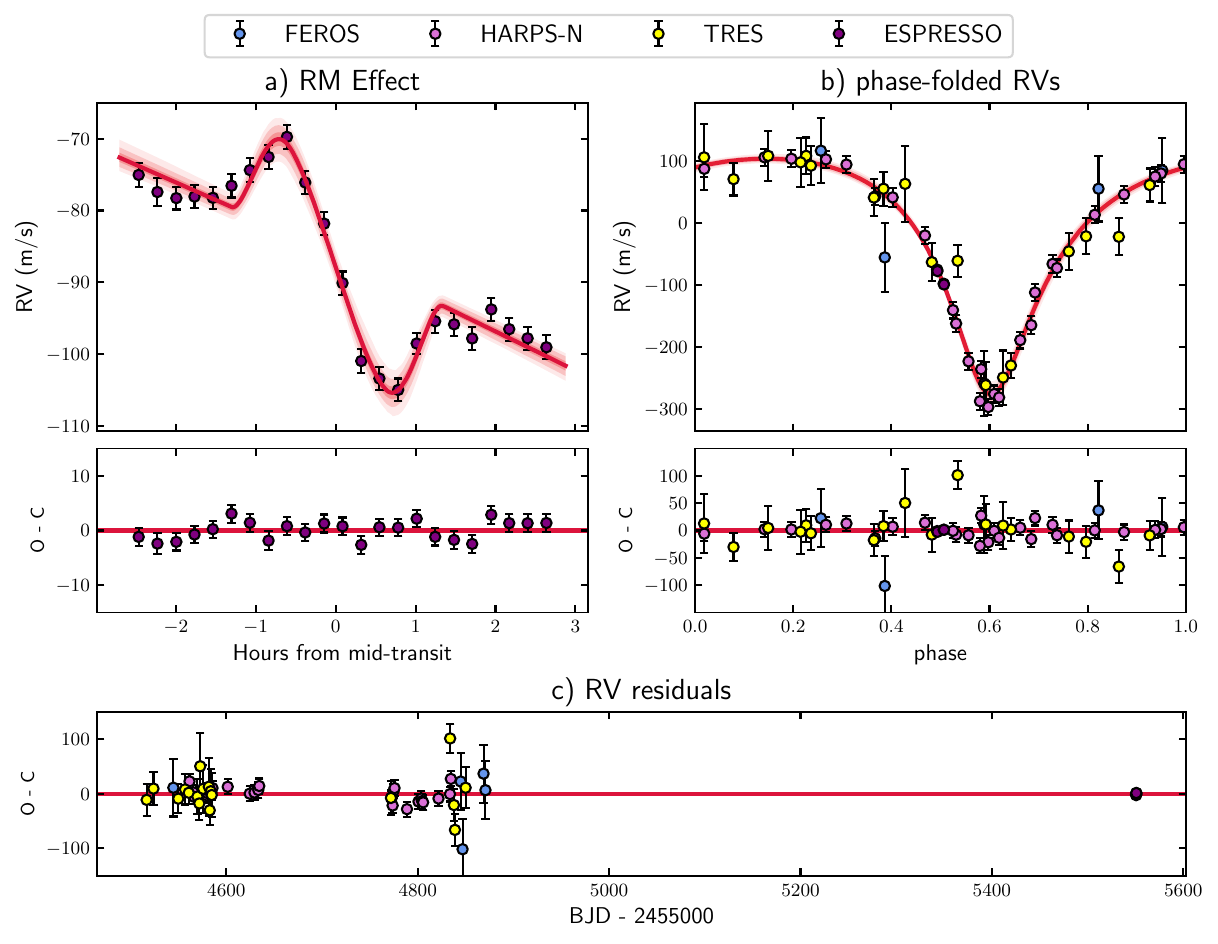}
    \includegraphics[scale=0.8]{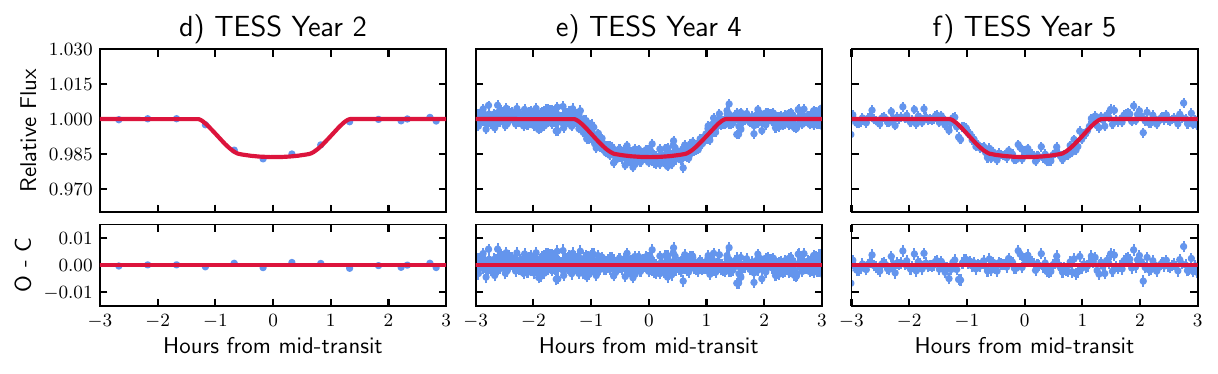}
    \includegraphics[scale=0.8]{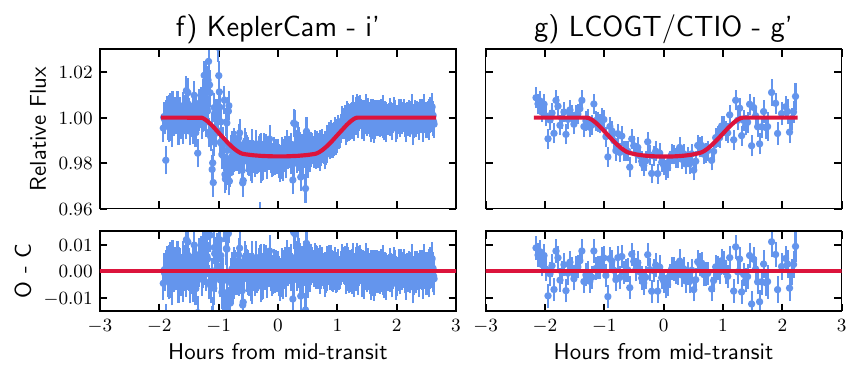}
    \caption{Same as Figure \ref{fig:K2-139} but for TOI-4515. Data used to create this figure is available as the data behind the figure.}
    \label{fig:TOI-4515}
\end{figure*}

\begin{figure*}[h!]
    \centering
    \vspace{-1cm}
    \includegraphics[scale=0.8]{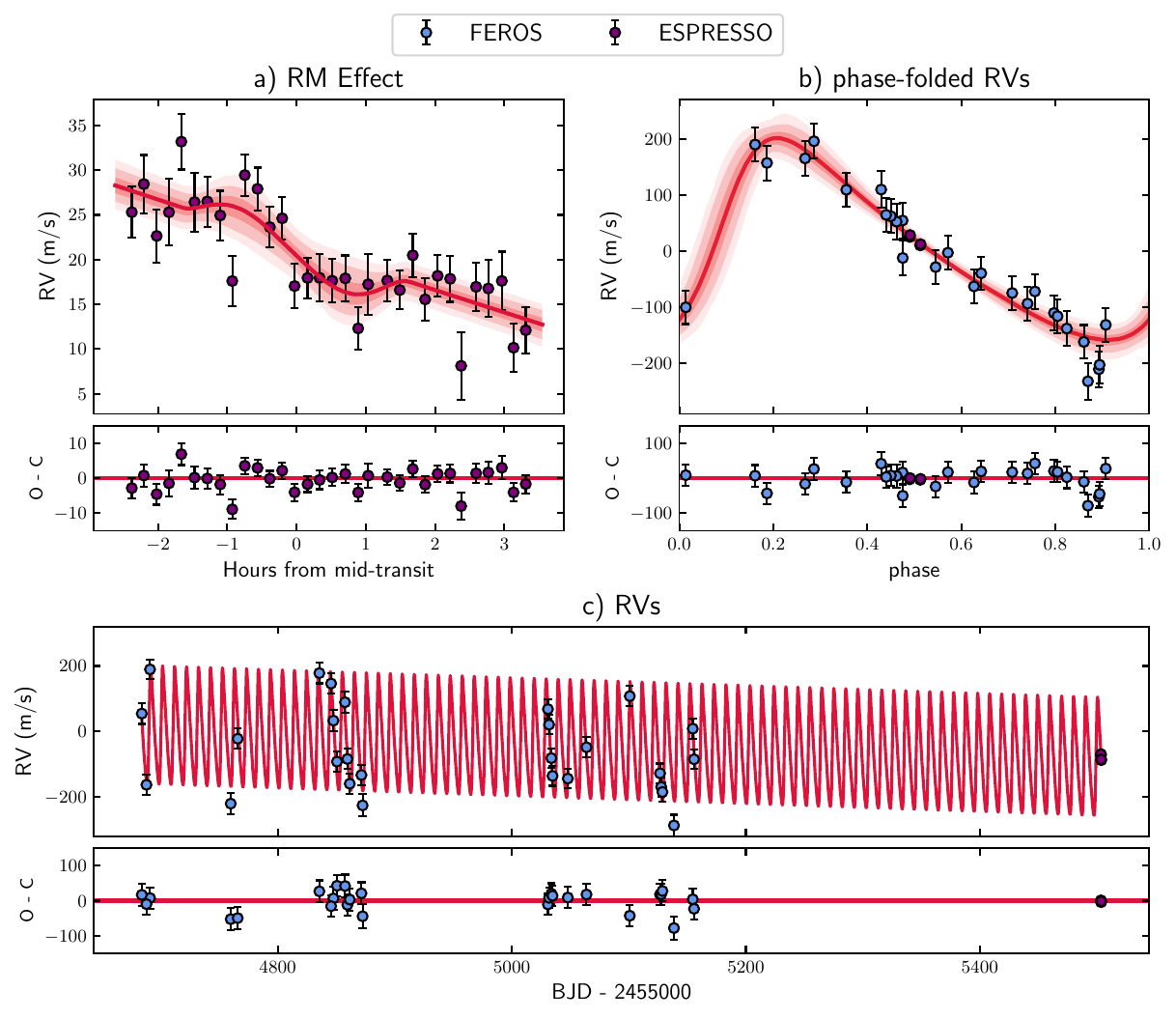}
    \includegraphics[scale=0.8]{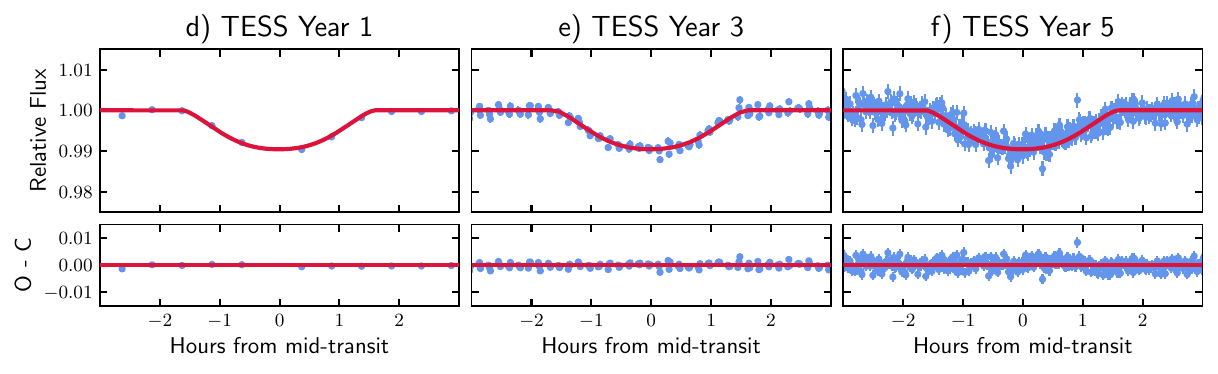}
    \includegraphics[scale=0.8]{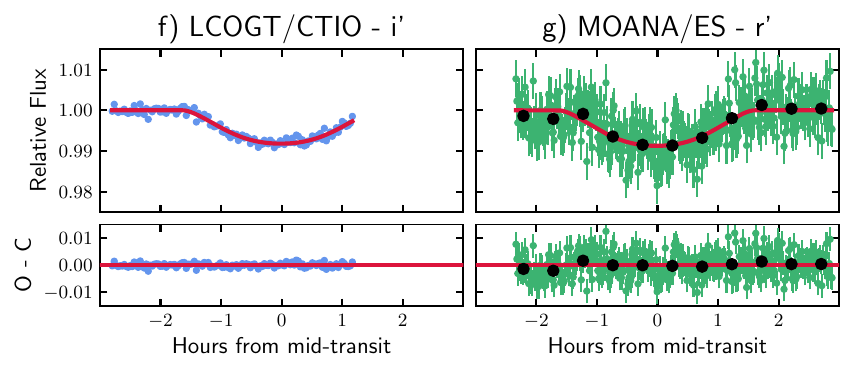}
    \caption{Same as Figure \ref{fig:K2-139} but for TOI-5027. Panel c here also shows the RVs as a function of time to show the linear RV trend detected at $\sim3\sigma$. The simultaneous MOANA/ES light curve is shown in green, with binned data in black. Data used to create this figure is available as the data behind the figure.}
    \label{fig:TOI-5027}
\end{figure*}

\clearpage
\section{Stellar Parameters}\label{app:stellar}

\begin{rotatetable*}
\centerwidetable
\begin{deluxetable*}{lrrrrrrrc}
\tablecaption{Stellar properties$^a$ of the different targets$^b$.\label{tab:stprops}}
\tablecolumns{9}
\tablewidth{0pt}
\tablehead{
\colhead{Parameter} &
\colhead{WASP-130} &
\colhead{K2-139} &
\colhead{K2-329 A} &
\colhead{TOI-558} &
\colhead{WASP-106} &
\colhead{TOI-4515} &
\colhead{TOI-5027} &
\colhead{Reference} \\
}
\startdata
RA \dotfill (J2015.5)    &  13h32m25.44s &  19h16m16.01s &  23h24m32.49s & 02h49m09.96s  &  11h05m43.11s & 01h24m44.68s & 16h15m32.41s & Gaia DR3$^c$\\
DEC \dotfill (J2015.5)   & -42d28m30.98s & -17d54m38.57s & -05d09m50.92s & -58d01m28.86s & -05d04m46.15s & 21d30m47.03s & -69d13m01.55s & Gaia DR3\\
pm$^{\rm RA}$ \hfill (mas yr$^{-1}$)    &  6.11 $\pm$ 0.04 &  37.28 $\pm$ 0.08 &  14.17 $\pm$ 0.06 & 1.07 $\pm$ 0.04 & -24.82 $\pm$ 0.08 & -4.12 $\pm$ 0.09& -22.02 $\pm$ 0.03 & Gaia DR3\\
pm$^{\rm DEC}$ \dotfill (mas yr$^{-1}$) & -1.24 $\pm$ 0.04 & -10.47 $\pm$ 0.06 & -12.39 $\pm$ 0.05 & 3.86 $\pm$ 0.04 &  13.29 $\pm$ 0.06 & 3.59 $\pm$ 0.07 & -31.99 $\pm$ 0.04 & Gaia DR3\\
$\pi$ \dotfill (mas)     & 5.78 $\pm$ 0.05    & 6.47 $\pm$ 0.04    & 4.26 $\pm$ 0.04    & 2.45  $\pm$ 0.02   & 2.81 $\pm$ 0.05 & 5.14 $\pm$ 0.05 & 4.86 $\pm$ 0.02& Gaia DR3 \\ 
\hline
G  \dotfill (mag)        & 10.972 $\pm$ 0.002 & 11.463 $\pm$ 0.002 & 12.446 $\pm$ 0.002 & 11.325 $\pm$ 0.002 &  11.366 $\pm$ 0.002 & 11.812 $\pm$ 0.002 & 11.334 $\pm$ 0.002 & Gaia DR3\\
G$_{BP}$  \dotfill (mag) & 11.367 $\pm$ 0.005 & 11.930 $\pm$ 0.005 & 12.902 $\pm$ 0.005 & 11.576 $\pm$ 0.005 &  11.652 $\pm$ 0.005 & 12.232 $\pm$ 0.006 & 11.669 $\pm$ 0.005 & Gaia DR3\\
G$_{RP}$  \dotfill (mag) & 10.446 $\pm$ 0.003 & 10.866 $\pm$ 0.003 & 11.854 $\pm$ 0.003 & 10.944 $\pm$ 0.003 &  11.938 $\pm$ 0.003 & 11.250 $\pm$ 0.003 & 10.861 $\pm$ 0.003 & Gaia DR3\\
J  \dotfill (mag)        &  9.90  $\pm$ 0.02  & 10.18  $\pm$ 0.02  & 11.17  $\pm$ 0.02  & 10.58  $\pm$ 0.03  &  10.47  $\pm$ 0.002 & 10.63 $\pm$ 0.03 & 10.32 $\pm$ 0.03& 2MASS$^d$\\
H  \dotfill (mag)        &  9.55  $\pm$ 0.03  &  9.77  $\pm$ 0.02  & 10.80  $\pm$ 0.03  & 10.31  $\pm$ 0.03  &  10.23 $\pm$ 0.002  & 10.19 $\pm$ 0.03 & 10.04 $\pm$ 0.02& 2MASS\\
K$_s$  \dotfill (mag)    &  9.46  $\pm$ 0.03  &  9.66  $\pm$ 0.02  & 10.67  $\pm$ 0.02  & 10.26  $\pm$ 0.02  &  10.16 $\pm$ 0.003  & 10.16 $\pm$ 0.03 & 9.98 $\pm$ 0.02& 2MASS\\
\hline
\teff  \dotfill (K)                 & 5680 $\pm$ 80          & 5360 $\pm$ 70             &  5300 $\pm$ 60                      & 6500 $\pm$ 100                 & 6270 $\pm$ 100          & 5460 $\pm$ 80          & 6100 $\pm$ 100& This work\\
\logg \dotfill (dex)                & 4.46 $\pm$ 0.02        & 4.56 $\pm$ 0.02           &  4.56 $\pm$ 0.02                    & 4.18 $\pm$ 0.02                & 4.23 $\pm$ 0.02         & 4.54 $\pm$ 0.02        & 4.51 $\pm$ 0.02& This work\\
\feh \dotfill (dex)                 & +0.26 $\pm$ 0.05       & +0.16 $\pm$ 0.05          &  +0.06 $\pm$ 0.04                   & -0.11 $\pm$ 0.05               & +0.06 $\pm$ 0.05        & +0.04 $\pm$ 0.05       & -0.20 $\pm$ 0.05& This work\\
\vsini \dotfill (km s$^{-1}$)       & 3.5 $\pm$ 0.3          & 3.6 $\pm$ 0.3             &  2.9 $\pm$ 0.3                      & 7.6 $\pm$ 0.4                  & 7.5 $\pm$ 0.4           & 4.6 $\pm$ 0.3          & 3.8 $\pm$ 0.3& This work\\
\mstar \dotfill (\msun)             & 1.05 $\pm$ 0.03        & $0.91\pm0.02$             &  $0.88_{-0.027}^{+0.025}$           & $1.24 \pm 0.02$                & $1.21\pm0.03$           & $0.91\pm0.03$          & $0.99^{+0.02}_{-0.03}$ & This work\\
\rstar \dotfill (\rsun)             & 1.00 $\pm$ 0.01        & $0.83\pm0.007$            &  $0.810\pm0.009$                    & $1.50\pm0.02$                  & $1.41\pm0.03$           & $0.85\pm0.01$          & $0.919\pm0.008$  & This work\\
L$_{\star}$ \dotfill (L$_{\odot}$)  & $0.92_{-0.03}^{+0.04}$ & $0.48\pm0.02$             &  $0.45\pm0.02$                      & $3.6\pm0.02$                   & $2.7\pm0.02$            & $0.56\pm0.02$          & $0.99\pm0.04$ & This work\\
A$_{V}$ \dotfill (mag)              & $0.09_{-0.05}^{+0.06}$ & $0.09\pm0.05$             &  $0.08_{-0.06}^{+0.07}$             & $0.09_{-0.05}^{+0.07}$         & $0.14\pm0.09$           & $0.12^{+0.06}_{-0.07}$ & $0.16^{+0.05}_{-0.06}$& This work\\
Age \dotfill (Gyr)                  & $2.7_{-1.6}^{+1.9}$    & $2.7_{-1.8}^{+2.5}$       &  $2.8_{-2}^{+3}$                    & $2.8\pm0.4$                    & $3.2_{-0.7}^{+0.8}$     & $3.8^{+3.0}_{-2.0}$    & $1.2^{+1.7}_{-0.8}$& This work\\
$\rho_\star$ \dotfill (g cm$^{-3}$) & $1.50_{-0.09}^{+0.08}$ & $2.22\pm0.09$             &  $2.3\pm0.1$                        & $0.52\pm0.03$                  & $0.61\pm0.03$           & $2.1\pm0.1$            & $1.81^{+0.06}_{-0.08}$& This work\\
\enddata
\tablecomments{$^a$The stellar parameters computed in this work do not consider possible systematic differences among different stellar evolutionary models \citep{tayar2022} and have underestimated uncertainties. $^b$We do not include TOI-2179 here as the values reported in \citet{Schlecker2020} were derived following the same methodology described in Section \ref{sec:stellar}. $^c$\citet{gaia:dr3}. $^d$\citet{2mass}}.
\end{deluxetable*}
\end{rotatetable*}
\clearpage

\section{Priors \& Posteriors}\label{app:posteriors}


\begin{deluxetable*}{llcc}
\tablecaption{Summary of priors and resulting posteriors of the joint fit for K2-139 b. \label{tab:K2-139}}
\tablewidth{70pt}
\tabletypesize{\scriptsize}
\tablehead{Parameter & Description & Prior & Posterior}
\startdata 
$\psi$ & True 3D obliquity (deg) & - & $21.8_{-6.3}^{+10.2}$\\
$\lambda$ & Sky-projected obliquity (deg) & $U(-180,180)$ & $-14.5_{-2.8}^{+2.5}$\\
$v\sin{i_\star}$ & Projected rotational velocity (km/s) & - & $2.5\pm0.2$\\
\mstar & Stellar Mass (\msun)& $N(0.92,0.03)$ & $0.90\pm0.03$\\
\rstar & Stellar Radius (\rsun)& $N(0.849,0.008)$ & $0.852\pm0.008$\\
$\rho_\star$ & Stellar density (g/cm$^3$)& - & $2.06_{-0.08}^{+0.09}$\\
$P_{\rm rot}$ & Stellar Rotation Period (days) & $N(17.0,1.7)$ & $16.4_{-1.1}^{+1.3}$\\
$\cos{i_\star}$ & Cosine of Stellar Inclination & $U(0,1)$ & $0.3\pm0.2$\\
$P$ & Orbital period (days) & $N(28.38266,0.00001)$ & $28.382674\pm0.000005$\\
$t_0$ & Transit midpoint (BJD) & $N(2457297.43901,0.0009)$ & $2457297.4370\pm0.0002$\\
$b$ & Impact parameter & $U(0,1)$ & $0.39_{-0.04}^{+0.03}$\\
$R_p/R_\star$ & Radius ratio & $U(0,1)$ & $0.098\pm0.001$\\
$e$ & Eccentricity & 0 (fixed) & - \\
$\omega$ & Argument of periastron (deg) & 90 (fixed) & - \\
$K$ & RV semiamplitude (m/s) & $U(0,1000)$ & $24\pm3$\\
$a/R_\star$ & Scaled semimajor axis & - & $44.4\pm0.6$\\
$i$ & Orbital inclination (deg) & - & $89.5\pm0.05$\\
$a$ & Semimajor axis (au) & - & $0.176\pm0.002$ \\
$R_p$ & Planet radius ($R_J$) & - & $0.82\pm0.01$ \\
$M_p$ & Planet mass ($M_J$) & - & $0.34\pm0.04$ \\
$\rho_p$ & Planet mean density (g/cm$^{3}$) & - & $0.8\pm0.1$ \\
\hline
$q_1^{\rm ESPRESSO}$ & ESPRESSO linear limb darkening parameter & $U(0,1)$ & $0.89_{-0.13}^{+0.08}$\\
$q_2^{\rm ESPRESSO}$ & ESPRESSO quadratic limb darkening parameter & $U(0,1)$ & $0.64_{-0.15}^{+0.14}$\\
$\beta_{\rm ESPRESSO}$ & Intrinsic stellar line width (km/s) & $N(4,2)$ & $3.4_{-1.3}^{+1.8}$\\
$\gamma_{\rm ESPRESSO}$ & ESPRESSO RV offset (m/s) & $U(-31400,-31300)$ & $-31368.3\pm0.4$\\
$\sigma_{\rm ESPRESSO2}$ & ESPRESSO RV jitter (m/s) & $LU(10^{-3},100)$ & $0.4_{-0.3}^{+0.4}$\\
$\gamma_{\rm FIES}$ & FIES RV offset (m/s) & $U(-31400,-31200)$ & $-31356\pm4$\\
$\sigma_{\rm FIES}$ & FIES RV jitter (m/s) & $LU(10^{-3},100)$ & $0.17_{-0.17}^{+4.27}$\\
$\gamma_{\rm HARPS-N}$ & HARPS-N RV offset (m/s) & $U(-31200,-31000)$ & $-31186\pm3$\\
$\sigma_{\rm HARPS-N}$ & HARPS-N RV jitter (m/s) & $LU(10^{-3},100)$ & $0.1_{-0.1}^{+3.4}$\\
$\gamma_{\rm HARPS}$ & HARPS RV offset (m/s) & $U(-31200,-31000)$ & $-31192_{-5}^{+6}$\\
$\sigma_{\rm HARPS}$ & HARPS RV jitter (m/s) & $LU(10^{-3},100)$ & $13.5_{-3.6}^{+5.9}$\\
\hline
$q_1^{\rm Kepler}$ & Kepler linear limb darkening parameter & $U(0,1)$ &  $0.93_{-0.09}^{+0.05}$\\
$q_2^{\rm Kepler}$ & Kepler quadratic limb darkening parameter & $U(0,1)$ & $0.07_{-0.04}^{+0.05}$\\
$q_1^{R_c}$ & $R_c$ linear limb darkening parameter & $U(0,1)$ & $0.78_{-0.20}^{+0.15}$\\
$q_2^{R_c}$ & $R_c$ quadratic limb darkening parameter & $U(0,1)$ &  $0.26_{-0.15}^{+0.19}$\\
$q_1^{\rm CHEOPS}$ & CHEOPS linear limb darkening parameter & $U(0,1)$ & $0.29_{-0.06}^{+0.07}$\\
$q_2^{\rm CHEOPS}$ & CHEOPS quadratic limb darkening parameter & $U(0,1)$ &  $0.65_{-0.14}^{+0.15}$\\
$\sigma_{\rm K2}$ & K2 photometric jitter (ppm) & $LU(1,5\times10^7)$ &  $155_{-9}^{+11}$\\
$\sigma_{\rm Rc}^{\rm EVANS/ES}$ & EVANS/ES $R_c$ photometric jitter (ppm)& $LU(1,5\times10^7)$ & 
 $19_{-17}^{+116}$\\
 $\sigma_{\rm CHEOPS}$ & CHEOPS photometric jitter (ppm) & $LU(1,5\times10^7)$ &  $1085_{-27}^{+26}$\\
\enddata
\tablecomments{$U(a,b)$ denotes a uniform prior with a start value $a$ and end value $b$. $N(m,\sigma)$ denotes a normal prior with mean $m$, and standard deviation $\sigma$. $LU(a,b)$ denotes a log-uniform prior with a start value $a$ and end value $b$.}
\end{deluxetable*}


\begin{deluxetable*}{llcc}
\tablecaption{Summary of priors and resulting posteriors of the joint fit for K2-329 A b. \label{tab:K2-329}}
\tablewidth{70pt}
\tabletypesize{\scriptsize}
\tablehead{Parameter & Description & Prior & Posterior}
\startdata 
$\psi$ & True 3D obliquity (deg) & - & $33.3_{-8.3}^{+7.9}$\\
$\lambda$ & Sky-projected obliquity (deg) & $U(-180,180)$ & $26.7_{-8.0}^{+8.5}$\\
$v\sin{i_\star}$ & Projected rotational velocity (km/s) & - & $1.6\pm0.1$\\
\mstar & Stellar Mass (\msun)& $N(0.88,0.03)$ & $0.88\pm0.03$\\
\rstar & Stellar Radius (\rsun)& $N(0.81,0.009)$ & $0.81\pm0.009$\\
$\rho_\star$ & Stellar density (g/cm$^3$)& - & $2.34\pm0.1$\\
$P_{\rm rot}$ & Stellar Rotation Period (days) & $N(24.9,2.5)$ & $24.5_{-1.8}^{+1.7}$\\
$\cos{i_\star}$ & Cosine of Stellar Inclination & $U(0,1)$ & $0.3\pm0.2$\\
$P$ & Orbital period (days) & $N(12.455112,0.000009)$ & $12.455109\pm0.000004$\\
$t_0$ & Transit midpoint (BJD) & $N(2457735.7921,0.0003)$ & $2457735.7921\pm0.0001$\\
$b$ & Impact parameter & $U(0,1)$ & $0.29_{-0.07}^{+0.06}$\\
$R_p/R_\star$ & Radius ratio & $U(0,1)$ & $0.096\pm0.001$\\
$e$ & Eccentricity & $U(0,0.95)$ & $0.08\pm0.03$\\
$\omega$ & Argument of periastron (deg) & $U(0,360)$ &  $151_{-21}^{+15}$\\
$K$ & RV semiamplitude (m/s) & $U(0,1000)$ & $24_{-2}^{+1}$\\
$a/R_\star$ & Scaled semimajor axis & - & $26.8\pm0.4$\\
$i$ & Orbital inclination (deg) & - & $89.35_{-0.13}^{+0.15}$\\
$a$ & Semimajor axis (au) & - &  $0.101\pm0.001$\\
$R_p$ & Planet radius ($R_J$) & - &  $0.75\pm0.01$\\
$M_p$ & Planet mass ($M_J$) & - &  $0.25\pm0.02$\\
$\rho_p$ & Planet mean density (g/cm$^{3}$) & - &  $0.71\pm0.06$\\
\hline
$q_1^{\rm ESPRESSO}$ & ESPRESSO linear limb darkening parameter & $U(0,1)$ & $0.70_{-0.18}^{+0.19}$\\
$q_2^{\rm ESPRESSO}$ & ESPRESSO quadratic limb darkening parameter & $U(0,1)$ & $0.69_{-0.25}^{+0.20}$\\
$\beta_{\rm ESPRESSO}$ & Intrinsic stellar line width (km/s) & $N(5.2,2.0)$ & $5.6\pm1.9$\\
$\gamma_{\rm ESPRESSO}$ & ESPRESSO RV offset (m/s) & $U(-17000,-16970)$ & $-16980.5\pm0.9$\\
$\sigma_{\rm ESPRESSO1}$ & ESPRESSO1 RV jitter (m/s) & $LU(10^{-3},100)$ & $1.2\pm0.3$\\
$\sigma_{\rm ESPRESSO2}$ & ESPRESSO2 RV jitter (m/s) & $LU(10^{-3},100)$ & $0.8_{-0.7}^{+1.2}$\\
$\gamma_{\rm PFS}$ & PFS RV offset (m/s) & $U(-50,50)$ & $1\pm1$\\
$\sigma_{\rm PFS}$ & PFS RV jitter (m/s) & $LU(10^{-3},100)$ & $3.3_{-1.0}^{+1.5}$\\
$\gamma_{\rm FEROS}$ & FEROS RV offset (m/s) & $U(-17100,-16900)$ & $-17000\pm4$\\
$\sigma_{\rm FEROS}$ & FEROS RV jitter (m/s) & $LU(10^{-3},100)$ & $16.1_{-2.6}^{+3.0}$\\
$\gamma_{\rm HARPS}$ & HARPS RV offset (m/s) & $U(-17100,-16900)$ & $-16983\pm2$\\
$\sigma_{\rm HARPS}$ & HARPS RV jitter (m/s) & $LU(10^{-3},100)$ & $0.1_{-0.1}^{+1.9}$\\
\hline
$q_1^{\rm Kepler}$ & Kepler linear limb darkening parameter & $U(0,1)$ &  $0.21_{-0.05}^{+0.06}$\\
$q_2^{\rm Kepler}$ & Kepler quadratic limb darkening parameter & $U(0,1)$ & $0.6_{-0.12}^{+0.14}$\\
$q_1^{\rm TESS}$ & TESS linear limb darkening parameter & $U(0,1)$ & $0.12_{-0.08}^{+0.16}$ \\
$q_2^{\rm TESS}$ & TESS quadratic limb darkening parameter & $U(0,1)$ & $0.37_{-0.27}^{+0.39}$\\
$q_1^{R_c}$ & $R_c$ linear limb darkening parameter & $U(0,1)$ & $0.71_{-0.22}^{+0.20}$\\
$q_2^{R_c}$ & $R_c$ quadratic limb darkening parameter & $U(0,1)$ & $0.43_{-0.19}^{+0.25}$\\
$\sigma_{\rm K2}^{\rm S12}$ & K2 S12 photometric jitter (ppm) & $LU(1,5\times10^7)$ & $198_{-14}^{+16}$\\
$\sigma_{\rm K2}^{\rm S19}$ & K2 S19 photometric jitter (ppm)& $LU(1,5\times10^7)$ & $160_{-100}^{+54}$\\
$\sigma_{\rm TESS}^{\rm Y4}$ & TESS Year 4 photometric jitter (ppm)& $LU(1,5\times10^7)$ & $17_{-14}^{+137}$\\
$\sigma_{R_c}^{\rm PEST}$ & PEST $R_c$ photometric jitter (ppm)& $LU(1,5\times10^7)$ & $2216_{-382}^{+382}$\\
\enddata
\tablecomments{$U(a,b)$ denotes a uniform prior with a start value $a$ and end value $b$. $N(m,\sigma)$ denotes a normal prior with mean $m$, and standard deviation $\sigma$. $LU(a,b)$ denotes a log-uniform prior with a start value $a$ and end value $b$.}
\end{deluxetable*}


\begin{deluxetable*}{llcc}
\tablecaption{Summary of priors and resulting posteriors of the joint fit for WASP-106 b. \label{tab:WASP-106}}
\tablewidth{70pt}
\tabletypesize{\scriptsize}
\tablehead{Parameter & Description & Prior & Posterior}
\startdata 
$\lambda$ & Sky-projected obliquity (deg) & $U(-180,180)$ & $1\pm12$\\
$v\sin{i_\star}$ & Projected rotational velocity (km/s) & $U(0,15)$ & $6.7\pm0.6$\\
$P$ & Orbital period (days) & $N(9.289709,0.000007)$ & $9.289710\pm0.000004$\\
$t_0$ & Transit midpoint (BJD) & $N(2458535.3607,0.0007)$ & $2458535.3606\pm0.0004$\\
$\rho_\star$ & Stellar density (g/cm$^3$)& $N(0.62,0.04)$ & $0.64_{-0.03}^{+0.02}$\\
$b$ & Impact parameter & $U(0,1)$ & $0.09_{-0.07}^{+0.14}$\\
$R_p/R_\star$ & Radius ratio & $U(0,1)$ & $0.075\pm0.001$\\
$e$ & Eccentricity & 0 (fixed) & -\\
$\omega$ & Argument of periastron (deg) & 90 (fixed) & - \\
$K$ & RV semiamplitude (m/s) & $U(0,1000)$ & $161\pm3$\\
$a/R_\star$ & Scaled semimajor axis & - & $14.3_{-0.3}^{+0.1}$\\
$i$ & Orbital inclination (deg) & - & $89.7_{-0.6}^{+0.3}$\\
$a$ & Semimajor axis (au) & - &  $0.094\pm0.002$\\
$R_p$ & Planet radius ($R_J$) & - & $1.03\pm0.02$\\
$M_p$ & Planet mass ($M_J$) & - & $1.91_{-0.05}^{+0.06}$\\
$\rho_p$ & Planet mean density (g/cm$^{3}$) & - &  $2.2\pm0.1$\\
\hline
$q_1^{\rm ESPRESSO}$ & ESPRESSO linear limb darkening parameter & $U(0,1)$ & $0.49_{-0.12}^{+0.15}$\\
$q_2^{\rm ESPRESSO}$ & ESPRESSO quadratic limb darkening parameter & $U(0,1)$ & $0.73_{-0.28}^{+0.19}$\\
$\beta_{\rm ESPRESSO}$ & Intrinsic stellar line width (km/s) & $N(5.9,2.0)$ & $5.2_{-1.9}^{+2.0}$\\
$\gamma_{\rm ESPRESSO}$ & ESPRESSO RV offset (m/s) & $U(17000,17500)$ & $17135.5\pm0.6$\\
$\sigma_{\rm ESPRESSO}$ & ESPRESSO RV jitter (m/s) & $LU(10^{-3},100)$ & $1.3_{-1.2}^{+0.6}$\\
$\gamma_{\rm CORALIE}$ & CORALIE RV offset (m/s) & $U(17000,17500)$ & $17248\pm4$\\
$\sigma_{\rm CORALIE}$ & CORALIE RV jitter (m/s) & $LU(10^{-3},100)$ & $0.1_{-0.1}^{+1.7}$\\
$\gamma_{\rm SOPHIE}$ & SOPHIE RV offset (m/s) & $U(17000,17500)$ & $17188\pm6$\\
$\sigma_{\rm SOPHIE}$ & SOPHIE RV jitter (m/s) & $LU(10^{-3},100)$ & $0.1_{-0.1}^{+3.7}$\\
$\gamma_{\rm HARPS}$ & HARPS RV offset (m/s) & $U(17000,17500)$ & $17248\pm4$\\
$\sigma_{\rm HARPS}$ & HARPS RV jitter (m/s) & $LU(10^{-3},100)$ & $12.2_{-2.8}^{+4.2}$\\
\hline
$q_1^{\rm TESS}$ & TESS linear limb darkening parameter & $U(0,1)$ &  $0.11_{-0.05}^{+0.08}$\\
$q_2^{\rm TESS}$ & TESS quadratic limb darkening parameter & $U(0,1)$ & $0.34_{-0.21}^{+0.32}$\\
$q_1^{r'}$ & $r'$ linear limb darkening parameter & $U(0,1)$ & $0.44_{-0.19}^{+0.25}$\\
$q_2^{r'}$ & $r'$ quadratic limb darkening parameter & $U(0,1)$ & $0.29_{-0.19}^{+0.31}$\\
$\sigma_{\rm TESS}^{\rm Y1}$ & TESS Year 1 photometric jitter (ppm) & $LU(1,5\times10^7)$ & $15_{-13}^{+82}$\\
$\sigma_{\rm TESS}^{\rm Y3}$ & TESS Year 3 photometric jitter (ppm)& $LU(1,5\times10^7)$ & $13_{-11}^{+67}$\\
$\sigma_{\rm TESS}^{\rm Y4}$ & TESS Year 4 photometric jitter (ppm)& $LU(1,5\times10^7)$ & $12_{-10}^{+61}$\\
$\sigma_{r'}^{\rm MOANA/ES}$ & MOANA/ES $r'$ photometric jitter (ppm)& $LU(1,5\times10^7)$ & $1661_{-117}^{+116}$\\
\enddata
\tablecomments{$U(a,b)$ denotes a uniform prior with a start value $a$ and end value $b$. $N(m,\sigma)$ denotes a normal prior with mean $m$, and standard deviation $\sigma$. $LU(a,b)$ denotes a log-uniform prior with a start value $a$ and end value $b$.}
\end{deluxetable*}


\begin{deluxetable*}{llcc}
\tablecaption{Summary of priors and resulting posteriors of the joint fit for WASP-130 b. \label{tab:WASP-130}}
\tablewidth{70pt}
\tabletypesize{\scriptsize}
\tablehead{Parameter & Description & Prior & Posterior}
\startdata 
$\lambda$ & Sky-projected obliquity (deg) & $U(-180,180)$ & $-5.8_{-2.2}^{+2.1}$\\
$v\sin{i_\star}$ & Projected rotational velocity (km/s) & $U(0,15)$ & $2.0_{-0.2}^{+0.1}$\\
$P$ & Orbital period (days) & $N(11.550968,0.000004)$ & $11.550968\pm0.000003$\\
$t_0$ & Transit midpoint (BJD) & $N(2458596.0348,0.0004)$ & $2458596.0348\pm0.0002$\\
$\rho_\star$ & Stellar density (g/cm$^3$)& $N(1.5,0.1)$ & $1.5\pm0.1$\\
$b$ & Impact parameter & $U(0,1)$ & $0.55\pm0.03$\\
$R_p/R_\star$ & Radius ratio & $U(0,1)$ & $0.097\pm0.001$\\
$e$ & Eccentricity & 0 (fixed) & -\\
$\omega$ & Argument of periastron (deg) & 90 (fixed) & - \\
$K$ & RV semiamplitude (m/s) & $U(0,1000)$ & $102\pm2$\\
$a/R_\star$ & Scaled semimajor axis & - & $22.1\pm0.4$\\
$i$ & Orbital inclination (deg) & - & $88.6\pm0.1$\\
$a$ & Semimajor axis (au) & - &  $0.103\pm0.002$\\
$R_p$ & Planet radius ($R_J$) & - & $0.95\pm0.02$\\
$M_p$ & Planet mass ($M_J$) & - & $1.17\pm0.03$\\
$\rho_p$ & Planet mean density (g/cm$^{3}$) & - &  $1.7\pm0.1$\\
\hline
$q_1^{\rm ESPRESSO}$ & ESPRESSO linear limb darkening parameter & $U(0,1)$ & $0.88_{-0.12}^{+0.08}$\\
$q_2^{\rm ESPRESSO}$ & ESPRESSO quadratic limb darkening parameter & $U(0,1)$ & $0.71_{-0.19}^{+0.16}$\\
$\beta_{\rm ESPRESSO}$ & Intrinsic stellar line width (km/s) & $N(4.7,2.0)$ & $4.9\pm1.9$\\
$\gamma_{\rm ESPRESSO}$ & ESPRESSO RV offset (m/s) & $U(1300,1400)$ & $1356.8\pm0.3$\\
$\sigma_{\rm ESPRESSO}$ & ESPRESSO RV jitter (m/s) & $LU(10^{-3},100)$ & $0.8_{-0.4}^{+0.3}$\\
$\gamma_{\rm CORALIE1}$ & CORALIE1 RV offset (m/s) & $U(1400,1600)$ & $1462\pm3$\\
$\sigma_{\rm CORALIE1}$ & CORALIE1 RV jitter (m/s) & $LU(10^{-3},100)$ & $7.4_{-7.2}^{+4.2}$\\
$\gamma_{\rm CORALIE2}$ & CORALIE2 RV offset (m/s) & $U(1400,1600)$ & $1494\pm5$\\
$\sigma_{\rm CORALIE2}$ & CORALIE2 RV jitter (m/s) & $LU(10^{-3},100)$ & $12.3_{-3.8}^{+5.0}$\\
$\gamma_{\rm HARPS}$ & HARPS RV offset (m/s) & $U(1400,1700)$ & $1502\pm2$\\
$\sigma_{\rm HARPS}$ & HARPS RV jitter (m/s) & $LU(10^{-3},100)$ & $5.6_{-1.3}^{+1.6}$\\
\hline
$q_1^{\rm TESS}$ & TESS linear limb darkening parameter & $U(0,1)$ & $0.29_{-0.10}^{+0.13}$ \\
$q_2^{\rm TESS}$ & TESS quadratic limb darkening parameter & $U(0,1)$ & $0.25_{-0.16}^{+0.25}$\\
$q_1^{r'}$ & $r'$ linear limb darkening parameter & $U(0,1)$ & $0.35_{-0.08}^{+0.12}$\\
$q_2^{r'}$ & $r'$ quadratic limb darkening parameter & $U(0,1)$ & $0.76_{-0.24}^{+0.17}$\\
$\sigma_{\rm TESS}^{\rm Y1}$ & TESS Year 1 photometric jitter (ppm) & $LU(1,5\times10^7)$ & $15_{-13}^{+93}$\\
$\sigma_{\rm TESS}^{\rm Y3}$ & TESS Year 3 photometric jitter (ppm)& $LU(1,5\times10^7)$ & $14_{-11}^{+66}$\\
$\sigma_{\rm TESS}^{\rm Y5}$ & TESS Year 5 photometric jitter (ppm)& $LU(1,5\times10^7)$ & $13_{-10}^{+56}$\\
$\sigma_{r'}^{\rm MOANA/ES}$ & MOANA/ES $r'$ photometric jitter (ppm)& $LU(1,5\times10^7)$ & $17_{-14}^{+103}$\\
\enddata
\tablecomments{$U(a,b)$ denotes a uniform prior with a start value $a$ and end value $b$. $N(m,\sigma)$ denotes a normal prior with mean $m$, and standard deviation $\sigma$. $LU(a,b)$ denotes a log-uniform prior with a start value $a$ and end value $b$.}
\end{deluxetable*}


\begin{deluxetable*}{llcc}
\tablecaption{Summary of priors and resulting posteriors of the joint fit for TOI-558 b. \label{tab:TOI-558}}
\tablewidth{70pt}
\tabletypesize{\scriptsize}
\tablehead{Parameter & Description & Prior & Posterior}
\startdata 
$\lambda$ & Sky-projected obliquity (deg) & $U(-180,180)$ & $11.7_{-4.8}^{+7.5}$\\
$v\sin{i_\star}$ & Projected rotational velocity (km/s) & $U(0,15)$ & $5.5_{-1.6}^{+1.8}$\\
$P$ & Orbital period (days) & $N(14.57408,0.00002)$ & $14.57409\pm0.00001$\\
$t_0$ & Transit midpoint (BJD) & $N(2458346.4068,0.0009)$ & $2458346.4067\pm0.0005$\\
$\rho_\star$ & Stellar density (g/cm$^3$)& $N(0.52,0.03)$ & $0.55\pm0.03$\\
$b$ & Impact parameter & $U(0,1)$ & $0.92\pm0.01$\\
$R_p/R_\star$ & Radius ratio & $U(0,1)$ & $0.075_{-0.002}^{+0.003}$\\
$e$ & Eccentricity & $U(0,0.95)$ & $0.31\pm0.02$\\
$\omega$ & Argument of periastron (deg) & $U(0,360)$ & $131\pm3$ \\
$K$ & RV semiamplitude (m/s) & $U(0,1000)$ & $257_{-6}^{+5}$\\
$a/R_\star$ & Scaled semimajor axis & - & $18.3\pm0.3$\\
$i$ & Orbital inclination (deg) & - & $86.1\pm0.1$\\
$a$ & Semimajor axis (au) & - & $0.128\pm0.003$ \\
$R_p$ & Planet radius ($R_J$) & - &  $1.09\pm0.04$\\
$M_p$ & Planet mass ($M_J$) & - & $3.4\pm0.1$\\
$\rho_p$ & Planet mean density (g/cm$^{3}$) & - &  $3.2_{-0.4}^{+0.3}$\\
\hline
$q_1^{\rm ESPRESSO}$ & ESPRESSO linear limb darkening parameter & $U(0,1)$ & $0.45_{-0.31}^{+0.35}$\\
$q_2^{\rm ESPRESSO}$ & ESPRESSO quadratic limb darkening parameter & $U(0,1)$ & $0.46_{-0.32}^{+0.36}$\\
$\beta_{\rm ESPRESSO}$ & Intrinsic stellar line width (km/s) & $N(6.6,2.0)$ & $7.0\pm1.9$\\
$\gamma_{\rm ESPRESSO}$ & ESPRESSO RV offset (m/s) & $U(29000,29500)$ & $29300\pm4$\\
$\sigma_{\rm ESPRESSO}$ & ESPRESSO RV jitter (m/s) & $LU(10^{-3},100)$ & $0.04_{-0.04}^{+0.55}$\\
$\gamma_{\rm PFS}$ & PFS RV offset (m/s) & $U(-500,500)$ & $-59\pm4$\\
$\sigma_{\rm PFS}$ & PFS RV jitter (m/s) & $LU(10^{-3},100)$ & $13_{-3}^{+4}$\\
\hline
$q_1^{\rm TESS}$ & TESS linear limb darkening parameter & $U(0,1)$ & $0.22_{-0.10}^{+0.13}$\\
$q_2^{\rm TESS}$ & TESS quadratic limb darkening parameter & $U(0,1)$ & $0.49_{-0.33}^{+0.34}$\\
$q_1^{i'}$ & $i'$ linear limb darkening parameter & $U(0,1)$ & $0.79_{-0.20}^{+0.15}$\\
$q_2^{i'}$ & $i'$ quadratic limb darkening parameter & $U(0,1)$ & $0.19_{-0.14}^{+0.22}$\\
$q_1^{z'}$ & $z'$ linear limb darkening parameter & $U(0,1)$ & $0.15_{-0.09}^{+0.14}$\\
$q_2^{z'}$ & $z'$ quadratic limb darkening parameter & $U(0,1)$ & $0.39_{-0.28}^{+0.38}$\\
$q_1^{B}$ & $B$ linear limb darkening parameter & $U(0,1)$ & $0.31_{-0.13}^{+0.19}$\\
$q_2^{B}$ & $B$ quadratic limb darkening parameter & $U(0,1)$ & $0.45_{-0.32}^{+0.36}$\\
$\sigma_{\rm TESS}^{\rm Y1}$ & TESS Year 1 photometric jitter (ppm) & $LU(1,5\times10^7)$ & $18_{-15}^{+94}$\\
$\sigma_{\rm TESS}^{\rm Y3}$ & TESS Year 3 photometric jitter (ppm)& $LU(1,5\times10^7)$ & $13_{-11}^{+58}$\\
$\sigma_{i'}^{\rm MOANA/ES}$ & MOANA/ES $i'$ photometric jitter (ppm)& $LU(1,5\times10^7)$ & $1722_{-141}^{+145}$\\
$\sigma_{i'}^{\rm LCOGT/CTIO}$ & LCOGT/CTIO $i'$ photometric jitter (ppm)& $LU(1,5\times10^7)$ & $1062_{-106}^{+106}$\\
$\sigma_{r'}^{\rm LCOGT/SAAO}$ & LCOGT/SAAO $r'$ photometric jitter (ppm)& $LU(1,5\times10^7)$ & $589_{-130}^{+112}$\\
$\sigma_{B}^{\rm LCOGT/CTIO}$ & LCOGT/CTIO $B$ photometric jitter (ppm)& $LU(1,5\times10^7)$ & $1254_{-101}^{+104}$\\
\enddata
\tablecomments{$U(a,b)$ denotes a uniform prior with a start value $a$ and end value $b$. $N(m,\sigma)$ denotes a normal prior with mean $m$, and standard deviation $\sigma$. $LU(a,b)$ denotes a log-uniform prior with a start value $a$ and end value $b$.}
\end{deluxetable*}


\begin{deluxetable*}{llcc}
\tablecaption{Summary of priors and resulting posteriors of the joint fit for TOI-2179 b. \label{tab:TOI-2179}}
\tablewidth{70pt}
\tabletypesize{\scriptsize}
\tablehead{Parameter & Description & Prior & Posterior}
\startdata 
$\lambda$ & Sky-projected obliquity (deg) & $U(-180,180)$ & $22.9_{-44.9}^{+84.9}$\\
$v\sin{i_\star}$ & Projected rotational velocity (km/s) & $U(0,15)$ & $0.7_{-0.5}^{+0.9}$\\
$P$ & Orbital period (days) & $N(15.168873,0.000008)$ & $15.168873\pm0.000005$\\
$t_0$ & Transit midpoint (BJD) & $N(2458319.1504,0.0006)$ & $2458319.1504\pm0.0004$\\
$\rho_\star$ & Stellar density (g/cm$^3$)& $N(1.12,0.11)$ & $1.08\pm0.08$\\
$b$ & Impact parameter & $U(0,1)$ & $0.91\pm0.01$\\
$R_p/R_\star$ & Radius ratio & $U(0,1)$ & $0.099_{-0.003}^{+0.004}$\\
$e$ & Eccentricity & $U(0,0.95)$ & $0.57\pm0.01$\\
$\omega$ & Argument of periastron (deg) & $U(0,360)$ & $22\pm2$\\
$K$ & RV semiamplitude (m/s) & $U(0,1000)$ & $191\pm6$\\
$a/R_\star$ & Scaled semimajor axis & - & $23.6\pm0.6$\\
$i$ & Orbital inclination (deg) & - & $86.0\pm0.2$\\
$a$ & Semimajor axis (au) & - &  $0.120\pm0.003$\\
$R_p$ & Planet radius ($R_J$) & - & $1.05_{-0.05}^{+0.04}$\\
$M_p$ & Planet mass ($M_J$) & - & $1.93\pm0.08$\\
$\rho_p$ & Planet mean density (g/cm$^{3}$) & - &  $2.1_{-0.3}^{+0.2}$\\
$\dot{\gamma}$ & RV linear trend (m/s/days) & $U(0,1)$ & $-0.09\pm0.04$\\
$\ddot{\gamma}$ & RV quadratic trend (m/s/days$^2$) & $U(0,1)$ & $0.00008\pm0.00002$\\
\hline
$q_1^{\rm ESPRESSO}$ & ESPRESSO linear limb darkening parameter & $U(0,1)$ & $0.60_{-0.38}^{+0.29}$\\
$q_2^{\rm ESPRESSO}$ & ESPRESSO quadratic limb darkening parameter & $U(0,1)$ & $0.58_{-0.37}^{+0.30}$\\
$\beta_{\rm ESPRESSO}$ & Intrinsic stellar line width (km/s) & $N(4.5,2.0)$ & $4.6_{-1.9}^{+2.0}$\\
$\gamma_{\rm ESPRESSO}$ & ESPRESSO RV offset (m/s) & $U(29400,30000)$ & $29537\pm17$\\
$\sigma_{\rm ESPRESSO}$ & ESPRESSO RV jitter (m/s) & $LU(10^{-3},100)$ & $2.8_{-1.3}^{+1.2}$\\
$\gamma_{\rm FEROS}$ & FEROS RV offset (m/s) & $U(29400,30000)$ & $29604\pm6$\\
$\sigma_{\rm FEROS}$ & FEROS RV jitter (m/s) & $LU(10^{-3},100)$ & $13.9_{-2.5}^{+3.1}$\\
\hline
$q_1^{\rm TESS}$ & TESS linear limb darkening parameter & $U(0,1)$ & $0.24_{-0.12}^{+0.20}$\\
$q_2^{\rm TESS}$ & TESS quadratic limb darkening parameter & $U(0,1)$ & $0.41_{-0.27}^{+0.34}$\\
$q_1^{i'}$ & $i'$ linear limb darkening parameter & $U(0,1)$ & $0.89_{-0.10}^{+0.07}$\\
$q_2^{i'}$ & $i'$ quadratic limb darkening parameter & $U(0,1)$ & $0.87_{-0.13}^{+0.09}$\\
$q_1^{r'}$ & $r'$ linear limb darkening parameter & $U(0,1)$ & $0.18_{-0.11}^{+0.19}$\\
$q_2^{r'}$ & $r'$ quadratic limb darkening parameter & $U(0,1)$ & $0.35_{-0.25}^{+0.36}$\\
$\sigma_{\rm TESS}^{\rm Y1}$ & TESS Year 1 photometric jitter (ppm) & $LU(1,5\times10^7)$ & $11_{-9}^{+57}$\\
$\sigma_{\rm TESS}^{\rm Y3}$ & TESS Year 3 photometric jitter (ppm)& $LU(1,5\times10^7)$ & $14_{-11}^{+66}$\\
$\sigma_{\rm TESS}^{\rm Y5}$ & TESS Year 5 photometric jitter (ppm)& $LU(1,5\times10^7)$ & $13_{-11}^{+65}$\\
$\sigma_{r'}^{\rm MOANA/ES}$ & MOANA/ES $r'$ photometric jitter (ppm)& $LU(1,5\times10^7)$ & $3205_{-117}^{+120}$\\
$\sigma_{i'}^{\rm LCOGT}$ & LCOGT $i'$ photometric jitter (ppm)& $LU(1,5\times10^7)$ & $476_{-78}^{+83}$\\
$\sigma_{i'}^{\rm CHAT}$ & CHAT $i'$ photometric jitter (ppm)& $LU(1,5\times10^7)$ & $1586_{-137}^{+152}$\\
\enddata
\tablecomments{$U(a,b)$ denotes a uniform prior with a start value $a$ and end value $b$. $N(m,\sigma)$ denotes a normal prior with mean $m$, and standard deviation $\sigma$. $LU(a,b)$ denotes a log-uniform prior with a start value $a$ and end value $b$.}
\end{deluxetable*}


\begin{deluxetable*}{llcc}
\tablecaption{Summary of priors and resulting posteriors of the joint fit for TOI-4515 b. \label{tab:TOI-4515}}
\tablewidth{70pt}
\tabletypesize{\scriptsize}
\tablehead{Parameter & Description & Prior & Posterior}
\startdata 
$\psi$ & True 3D obliquity (deg) & - & $16_{-11}^{+16}$\\
$\lambda$ & Sky-projected obliquity (deg) & $U(-180,180)$ & $2.6_{-1.6}^{+1.7}$\\
$v\sin{i_\star}$ & Projected rotational velocity (km/s) & - & $2.6\pm0.3$\\
\mstar & Stellar Mass (\msun)& $N(0.91,0.03)$ & $0.93\pm0.03$\\
\rstar & Stellar Radius (\rsun)& $N(0.85,0.01)$ & $0.843\pm0.009$\\
$\rho_\star$ & Stellar density (g/cm$^3$)& - & $2.19\pm0.09$\\
$P_{\rm rot}$ & Stellar Rotation Period (days) & $N(15.6,1.6)$ & $15.3\pm1.4$\\
$\cos{i_\star}$ & Cosine of Stellar Inclination & $U(0,1)$ & $0.3_{-0.2}^{+0.3}$\\
$P$ & Orbital period (days) & $N(15.26645,0.00001)$ & $15.266449\pm0.000007$\\
$t_0$ & Transit midpoint (BJD) & $N(2458764.6321,0.0007)$ & $2458764.6322\pm0.0004$\\
$b$ & Impact parameter & $U(0,1)$ & $0.79\pm0.01$\\
$R_p/R_\star$ & Radius ratio & $U(0,1)$ & $0.131\pm0.001$\\
$e$ & Eccentricity & $U(0,0.95)$ &  $0.46\pm0.01$\\
$\omega$ & Argument of periastron (deg) & $U(0,360)$ &  $172\pm2$\\
$K$ & RV semiamplitude (m/s) & $U(0,1000)$ & $191_{-4}^{+3}$\\
$a/R_\star$ & Scaled semimajor axis & - & $30.0\pm0.4$\\
$i$ & Orbital inclination (deg) & - & $87.98\pm0.05$\\
$a$ & Semimajor axis (au) & - & $0.118\pm0.001$ \\
$R_p$ & Planet radius ($R_J$) & - &  $1.08\pm0.02$\\
$M_p$ & Planet mass ($M_J$) & - &  $1.97\pm0.06$\\
$\rho_p$ & Planet mean density (g/cm$^{3}$) & - &  $2.0\pm0.1$\\
\hline
$q_1^{\rm ESPRESSO}$ & ESPRESSO linear limb darkening parameter & $U(0,1)$ & $0.74_{-0.22}^{+0.17}$\\
$q_2^{\rm ESPRESSO}$ & ESPRESSO quadratic limb darkening parameter & $U(0,1)$ & $0.49\pm0.31$\\
$\beta_{\rm ESPRESSO}$ & Intrinsic stellar line width (km/s) & $N(4.1,2.0)$ & $3.3_{-1.5}^{+1.9}$\\
$\gamma_{\rm ESPRESSO}$ & ESPRESSO RV offset (m/s) & $U(12600,13300)$ & $12974_{-3}^{+2}$\\
$\sigma_{\rm ESPRESSO}$ & ESPRESSO RV jitter (m/s) & $LU(10^{-3},100)$ & $0.9_{-0.5}^{+0.6}$\\
$\gamma_{\rm FEROS}$ & FEROS RV offset (m/s) & $U(12600,13300)$ & $13062_{-25}^{+24}$\\
$\sigma_{\rm FEROS}$ & FEROS RV jitter (m/s) & $LU(10^{-3},100)$ & $52_{-15}^{+23}$\\
$\gamma_{\rm HARPS-N}$ & HARPS-N RV offset (m/s) & $U(12600,13300)$ & $13106\pm3$\\
$\sigma_{\rm HARPS-N}$ & HARPS-N RV jitter (m/s) & $LU(10^{-3},100)$ & $13.2_{-1.9}^{+2.6}$\\
$\gamma_{\rm TRES}$ & TRES RV offset (m/s) & $U(-400,200)$ & $-41\pm8$\\
$\sigma_{\rm TRES}$ & TRES RV jitter (m/s) & $LU(10^{-3},100)$ & $1.3_{-1.3}^{+19.7}$\\
\hline
$q_1^{\rm TESS}$ & TESS linear limb darkening parameter & $U(0,1)$ & $0.20_{-0.07}^{+0.09}$\\
$q_2^{\rm TESS}$ & TESS quadratic limb darkening parameter & $U(0,1)$ & $0.74_{-0.29}^{+0.18}$\\
$q_1^{i'}$ & $i'$ linear limb darkening parameter & $U(0,1)$ & $0.17_{-0.11}^{+0.20}$\\
$q_2^{i'}$ & $i'$ quadratic limb darkening parameter & $U(0,1)$ &  $0.25_{-0.19}^{+0.35}$\\
$q_1^{\rm g'}$ & $g'$ linear limb darkening parameter & $U(0,1)$ & $0.42_{-0.24}^{+0.30}$\\
$q_2^{\rm g'}$ & $g'$ quadratic limb darkening parameter & $U(0,1)$ &  $0.15_{-0.11}^{+0.25}$\\
$\sigma_{\rm TESS}^{\rm Y2}$ & TESS Year 2 photometric jitter (ppm) & $LU(1,5\times10^7)$ & $14_{-11}^{+64}$ \\
$\sigma_{\rm TESS}^{\rm Y4}$ & TESS Year 4 photometric jitter (ppm) & $LU(1,5\times10^7)$ &  $18_{-15}^{+93}$\\
$\sigma_{\rm TESS}^{\rm Y5}$ & TESS Year 5 photometric jitter (ppm) & $LU(1,5\times10^7)$ &  $19_{-16}^{+94}$\\
$\sigma_{\rm i'}^{\rm KeplerCam}$ & KeplerCam $i'$ photometric jitter (ppm)& $LU(1,5\times10^7)$ & 
 $6014_{-192}^{+199}$\\
 $\sigma_{\rm g'}^{\rm LCOGT/CTIO}$ & LCOGT/CTIO $g'$ photometric jitter (ppm) & $LU(1,5\times10^7)$ & $3664_{-327}^{+362}$ \\
\enddata
\tablecomments{$U(a,b)$ denotes a uniform prior with a start value $a$ and end value $b$. $N(m,\sigma)$ denotes a normal prior with mean $m$, and standard deviation $\sigma$. $LU(a,b)$ denotes a log-uniform prior with a start value $a$ and end value $b$.}
\end{deluxetable*}


\begin{deluxetable*}{llcc}
\tablecaption{Summary of priors and resulting posteriors of the joint fit for TOI-5027 b. \label{tab:TOI-5027}}
\tablewidth{70pt}
\tabletypesize{\scriptsize}
\tablehead{Parameter & Description & Prior & Posterior}
\startdata 
$\lambda$ & Sky-projected obliquity (deg) & $U(-180,180)$ & $5.9_{-6.1}^{+8.3}$\\
$v\sin{i_\star}$ & Projected rotational velocity (km/s) & $U(0,15)$ & $1.7_{-0.7}^{+0.8}$\\
$P$ & Orbital period (days) & $N(10.243682,0.000009)$ & $10.243680\pm0.000006$\\
$t_0$ & Transit midpoint (BJD) & $N(2458639.215,0.001)$ & $2458639.2141\pm0.0006$\\
$\rho_\star$ & Stellar density (g/cm$^3$)& $N(1.8,0.7)$ & $1.81\pm0.07$\\
$b$ & Impact parameter & $U(0,1)$ & $0.91\pm0.02$\\
$R_p/R_\star$ & Radius ratio & $U(0,1)$ & $0.105_{-0.004}^{+0.005}$\\
$e$ & Eccentricity & $U(0,0.95)$ & $0.35\pm0.03$\\
$\omega$ & Argument of periastron (deg) & $U(0,360)$ & $290.0_{-4.8}^{+4.4}$ \\
$K$ & RV semiamplitude (m/s) & $U(0,1000)$ & $182\pm9$\\
$a/R_\star$ & Scaled semimajor axis & - & $21.6\pm0.3$\\
$i$ & Orbital inclination (deg) & - & $88.16\pm0.04$\\
$a$ & Semimajor axis (au) & - & $0.092\pm0.001$ \\
$R_p$ & Planet radius ($R_J$) & - & $0.94_{-0.03}^{+0.05}$ \\
$M_p$ & Planet mass ($M_J$) & - & $1.81\pm0.09$\\
$\rho_p$ & Planet mean density (g/cm$^{3}$) & - &  $2.7_{-0.4}^{+0.3}$\\
$\dot{\gamma}$ & RV linear trend (m/s/days) & $U(-1,1)$ & $-0.12\pm0.04$\\
\hline
$q_1^{\rm ESPRESSO}$ & ESPRESSO linear limb darkening parameter & $U(0,1)$ & $0.58_{-0.37}^{+0.30}$\\
$q_2^{\rm ESPRESSO}$ & ESPRESSO quadratic limb darkening parameter & $U(0,1)$ & $0.55_{-0.35}^{+0.31}$\\
$\beta_{\rm ESPRESSO}$ & Intrinsic stellar line width (km/s) & $N(4.7,2.0)$ & $5.0\pm1.9$\\
$\gamma_{\rm ESPRESSO}$ & ESPRESSO RV offset (m/s) & $U(-34700,-34200)$ & $-34449.3_{-31.6}^{+32.3}$\\
$\sigma_{\rm ESPRESSO}$ & ESPRESSO RV jitter (m/s) & $LU(10^{-3},100)$ & $0.6_{-0.6}^{+1.3}$\\
$\gamma_{\rm FEROS}$ & FEROS RV offset (m/s) & $U(-34700,-34200)$ & $-34391_{-12}^{+13}$\\
$\sigma_{\rm FEROS}$ & FEROS RV jitter (m/s) & $LU(10^{-3},100)$ & $28.81_{-4.45}^{+5.75}$\\
\hline
$q_1^{\rm TESS}$ & TESS linear limb darkening parameter & $U(0,1)$ & $0.21_{-0.14}^{+0.22}$\\
$q_2^{\rm TESS}$ & TESS quadratic limb darkening parameter & $U(0,1)$ & $0.39_{-0.27}^{+0.35}$\\
$q_1^{i'}$ & $i'$ linear limb darkening parameter & $U(0,1)$ & $0.44_{-0.14}^{+0.17}$\\
$q_2^{i'}$ & $i'$ quadratic limb darkening parameter & $U(0,1)$ & $0.67_{-0.28}^{+0.22}$\\
$q_1^{r'}$ & $r'$ linear limb darkening parameter & $U(0,1)$ & $0.48_{-0.24}^{+0.31}$\\
$q_2^{r'}$ & $r'$ quadratic limb darkening parameter & $U(0,1)$ & $0.29_{-0.21}^{+0.34}$\\
$\sigma_{\rm TESS}^{\rm Y1}$ & TESS Year 1 photometric jitter (ppm) & $LU(1,5\times10^7)$ & $13_{-10}^{+56}$\\
$\sigma_{\rm TESS}^{\rm Y3}$ & TESS Year 3 photometric jitter (ppm)& $LU(1,5\times10^7)$ & $24_{-21}^{+168}$\\
$\sigma_{\rm TESS}^{\rm Y5}$ & TESS Year 5 photometric jitter (ppm)& $LU(1,5\times10^7)$ & $20_{-18}^{+157}$\\
$\sigma_{i'}^{\rm LCOGT/CTIO}$ & LCOGT/CTIO $i'$ photometric jitter (ppm)& $LU(1,5\times10^7)$ & $666_{-57}^{+61}$\\
$\sigma_{r'}^{\rm MOANA/ES}$ & MOANA/ES $r'$ photometric jitter (ppm)& $LU(1,5\times10^7)$ & $3156_{-232}^{+244}$\\
\enddata
\tablecomments{$U(a,b)$ denotes a uniform prior with a start value $a$ and end value $b$. $N(m,\sigma)$ denotes a normal prior with mean $m$, and standard deviation $\sigma$. $LU(a,b)$ denotes a log-uniform prior with a start value $a$ and end value $b$.}
\end{deluxetable*}

\end{document}